\documentclass[letter]{emulateapj}

\usepackage{natbib}
\slugcomment{Accepted to ApJ}

\begin{document}


\title{Numerical Study of Turbulent Mixing Layers with 
Non-Equilibrium Ionization Calculations}


\author{Kyujin Kwak and Robin L. Shelton}
\affil{Department of Physics and Astronomy, University of Georgia, 
Athens, GA 30602}







\begin{abstract}

Highly ionized species such as \ion{C}{4}, \ion{N}{5}, and \ion{O}{6},
are commonly observed in diffuse gas in various places in the universe,
such as in our Galaxy's disk and halo, high velocity clouds (HVCs), external 
galaxies, and the intergalactic medium. These ions are 
often used to trace hot gas whose temperature is a few times $10^5~\mbox{K}$. 
One possible mechanism for producing high ions is 
turbulent mixing of cool gas (such as that in a high or intermediate 
velocity cloud) with hotter (a few times $10^6~\mbox{K}$) 
gas in locations where these gases slide past
each other. By using hydrodynamic simulations with radiative cooling 
and non-equilibrium ionization (NEI) calculations, 
we investigate the physical properties of turbulent mixing layers 
and the production of high ions (\ion{C}{4}, \ion{N}{5}, and \ion{O}{6}). 
We find that most of the mixing occurs on the hot side of the hot/cool 
interface, where denser cool gas is entrained and mixed into the 
hotter, more diffuse gas. Our simulations reveal that the mixed region 
separates into a tepid zone containing 
radiatively cooled, \ion{C}{4}--rich gas and a hotter zone 
which is rich in \ion{C}{4}, \ion{N}{5}, and \ion{O}{6}. 
The hotter zone contains a mixture of low 
and intermediate ions contributed by the 
cool gas and intermediate and high--stage ions contributed by the hot 
gas. Mixing occurs faster than ionization or recombination, 
making the mixed gas a better source of \ion{C}{4}, \ion{N}{5}, and 
\ion{O}{6} in our NEI simulations than in 
our collisional ionization equilibrium (CIE) simulations. 
In addition, the gas radiatively cools faster than the ions recombine, 
which also allows large numbers of \ion{C}{4}, \ion{N}{5}, 
and \ion{O}{6} ions to linger in the NEI simulations. 
For these reasons, our NEI calculations predict more \ion{C}{4}, 
\ion{N}{5}, and \ion{O}{6} than our CIE calculations predict. 
We also simulate various initial configurations and 
find that more \ion{C}{4} is produced when 
the shear speed is smaller or 
the hot gas has a higher temperature. We find no significant 
differences between simulations having different perturbation 
amplitudes in the initial boundary between the hot and cool gas. 
We discuss the results of our simulations, compare with 
observations of the Galactic halo and highly ionized HVCs, 
and compare with other models, including other turbulent mixing 
calculations. The ratios of \ion{C}{4} to \ion{N}{5} and 
\ion{N}{5} to \ion{O}{6} are in reasonable agreement with the 
averages calculated from observations of the halo. There is a 
great deal of variation from sightline to sightline and with time 
in our simulations. Such spatial and temporal variation may 
explain some of the variation seen among observations.  

\end{abstract}





\section{Introduction}

Highly ionized species such as \ion{C}{4}, \ion{N}{5}, and
\ion{O}{6} are often used as tracers 
for diffuse gas within the temperature range 
of $(1-3)\times10^5~\mbox{K}$. These high ions are found by 
absorption line measurements in various places in the 
universe, including the disk \citep{Bowenetal2008ApJS,
Sallmenetal2008ApJ,Savageetal2001ApJS,SavageMassa1987ApJ,
Cowieetal1981ApJ,Jenkins1978ApJa,Jenkins1978ApJb} 
and the halo of the Milky Way 
\citep{Gangulyetal2005ApJS,Savageetal2003ApJS,Zsargoetal2003ApJ,
Sterlingetal2002ApJ,Savageetal2001ApJ,Savageetal1997AJ,
Sembachetal1997ApJ,SavageSembach1994ApJ,SembachSavage1992ApJS}. 
The emission lines of these high 
ions are also found in the Galaxy \citep{Sheltonetal2001ApJ,
Sheltonetal2007ApJ,Dixonetal2006ApJ,OtteDixon2006ApJ,
Korpelaetal2006ApJ,Welshetal2007AA}. 
Most of the observed high ions in the Galactic disk and halo 
are nearly stationary.

There is also a population of high ions moving 
at velocities of a few hundreds of $\mbox{km}~\mbox{s}^{-1}$ with 
respect to the local standard of rest (LSR) 
\citep{Collinsetal2007ApJ,Foxetal2005ApJ,
Foxetal2004ApJ,Sembachetal2003ApJS}. The distances to some of 
these ions are unknown. 
Most of these ions are on sightlines that intersect \ion{H}{1} 
high velocity clouds (HVCs), but some are not. Thus, these latter 
ions are due to ionized HVCs.

High ions are also found in external galaxies via absorption line 
measurements \citep[LMC:][]{LehnerHowk2007MNRAS,
DanforthBlair2006ApJ,Sankritetal2004AJ} and 
\ion{O}{6} emission measurements 
\citep{Bregmanetal2006ApJb,Gangulyetal2006ApJ}. 
High ions are detected not only in nearby galaxies 
but also in damped Lyman-$\alpha$ systems for distant 
galaxies \citep{Foxetal2009AA,Foxetal2007AA} and 
gamma-ray burst (GRB) host galaxies 
\citep{Prochaskaetal2008ApJ,Foxetal2008AA}.
Even the intergalactic medium contains high ions. 
In the nearby intergalactic space, high ions are
detected along sightlines to distant QSO's
\citep{Trippetal2008ApJS}. In the cooling flows of clusters of
galaxies such as Abell 426, Abell 1795, and Abell 2597, high ions are
observed in the flow gas cooling from hotter, X-ray emitting gas 
\citep{Bregmanetal2006ApJ,Oegerleetal2001ApJ,
Dixonetal1996ApJ}.

Observations of \ion{C}{4}, \ion{N}{5}, and \ion{O}{6} 
in various places in the universe indicate 
that these ions are very commonly produced. 
How they are produced and what their existence implies about 
the local physical conditions have been longstanding questions. 
For example, in order to trace the detailed physical 
condition of the plasma in the Galactic disk and halo, 
ratios between the quantities of 
different ions, either their absorption column density or their 
emission intensity, are measured and 
compared with various model predictions 
\citep{GnatSternberg2007ApJS,
IndebetouwShull2004aApJ,IndebetouwShull2004bApJ,
ShullSlavin1994ApJ,SlavinShullBegelman1993ApJ,SlavinCox1992ApJ,
Borkowskietal1990ApJ,EdgarChevalier1986ApJL}.

One of several possible production 
mechanisms for the high ions is 
turbulent mixing in places where hot and cool
gas flow past each other. 
This idea is supported by the fact that the universe is also
rich in hot, X-ray emitting gas and 
that the regions containing high ions 
are sometimes correlated with hotter, X-ray emitting regions. 
Early analytic models of mixing layers were made by 
\citet{BegelmanFabian1990MNRAS}. They estimated the temperature of 
the mixed layer as the geometrical density--weighted 
mean of the hot and cool gas 
temperatures under the assumption that two gases mix due to
turbulence and that mixing is efficient. 
\citet{SlavinShullBegelman1993ApJ} further developed 
\citet{BegelmanFabian1990MNRAS}'s idea and analytically 
calculated the emission spectra and column densities of high ions 
under the assumption that mixing reaches steady state. 
Their calculated ratios between the quantities of 
different ions are often used as diagnostics for observations. 
The results of \citet{SlavinShullBegelman1993ApJ} were tested 
by \citet{Esquiveletal2006ApJ} who used 3-D 
magneto-hydrodynamic (MHD) simulations. They assumed collisional 
ionization equilibrium (CIE) for their column density calculations. 
They simulated the mixing layer on a 10 pc wide computational domain 
for durations up to 3 Myr, but they found that their 
simulations did not reach steady state as 
\citet{SlavinShullBegelman1993ApJ} had assumed. 
Although their calculated ion ratios can differ significantly 
with time and sightline, their ratios are more similar to the 
analytic results of \citet{SlavinShullBegelman1993ApJ} than 
the average ratios calculated for other phenomena such as 
radiative cooling.

The next logical step in the progression is to calculate the ion 
content using non-equilibrium ionization algorithms. In this paper, 
we do that. We run detailed hydrodynamic simulations that include 
non-equilibrium ionization (NEI)
calculations of interesting ions. We also perform CIE calculations 
for comparison. In order to offset the memory demands of the NEI 
calculations, we reduce the spatial dimensions from three to two. 
Our NEI results provide better predictions for high
ion column densities and ion ratios and can be used as updated
diagnostics for comparison with observations.
Our simulations reveal that NEI 
calculations predict more high ions than CIE calculations, although the
ion ratios do not change dramatically. 
We examine the reason for the difference
between NEI and CIE by looking into the distribution of 
ionization levels of interesting atoms (\S \ref{neicie_S}).

We also consider various physical conditions that form turbulent
mixing layers and find that the detailed configuration of the mixing
layer, such as the perturbation amplitude of the initial 
boundary between the hot and 
cool gas, does not affect the ion ratios 
as long as mixing is efficient. 
However, the initial velocity difference between the hot and the 
cool gas and the temperature of the hot gas affect the 
ion ratios. When the initial speed has a slower value 
($50~\mbox{km}~\mbox{s}^{-1}$) and the hot gas has a higher 
temperature ($3\times10^6~\mbox{K}$), the ion ratios show 
different behavior than when the initial speed is 
$100~\mbox{km}~\mbox{s}^{-1}$ and the hot gas is 
$1\times10^6~\mbox{K}$ in temperature. 
These trends may be used as diagnostics that shed light on 
the conditions of the mixing gas (\S \ref{diagnostics_S}).

When compared with analytic turbulent mixing models by 
\citet{SlavinShullBegelman1993ApJ}, our model simulations 
both from NEI and CIE calculations produce somewhat smaller 
ratios of \ion{C}{4} column density to \ion{O}{6} column 
density, N(\ion{C}{4})/N(\ion{O}{6}), and/or somewhat 
larger N(\ion{N}{5})/N(\ion{O}{6}) ratios. Our ion ratios 
are more similar to the CIE simulations done by 
\citet{Esquiveletal2006ApJ} in which radiative cooling 
was allowed. We find that our average simulated ion ratios 
are close to the average values from halo observations. 
Individual observations can vary greatly from the average. 
Some of the variation in the observed ratios may be due to 
line of sight geometry or age of the mixing layer, given that 
the ion ratios calculated for individual sightlines in our 
domain vary greatly with time and sightline location. 
We also compare our simulated ion ratios 
with those observed in Complex~C and find 
that our model simulations (and other turbulent mixing models) are 
more likely to produce the observed high ion ratios in Complex~C 
when the low metallicity of Complex~C is considered.

This paper is organized as follows. The next section provides the numerical
methods and physical parameters used in our simulations. Section 3
explains the results of our numerical study and Section 4 compares 
them with observations of the halo and Complex~C. 
Section 5 presents the summary.

\section{Numerical Methods and Physical Parameters} \label{method_S}

\begin{deluxetable*}{ccccccccccc}

\tablewidth{0pt}
\tabletypesize{\footnotesize}
\tablecaption{Models \label{models_T} }
\tablecolumns{11}

\tablehead{
\colhead{} &  
\multicolumn{2}{c}{Domain} &
\colhead{Maximum} &
\multicolumn{3}{c}{Hot Gas} &
\colhead{Initial Interface} &
\multicolumn{3}{c}{Cool Gas} \\
\colhead{Model} &
\multicolumn{2}{c}{-----------------} &
\colhead{Refinement} &
\multicolumn{3}{c}{--------------------------------} &
\colhead{---------------------} &
\multicolumn{3}{c}{--------------------------------} \\
\colhead{} &
\colhead{$x$} &
\colhead{$y$} &
\colhead{Level} &
\colhead{$n_H$} &
\colhead{T} &
\colhead{$v_x$} &
\colhead{$y=f(x)$} &
\colhead{$n_H$} &
\colhead{T} &
\colhead{$v_x$} \\
\colhead{} & \colhead{(pc)} & \colhead{(pc)} & \colhead{} &
\colhead{($\mbox{cm}^{-3}$)} & \colhead{(K)} & 
\colhead{($\mbox{km}~\mbox{s}^{-1}$)} & \colhead{} & 
\colhead{($\mbox{cm}^{-3}$)} & \colhead{(K)} & 
\colhead{($\mbox{km}~\mbox{s}^{-1}$)}
}

\startdata
A \tablenotemark{a} & [0, 100] & [-250, 50] & 6 & $10^{-4}$ & $10^6$ &
0 & $y=(2.5\mbox{pc})~\sin (\frac{2 \pi x}{100\mbox{pc}})$ & 0.1 &
$10^3$ & 100 \\
B & [0, 100] & [-250, 50] & 7 \tablenotemark{b} & 
$10^{-4}$ & $10^6$ &
0 & $y=(2.5\mbox{pc})~\sin (\frac{2 \pi x}{100\mbox{pc}})$ & 0.1 &
$10^3$ & 100 \\
C & [0, 10] \tablenotemark{c} & [-25, 5] & 6 & $10^{-4}$ & $10^6$ &
0 & $y=(0.25\mbox{pc})~\sin (\frac{2 \pi x}{10\mbox{pc}})$ & 0.1 &
$10^3$ & 100 \\
D & [0, 100] & [-250, 50] & 6 & $10^{-4}$ & $10^6$ &
0 & $y=(2.5\mbox{pc})~\sin (\frac{2 \pi x}{100\mbox{pc}})$ & 0.1 &
$10^3$ & 50 \tablenotemark{d} \\
E & [0, 100] & [-250, 50] & 6 & $10^{-4}$ & $10^6$ & 0 & 
$y=(5.0\mbox{pc})~\sin (\frac{2 \pi x}{100\mbox{pc}})$
\tablenotemark{e} & 0.1 & $10^3$ & 100 \\
F & [0, 100] & [-250, 50] & 6 &
$\frac{1}{3}\times10^{-4}$ & $3\times10^6$ \tablenotemark{f} &
0 & $y=(2.5\mbox{pc})~\sin (\frac{2 \pi x}{100\mbox{pc}})$ & 0.1 &
$10^3$ & 100 
\enddata

\tablenotetext{a}{Reference simulation}
\tablenotetext{b}{Higher spatial resolution by one refinement level
  (factor 2) than model A}
\tablenotetext{c}{The computational domain is 1/10 of model A}
\tablenotetext{d}{Cool gas has half initial speed of model A}
\tablenotetext{e}{Amplitude of initial interface between hot and cool
  gas is twice that of model A}
\tablenotetext{f}{The temperature of the hot gas is three million Kelvin}

\end{deluxetable*}

\begin{figure*}

\centering

\includegraphics[scale=0.8]{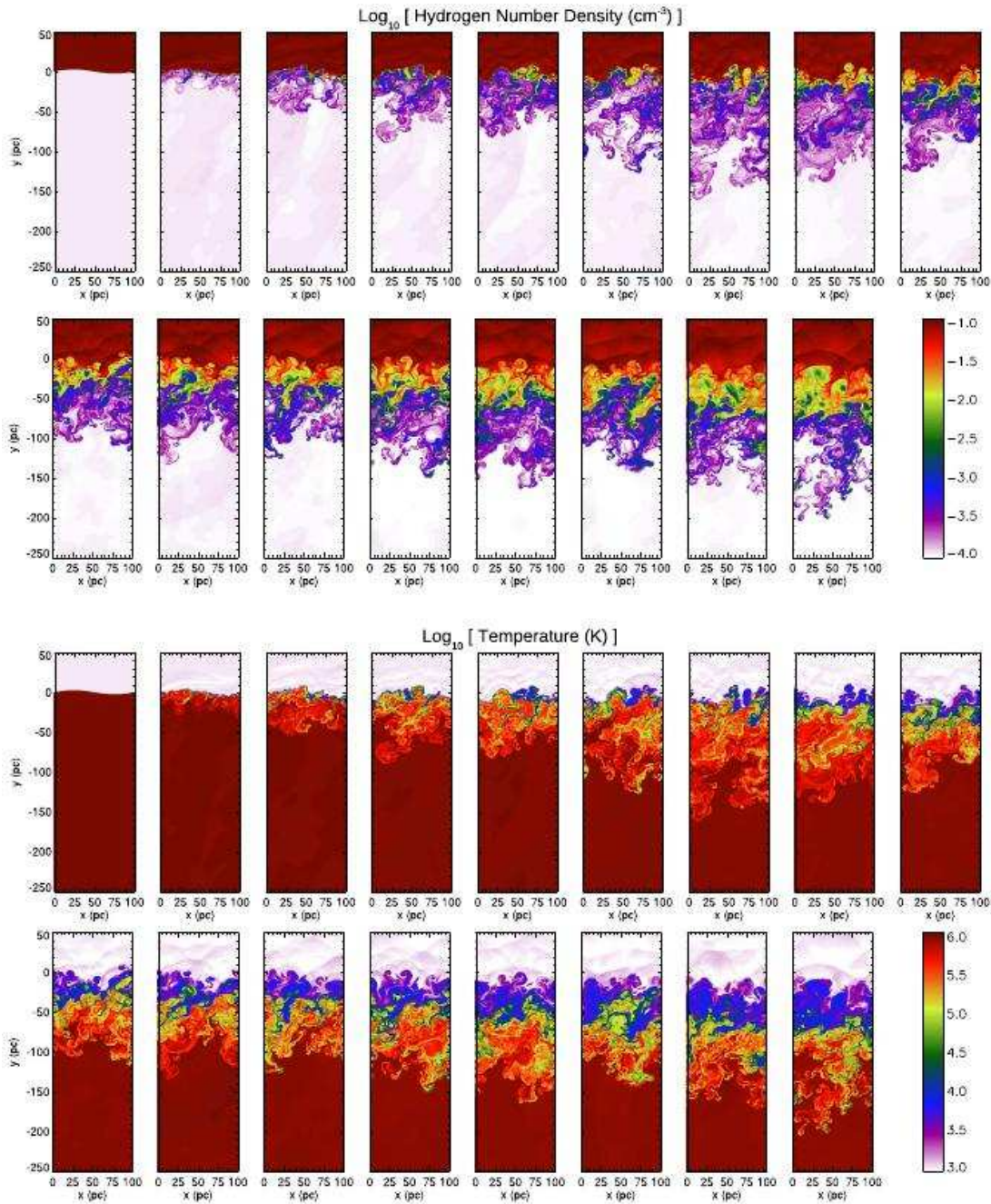}

\caption{Model A: the top two panels depict the log of the 
  hydrogen number density 
  and the bottom two panels depict the log of the 
  temperature. The leftmost plots 
  in the first and third rows show the model at $t=0$ Myr. A time 
  period of 5 Myr elapses between successive plots. 
\label{modelA_dens_temp_fig}}
\end{figure*} 

\begin{figure*}

\figurenum{\ref{modelA_dens_temp_fig}, continued}

\centering

\includegraphics[scale=0.8]{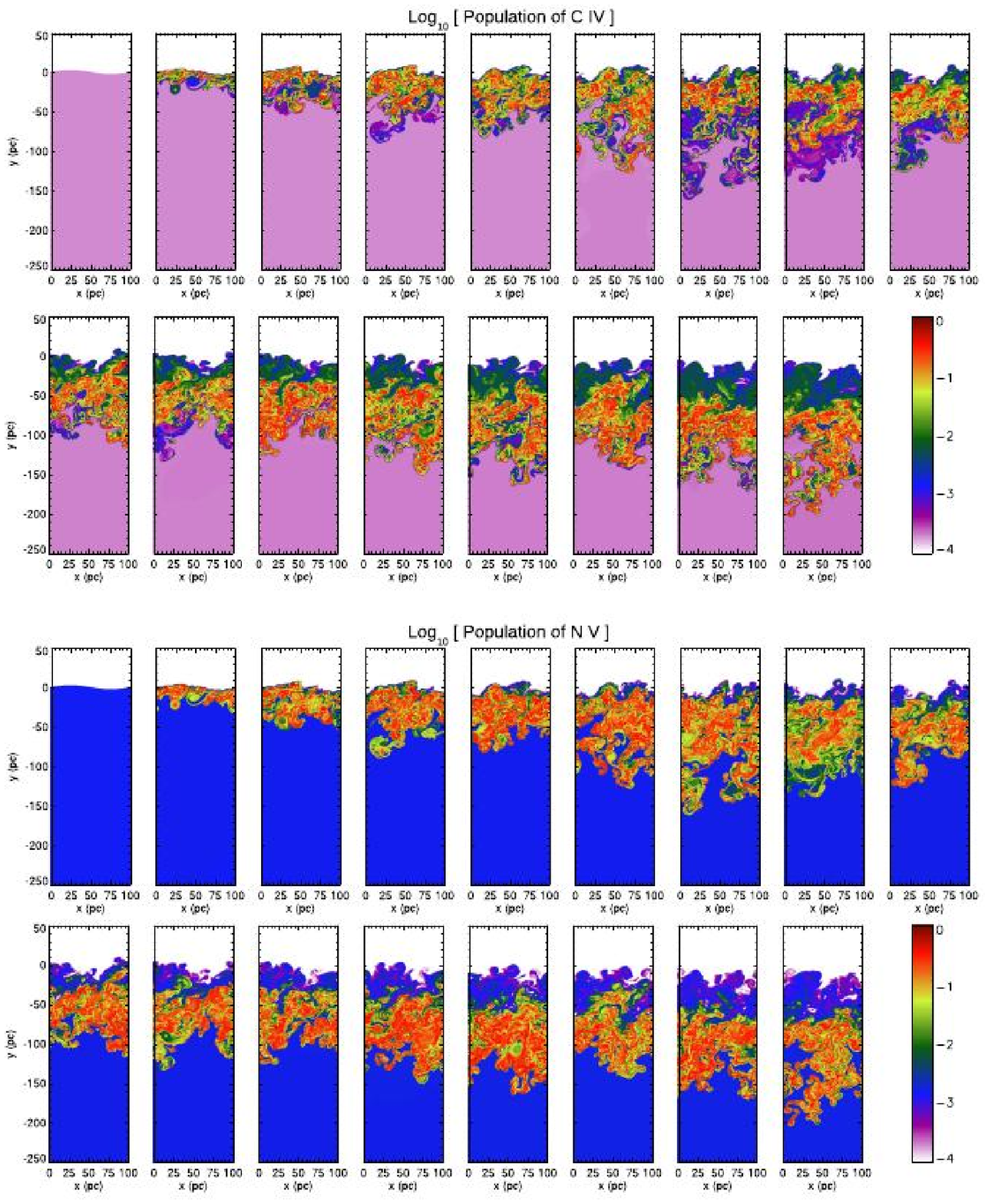}

\caption{Model A: population of \ion{C}{4} (top two panels) 
  and \ion{N}{5} (bottom two panels) in log scale 
  as a function of time obtained from NEI calculations. 
  In these and subsequent plots, the population of \ion{C}{4}, 
  \ion{N}{5}, or \ion{O}{6} refers to the fraction of ions 
  in the specified ionization stage compared to the total number 
  of ions of carbon, nitrogen, or oxygen, respectively.}
\end{figure*} 

\begin{figure*}

\figurenum{\ref{modelA_dens_temp_fig}, continued}

\centering

\includegraphics[scale=0.8]{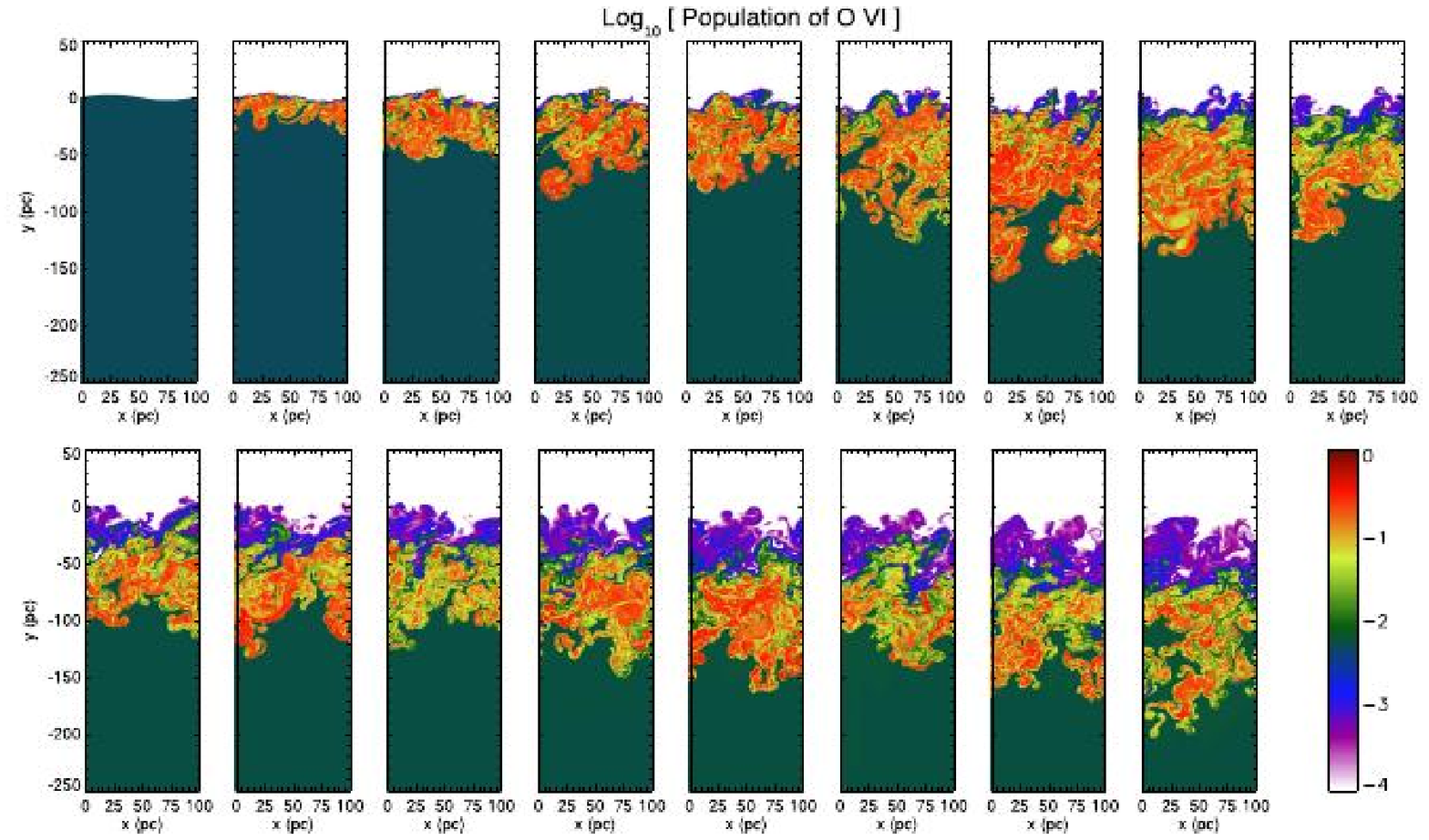}

\caption{Model A: population of \ion{O}{6} in log scale 
  as a function of time obtained from NEI calculations.}

\end{figure*}

We use FLASH version 2.5 for our simulations \citep{Fryxelletal2000ApJS}. 
We include radiative cooling by using relevant modules in FLASH. 
We also use the FLASH NEI module to track 
the degree of ionization of the carbon, 
nitrogen, and oxygen atoms in the gas in each zone in a time dependent 
fashion. In FLASH, each timestep's NEI calculation 
is done in two pieces. In the first piece, which is done as 
a part of the hydrodynamic update, the mass density of each ion is 
updated via the mass conservation equation. 
During the hydrodynamic update, the total mass density and
temperature are also updated.  They will be used as 
inputs for the second piece of the NEI calculation. In the second
piece, which occurs after the hydrodynamic update, the populations 
of the ionization levels within the atoms (of carbon, nitrogen, 
and oxygen) are updated by 
solving sets of ordinary differential equations which include
the ionization and recombination rates for each ion. 
In FLASH, the ionization and recombination rates include the effects 
of collisional ionization, auto-ionization, 
radiative recombination, and dielectronic recombination.

In the FLASH NEI calculations, 
we set the abundances of carbon, nitrogen, and oxygen to be 
consistent with those used in the Raymond and Smith code 
(\citet{RaymondSmith1977ApJS}; downloaded from HEASARC), 
which is used for the CIE calculations. In our CIE calculations, 
the Raymond and Smith code calculates the populations 
of the ionization levels in the atoms in each zone as a function of 
the zone temperature reported by the hydrodynamic simulations. 
The ionization fractions multiplied by the elemental abundance and 
the volume density reported by the hydrodynamic code yield the 
volume densities of ions for each zone. The column densities 
are obtained by integrating the volume densities along sightlines.

It is worth mentioning that our calculations are approximate 
because we do not calculate the effects of NEI ionization levels 
on the radiative energy loss rate. 
Calculating the total radiative loss rate from the sum of 
the loss rates from individual NEI ions would require very large 
computing resources. 
Instead, in our simulations, we use the CIE cooling curve 
calculated from the CIE cooling rates of all of the relevant 
elements. 
Assuming that the plasma radiatively cools 
according to the CIE cooling curve 
at each time step of the simulation allows us 
to save significant computing resources.

In order to verify the validity of this approximation, 
more complete future studies would be required, 
which compare simulations using NEI cooling with 
simulations using CIE cooling. 
The currently available comparison studies between CIE and NEI 
cooling rates show opposite trends depending upon whether 
the gas is in the process of ionizing (because its temperature 
has been raised) or is in the process of recombining 
(because its temperature has fallen). 
\citet{SutherlandDopita1993ApJS} and \citet{GnatSternberg2007ApJS} 
calculated the NEI radiation rates from all relevant ions 
(as well as their ionization states) for cooling gas. 
Both studies showed that the NEI cooling rates are lower than the
CIE cooling rates because the recombination of high stage ions 
in the cooling gas is delayed. 
In contrast, cool gas in the process of heating 
due to external heat sources shows 
delayed ionization and appears to have NEI cooling rates 
that are higher than the CIE cooling rate 
\citep{Gnatetal2010arXiv}. 
In turbulent mixing layers, we can see both delayed recombination 
and delayed ionization in the mixed gas 
(this is verified by our simulations: see \S \ref{neicie_S}). 
Thus the true cooling rate would be a complex combination of 
rates that sometimes exceed and sometimes fall below the CIE 
rate.

We run our simulations in 2-D Cartesian coordinates for two reasons. 
Firstly, the previous 3-D MHD study of \citet{Esquiveletal2006ApJ} 
showed that a 10 pc scale during 3 Myr 
was not long enough to see efficient mixing of two
gases. So, it is necessary to run larger
scale simulations to a later time with significant spatial 
resolution. The mixing zone expands over time, so long duration 
runs must also have larger domains. 
Secondly, tracing ion fractions of carbon, nitrogen, and oxygen 
requires more computing resources. 
We find that 2-D hydrodynamic simulations 
with $100~\mbox{pc} \times 300~\mbox{pc}$ domains running 
for several tens of Myr reveal the 
physical properties of high ions in mixing layers. In order to check 
the validity of our simulations, we also run a simulation with 
a 10 pc wide domain for 8 Myr.

In this paper, we present the results of 6 simulations, labeled 
Models A, B, C, D, E, and F. Model A is our reference simulation; 
each of the other models is made after varying one of Model A's 
parameters and runs in order to test the effect of that parameter. 
In Model A, the computational domain is 100 pc
$\times$ 300 pc and the maximum refinement level is 6. According to 
the FLASH adaptive mesh refinement (AMR) convention, this means 
that if our grid were fully refined, it would have $256 \times 768$ 
zones and the smallest zone would be $(0.4~\mbox{pc})^2$. 
At the beginning of the simulation, the hot gas 
($n_{H}=10^{-4}~\mbox{cm}^{-3}$, $T=10^6~\mbox{K}$) 
occupies the lower 5/6 of the domain 
and the cool gas ($n_{H}=0.1~\mbox{cm}^{-3}$, $T=10^3~\mbox{K}$) 
occupies the 
remainder. The gas is initially in pressure balance with 
$P/k_B=230~\mbox{cm}^{-3}~\mbox{K}$ (considering that 
the number of helium atoms is $10\%$ of that of hydrogen
atoms). 
Note that this low thermal pressure is characteristic of 
the halo but the thermal pressure in the Galactic disk 
is higher by at least an order of magnitude 
\citep{Ferriere1998ApJ,Jenkins2004ApSS,Cox2005ARAA}. 
However, as \citet{SlavinShullBegelman1993ApJ} 
pointed out, the column 
densities are independent of the thermal pressure 
although the emission intensity is proportional to the thermal 
pressure. The thermal pressure would also affect the 
speed at which the turbulent mixing layers develop.

In order to create shear, the cool gas moves to the right (toward the 
positive $x$-direction) at $100~\mbox{km}~\mbox{s}^{-1}$ 
throughout the simulation while the hot gas does not. 
Periodic boundary conditions on the left and right sides and 
outflow boundary conditions on the top and bottom sides enable 
such motion. In order to seed the turbulence, we add curvature 
to the initial boundary between the cool gas and the hot gas.
The shape of our boundary is 
$y=(2.5\mbox{pc})~\sin (\frac{2 \pi x}{100\mbox{pc}})$.

Model B is a higher spatial resolution analog of Model A. It 
has a maximum refinement level of 7, which would correspond 
to $512 \times 1536$ zones if the domain were fully resolved. 
Model C has a smaller computational domain (1/10 in height 
and width), Model D has a smaller initial speed 
difference ($50~\mbox{km}~\mbox{s}^{-1}$) between the cool 
gas and the hot gas, Model E has a larger ripple 
(maximum amplitude = 5 pc) in the boundary between the gases, 
and Model E has hotter ($T=3\times10^6~\mbox{K}$) and less 
dense ($n_H = \onethird \times 10^{-4}~\mbox{cm}^{-3}$) hot gas 
than Model A. Our model parameters are summarized in Table
\ref{models_T}.

\section{Results}





\subsection{Model A: Reference Simulation} \label{ModelA_S}


The sequences of snapshots of Model A presented in Figure 
\ref{modelA_dens_temp_fig} show the mixing zone between the hot and 
cool gas developing and growing in time. Five sequences are 
presented. They portray hydrogen number density, 
temperature, and indicators of the prevalence of 
\ion{C}{4}, \ion{N}{5}, and \ion{O}{6}. 
The first snapshot in each sequence shows the model at $t=0$ Myr, 
when it contains only hot, rarefied gas in the lower portion of the 
grid and cool, denser gas in the upper portion of the grid. 
Between this epoch and $t=5$ Myr (shown as the next snapshot in 
each sequence), the relative motion between the hot and cool 
layers has stirred the gas, creating a zone of mixed, intermediate 
temperature gas between the hot and cool layers. Most of the 
mixing shown in our simulations occurs on the hot side of the 
hot--cool interface, where dense cool gas is entrained and 
mixed into stationary hot, rarefied gas. The entrained gas 
expands due to the raising of its temperature. Whatever hot gas 
becomes entrained into the cool gas compresses as its temperature 
falls and moves off of the grid at roughly 
$100~\mbox{km}~\mbox{s}^{-1} \approx 100~\mbox{pc}~\mbox{Myr}^{-1}$ 
and so appears only fleetingly in the simulations.

When tendrils of cool gas first begin to intrude into and mix 
with the hot gas, the temperature of the entrained gas rises 
immediately to the geometric mean. Its ionization state, however, 
does not equilibrate as quickly and thus the freshly entrained gas 
contains a mixture of poorly ionized gas that is in the process 
of ionizing and highly ionized gas that is in the process of 
recombining. As the next several snapshots (each of which is 
spaced 5 Myr apart) show, the tendrils of intermediate temperature 
mixed gas intrude into the hot gas as time goes by, 
increasing the depth of the mixed zone.

The total depth of the mixed zone grows for the first 30 Myr then 
shrinks slightly, then begins to grow again around 50 Myr. Because 
the timescale of the growth spurts is a large fraction of the 
simulation time, it is not possible to conclude that the mixing 
layer has achieved a steady state, even after 80 Myr. Note that 
individual cool clouds in the Galaxy 
should be much shorter than 8 kpc, the 
distance through which the hot and cool material in Model A 
slide past each other in 80 Myr. Thus, carrying out the simulations 
to larger times would not be realistic.

Mixing plays an important role with regards to radiative cooling. 
Intermediate temperature ($T\sim10^5~\mbox{K}$) mixed gas cools 
far faster than hotter gas of the same density. Thus the mixing 
process seeds the hot gas with what it needs (cool gas) to cool 
quickly. Radiative cooling in the $T=1.5\times10^5~\mbox{K}$, 
$n=6.8\times10^{-4}~\mbox{cm}^{-3}$ mixed gas (such as that beneath 
the interface at $t=5$ Myr) occurs on the timescale 
$t_{cool} = \frac{3}{2} \frac{kT}{n\Lambda(T)} \approx 2~\mbox{Myr}$ 
(where $\Lambda(T)$ is the cooling coefficient). As a result, this 
gas quickly cools to $\sim 10^4 ~\mbox{K}$, 
the temperature at which cooling 
ceases to be applied by FLASH. As time progresses, neighboring 
mixed gas cools and joins it to make an ever--deepening layer of 
$T\sim10^4~\mbox{K}$ gas. This layer contains, but is not especially 
rich in \ion{C}{4} ions. It also contains \ion{N}{5} and \ion{O}{6} 
ions, but in smaller portions to their atomic abundances. The hotter 
mixed gas is richer in all three of these ions.

\subsubsection{Calculation of Column Densities}

\begin{figure*}
\centering

\includegraphics[scale=0.24]{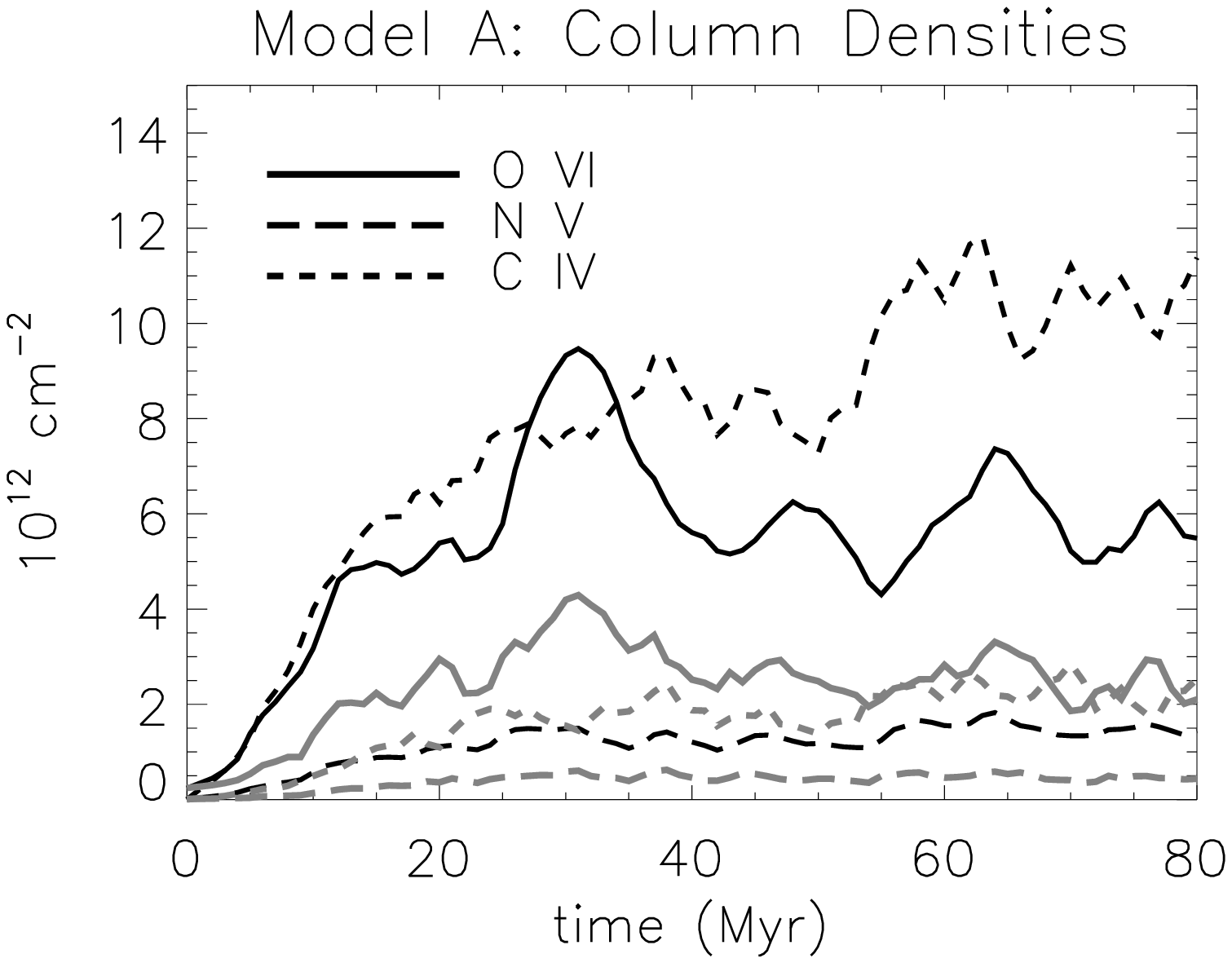}
\includegraphics[scale=0.24]{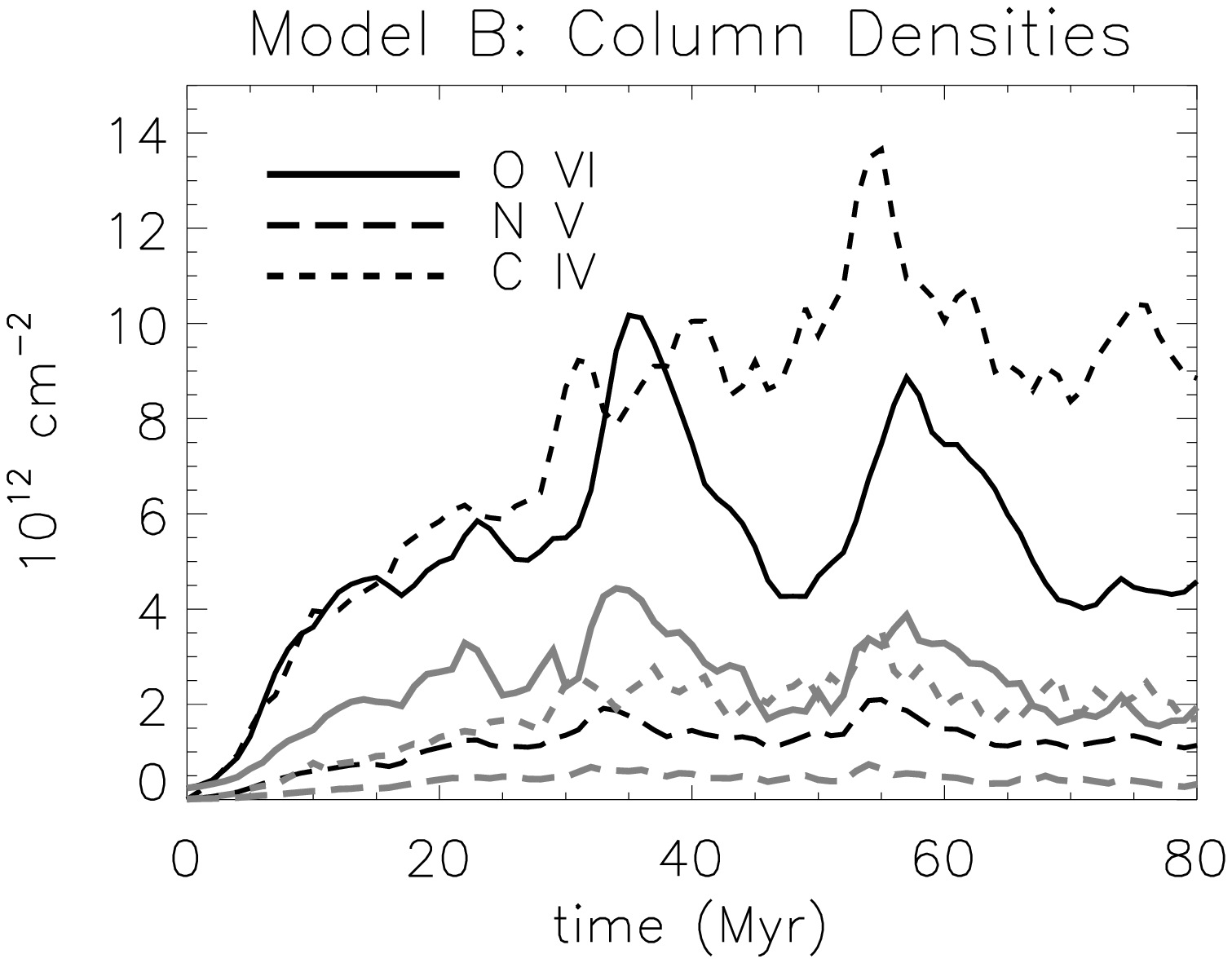}
\includegraphics[scale=0.24]{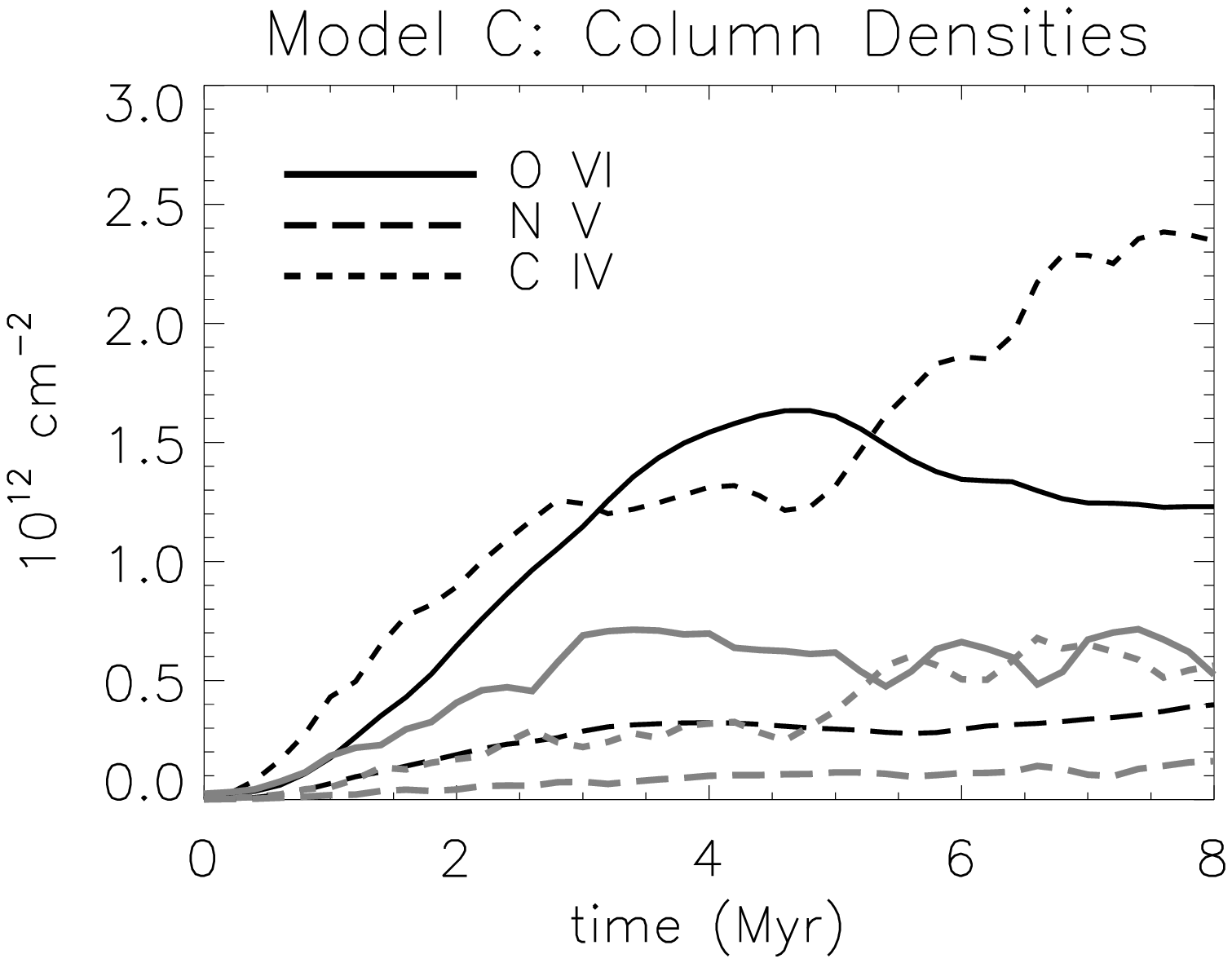} \\
\includegraphics[scale=0.24]{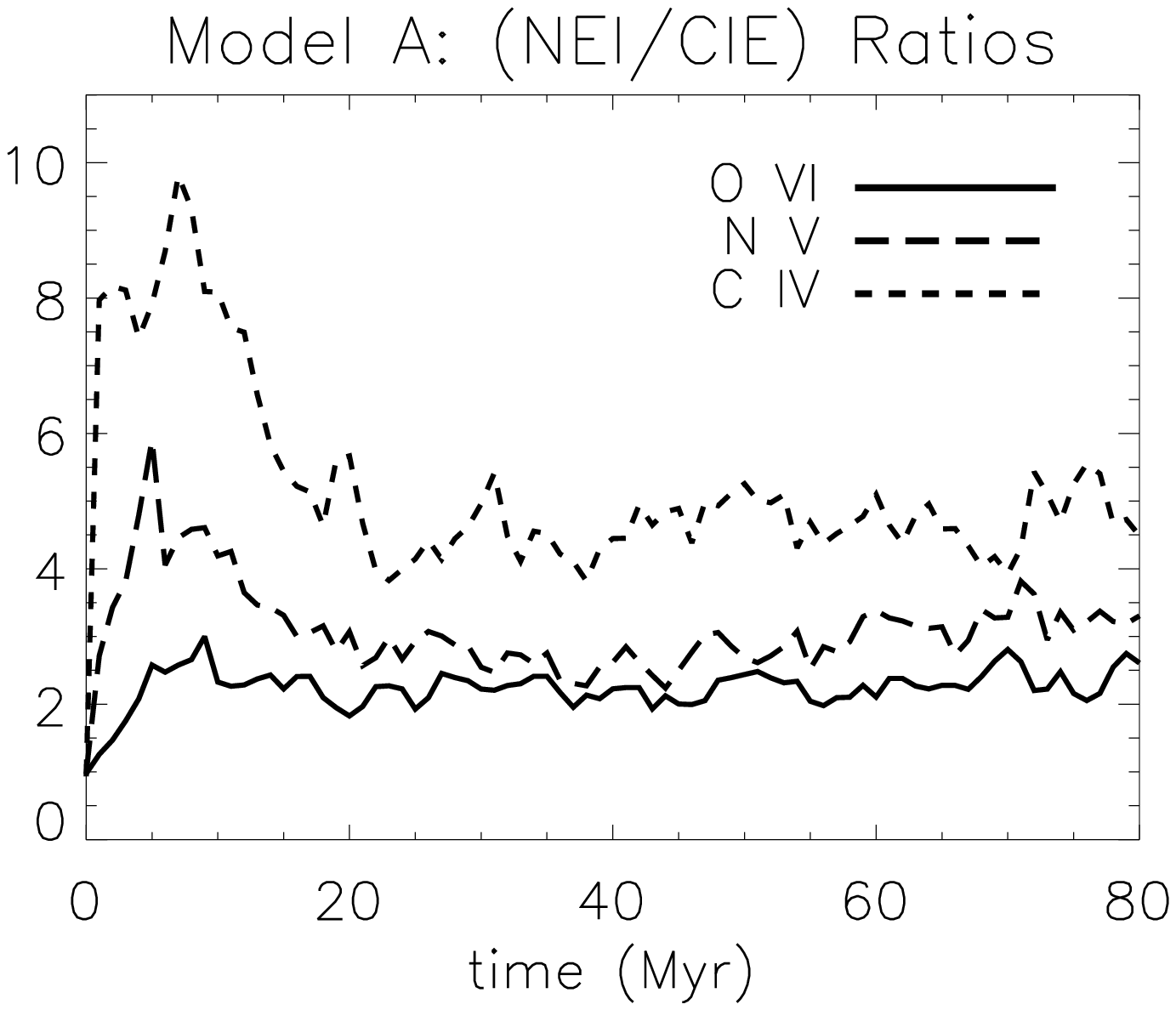}
\includegraphics[scale=0.24]{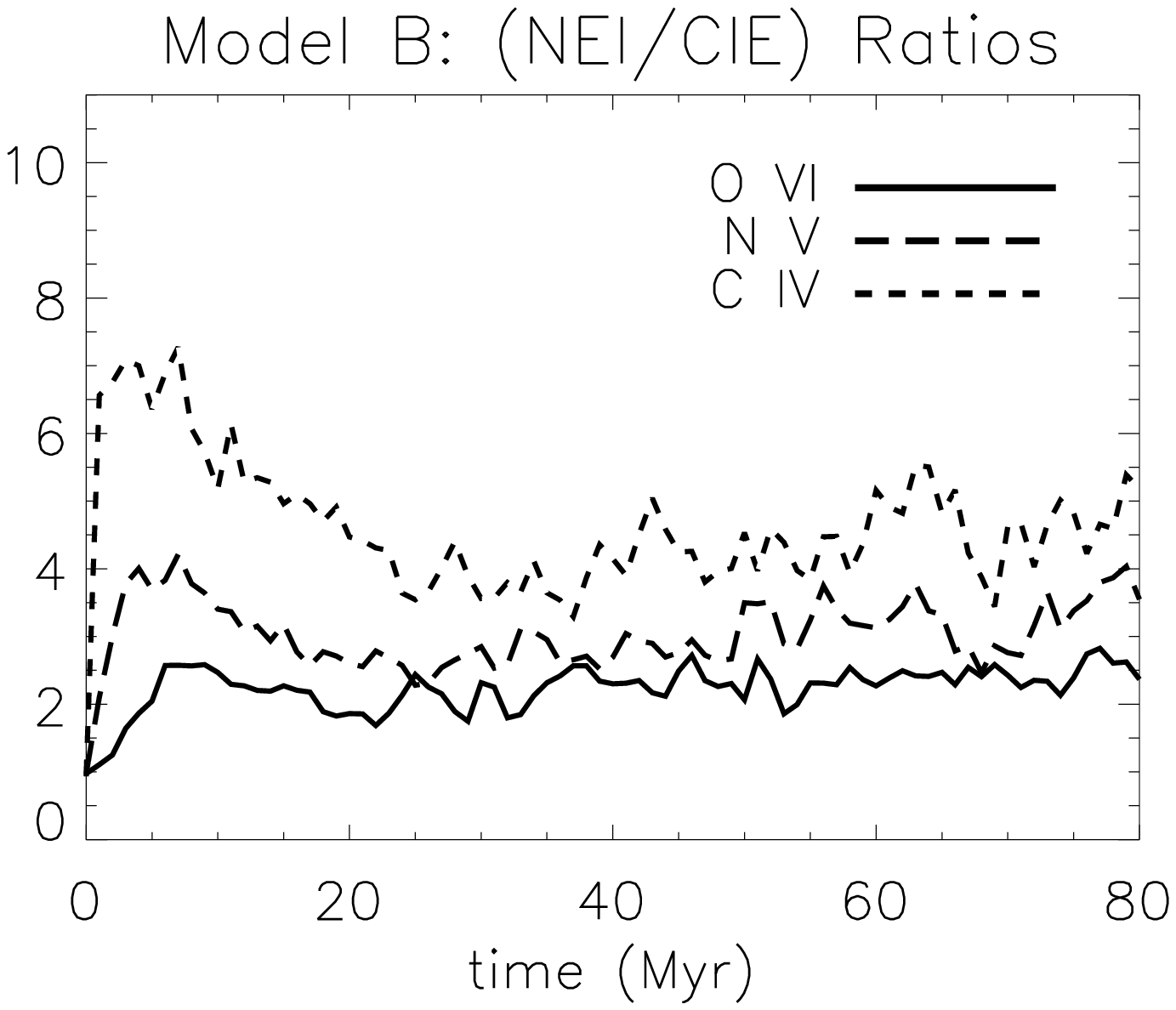}
\includegraphics[scale=0.24]{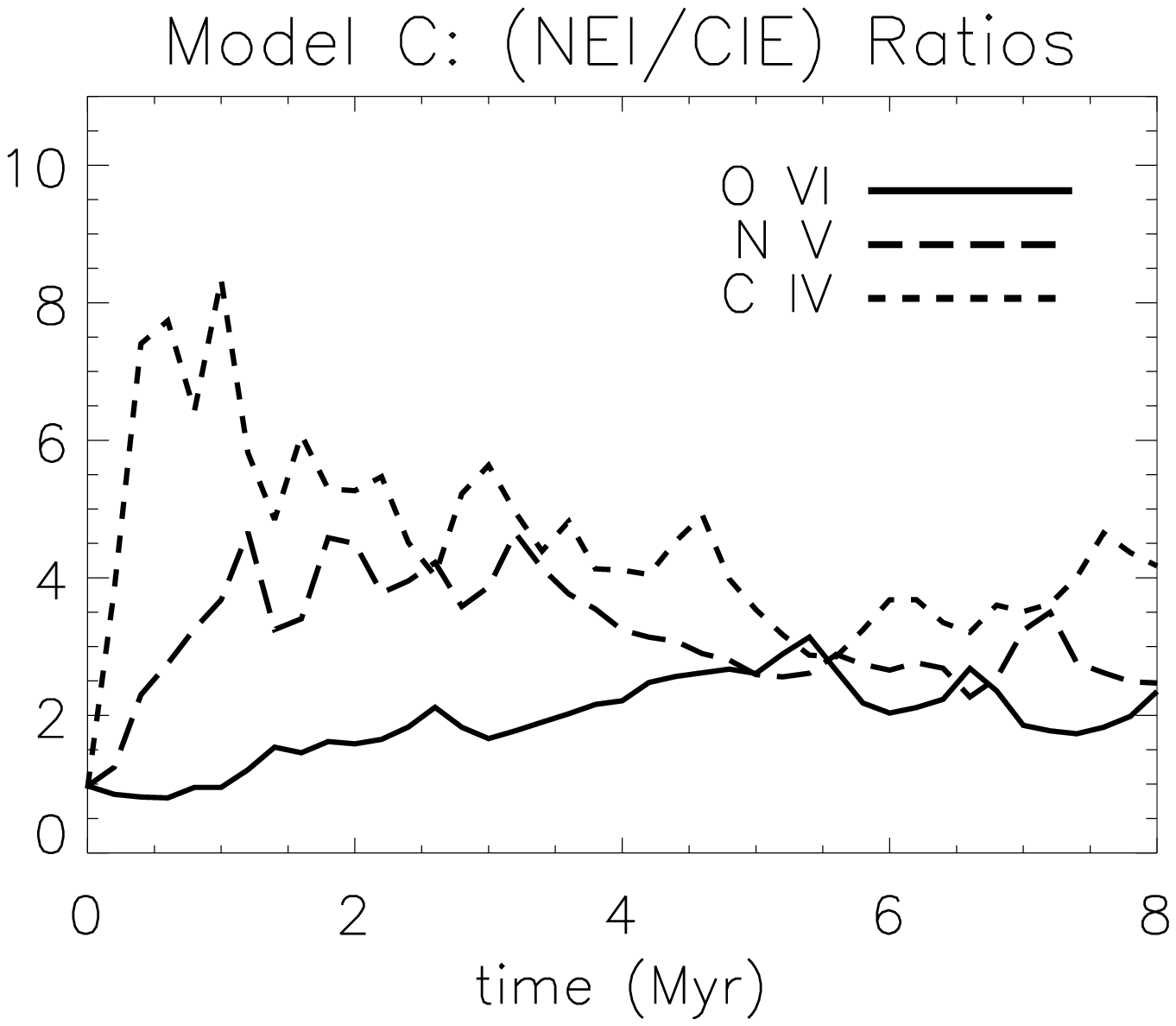} \\
\includegraphics[scale=0.24]{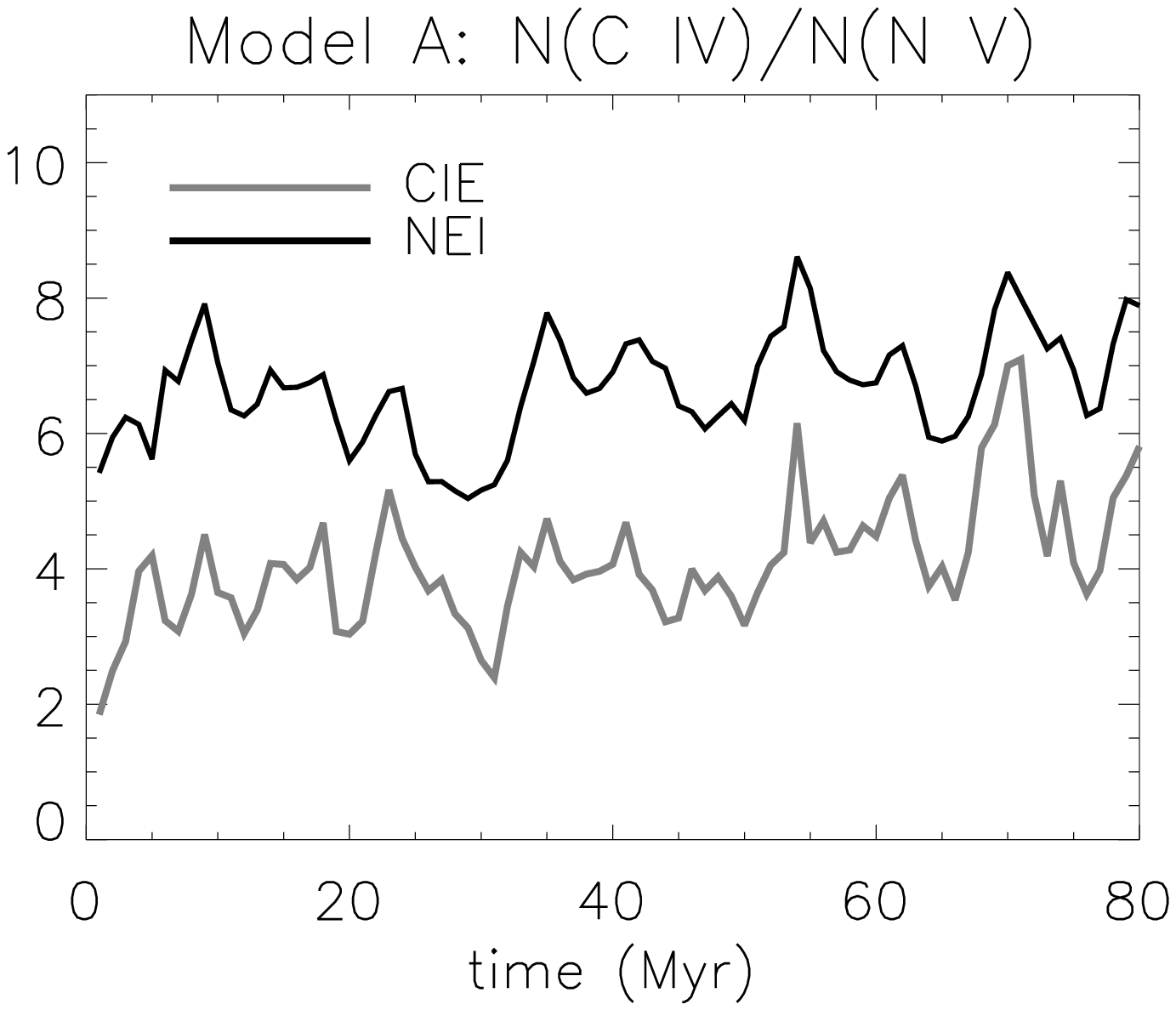}
\includegraphics[scale=0.24]{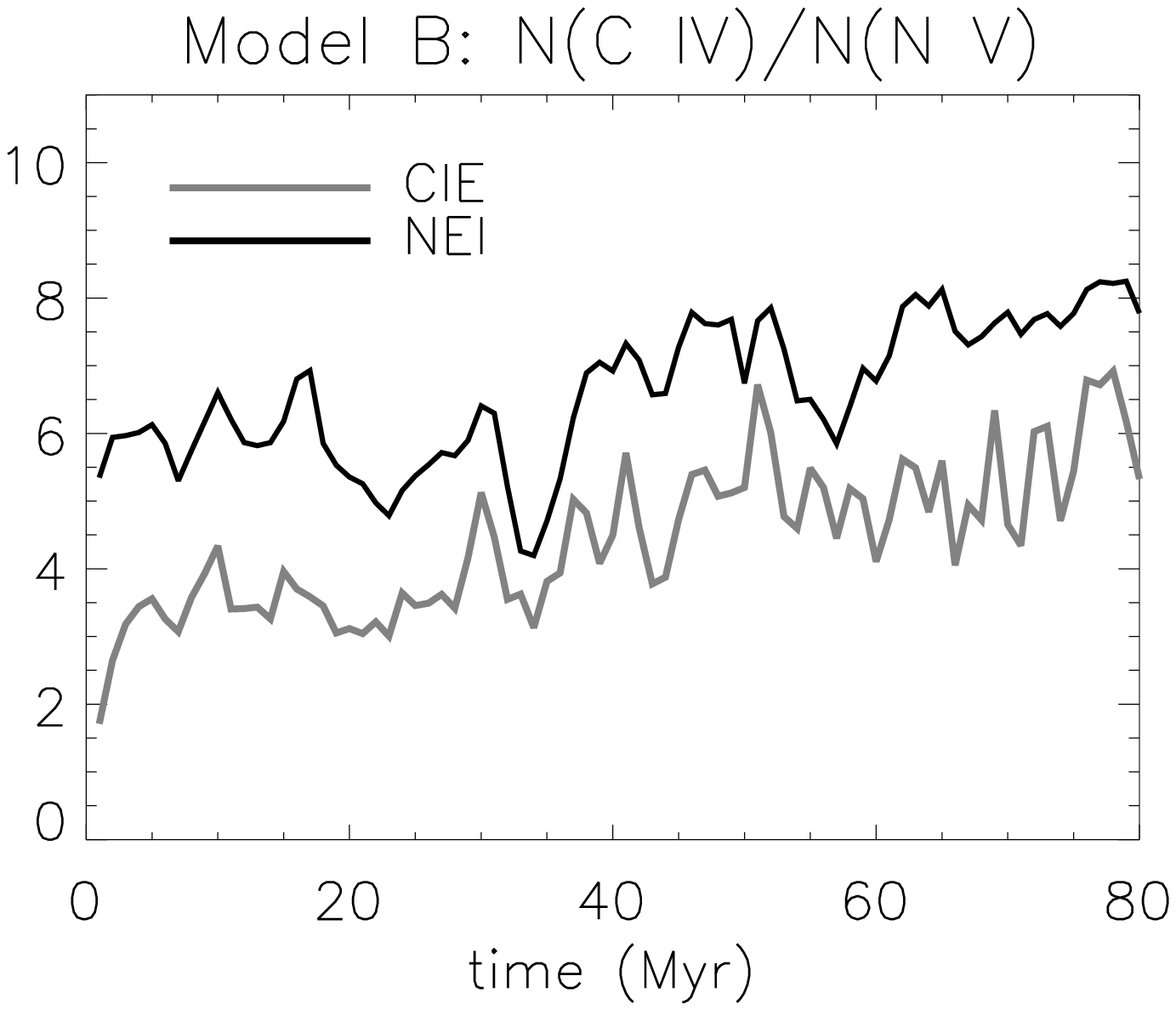}
\includegraphics[scale=0.24]{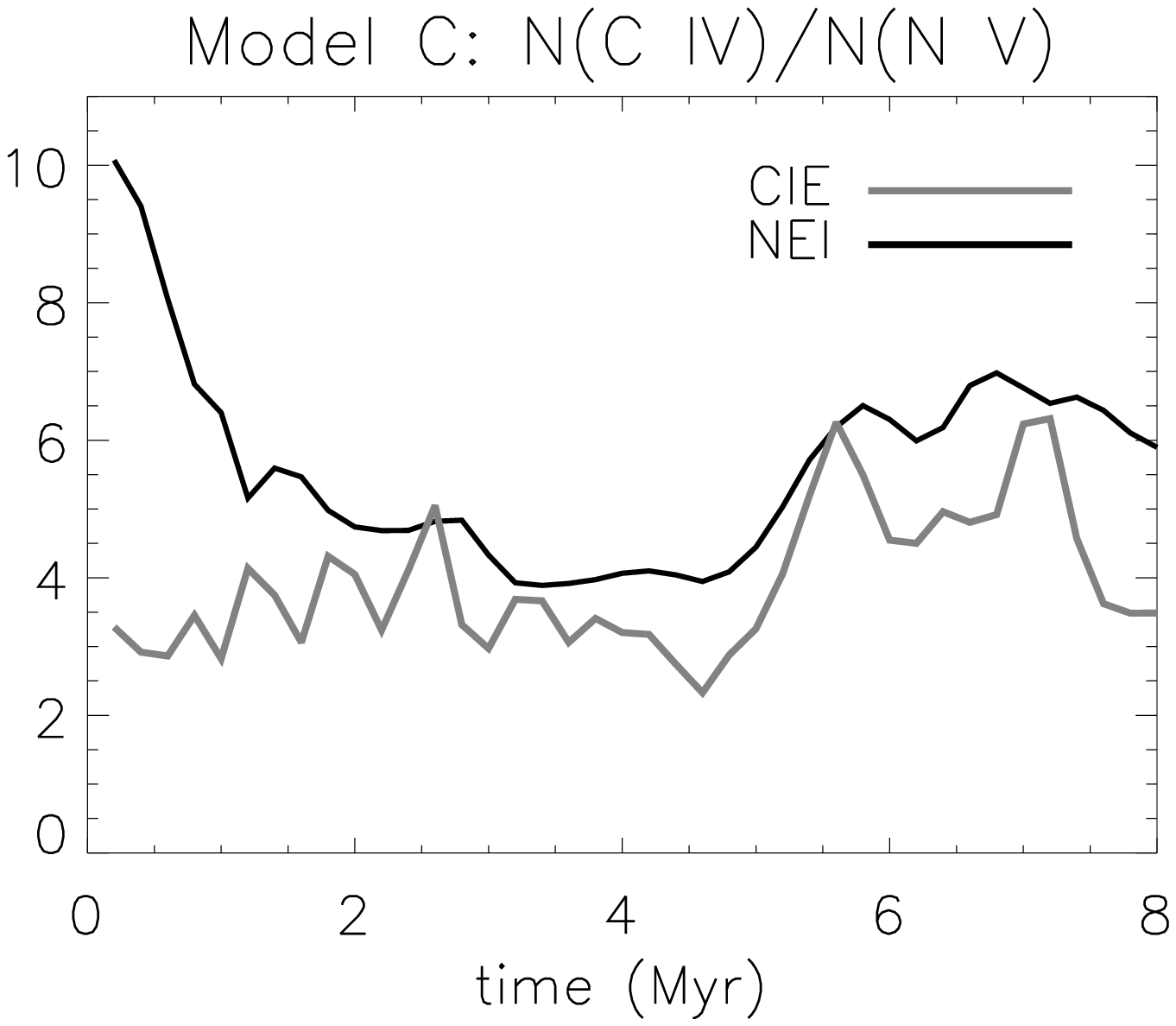} \\
\includegraphics[scale=0.24]{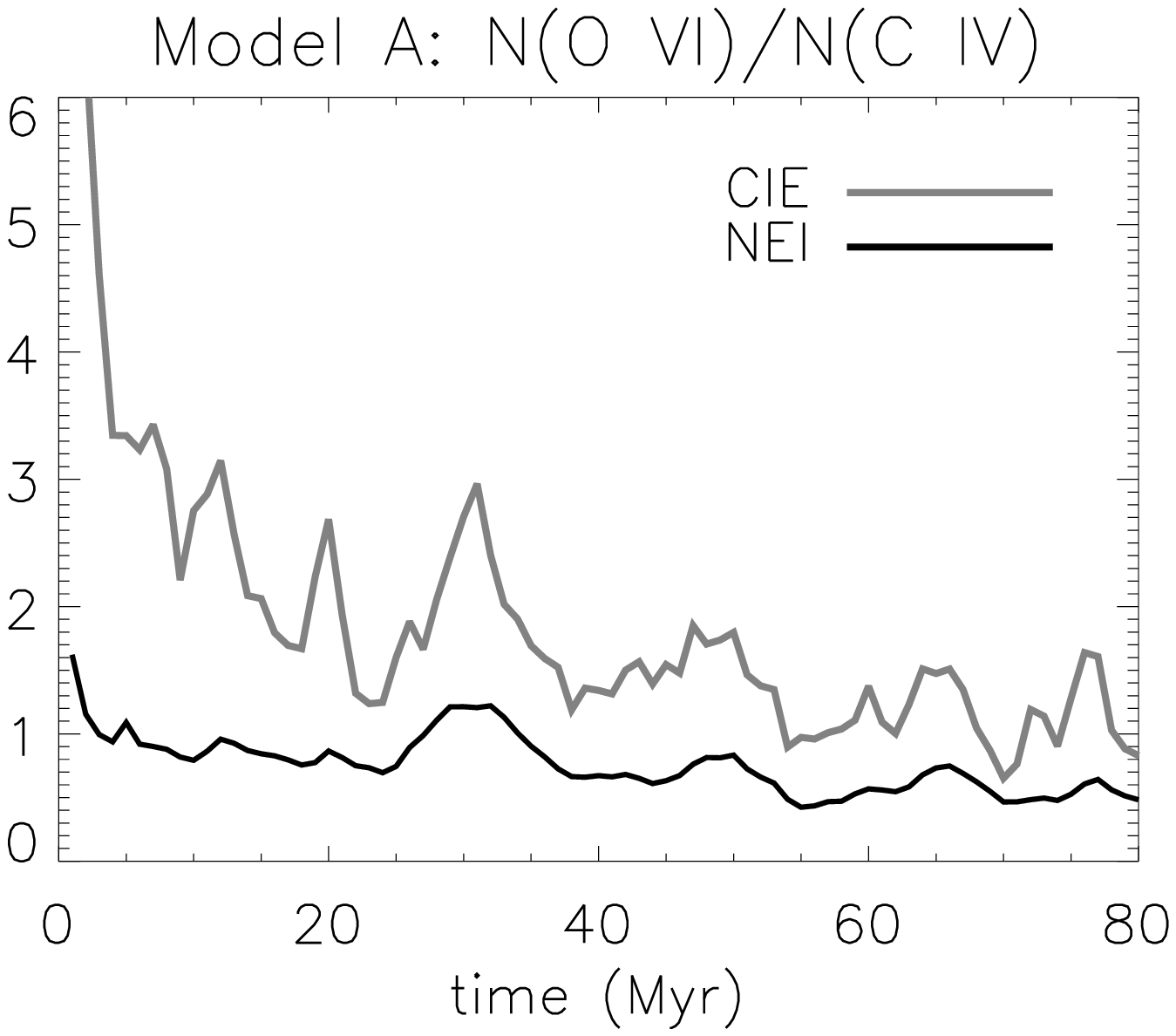}
\includegraphics[scale=0.24]{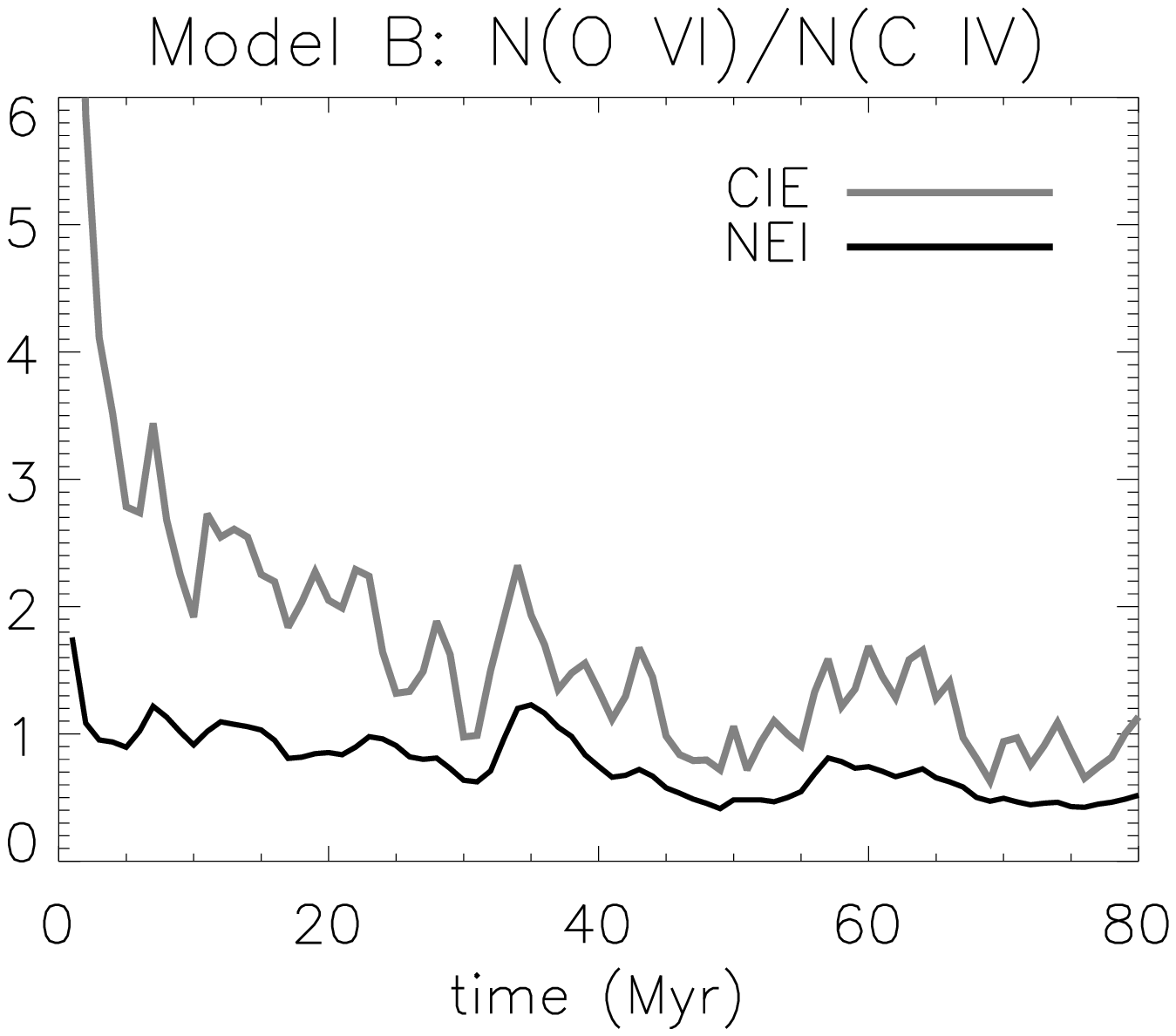}
\includegraphics[scale=0.24]{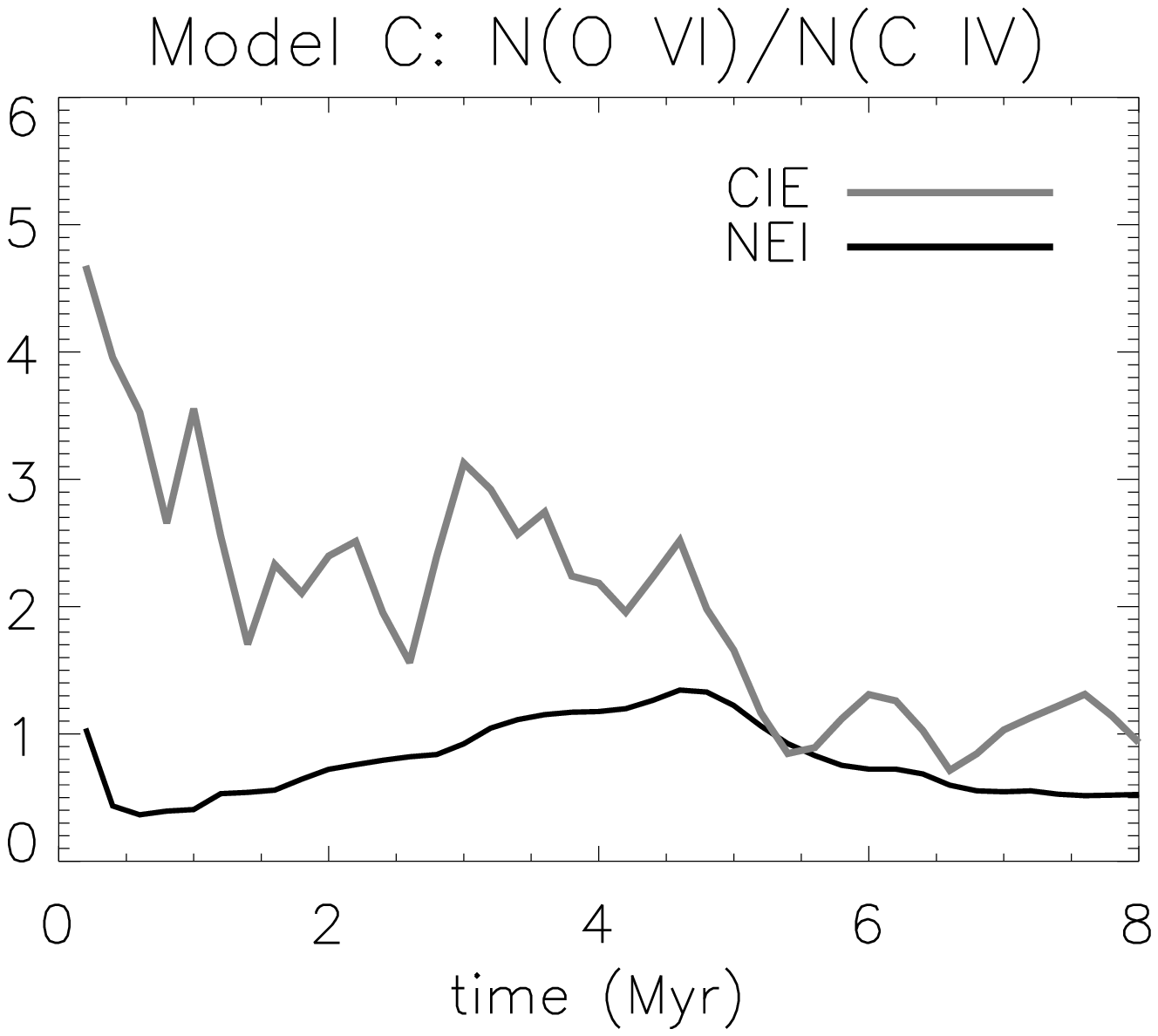} \\
\includegraphics[scale=0.24]{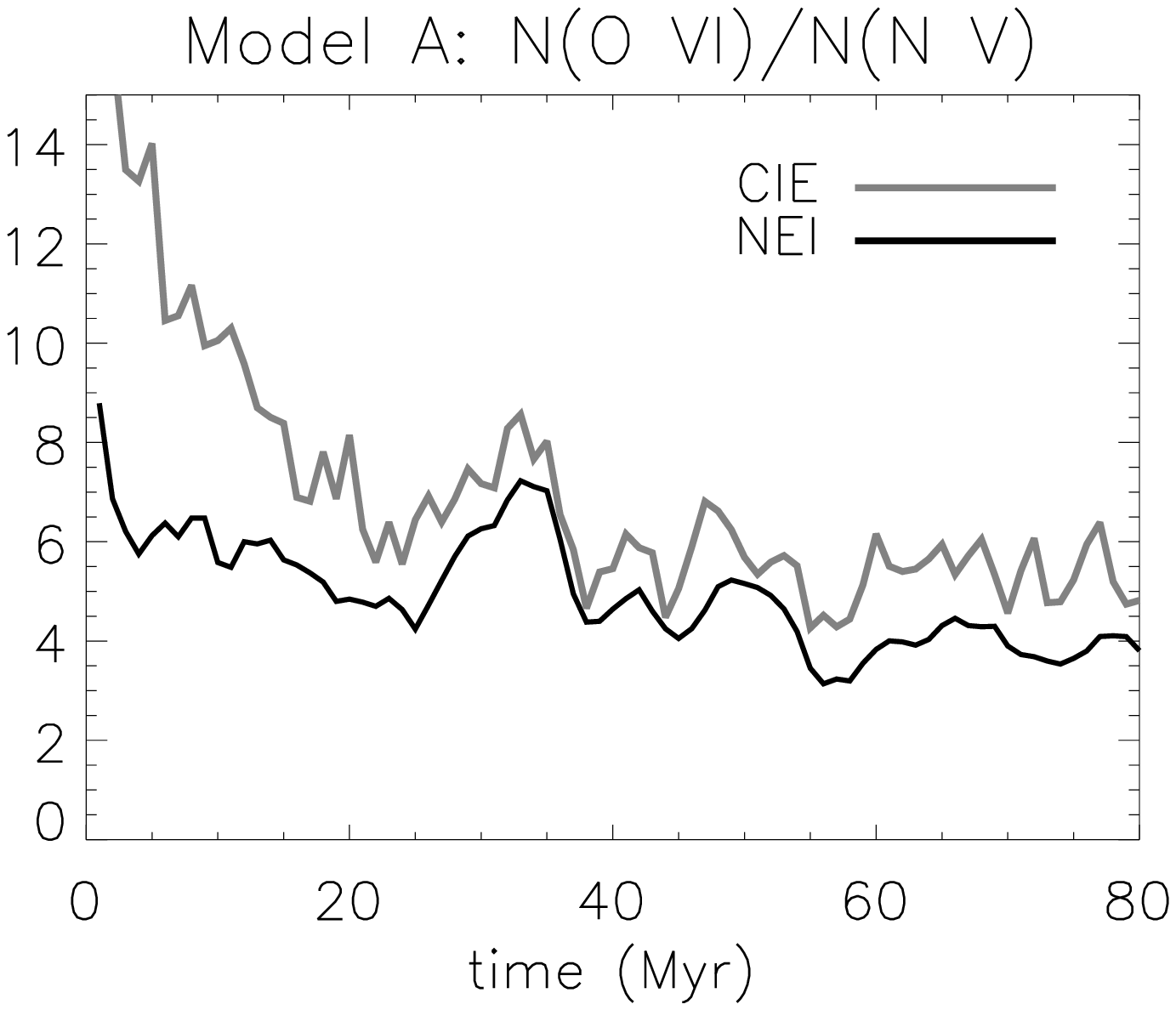}
\includegraphics[scale=0.24]{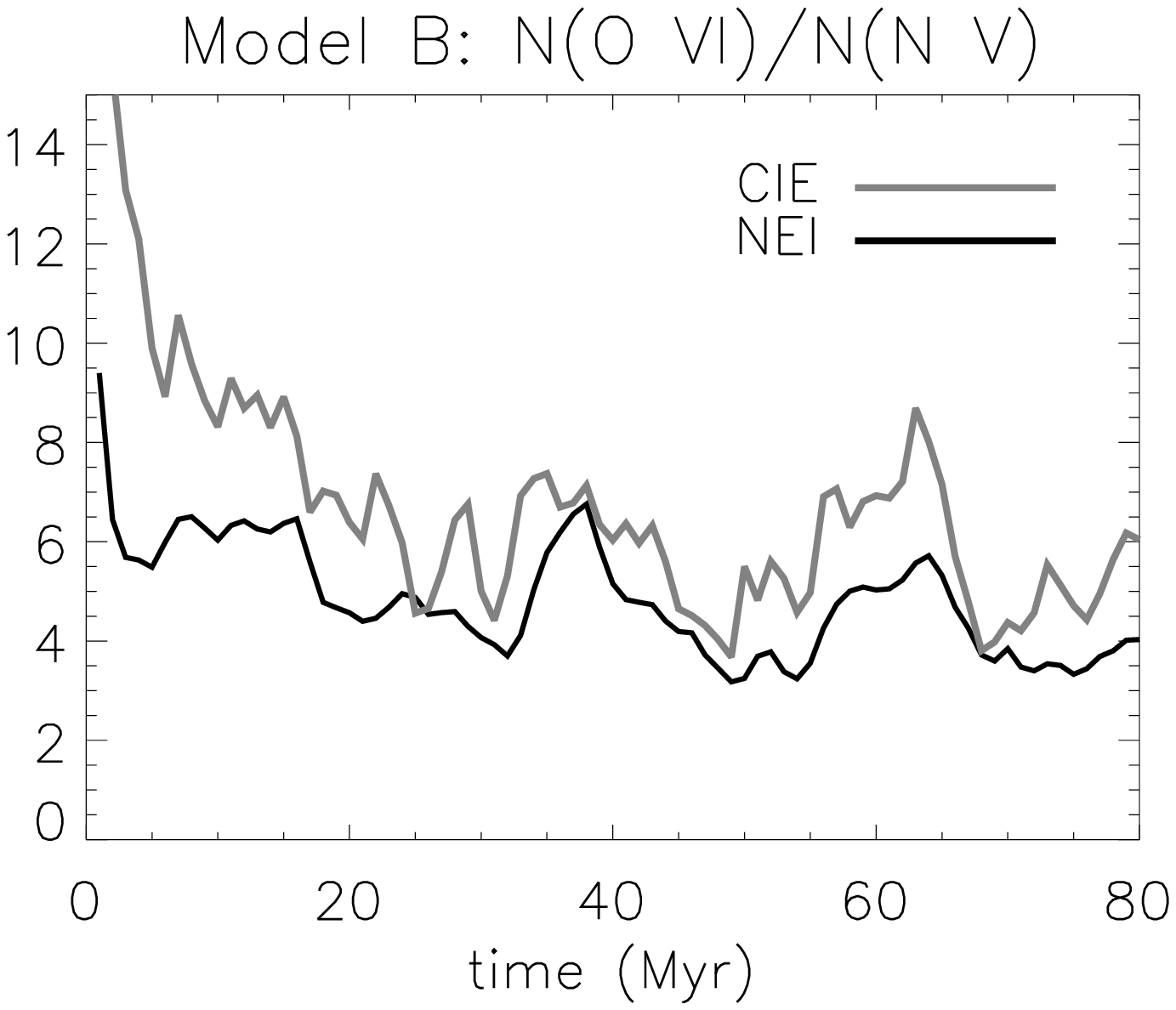}
\includegraphics[scale=0.24]{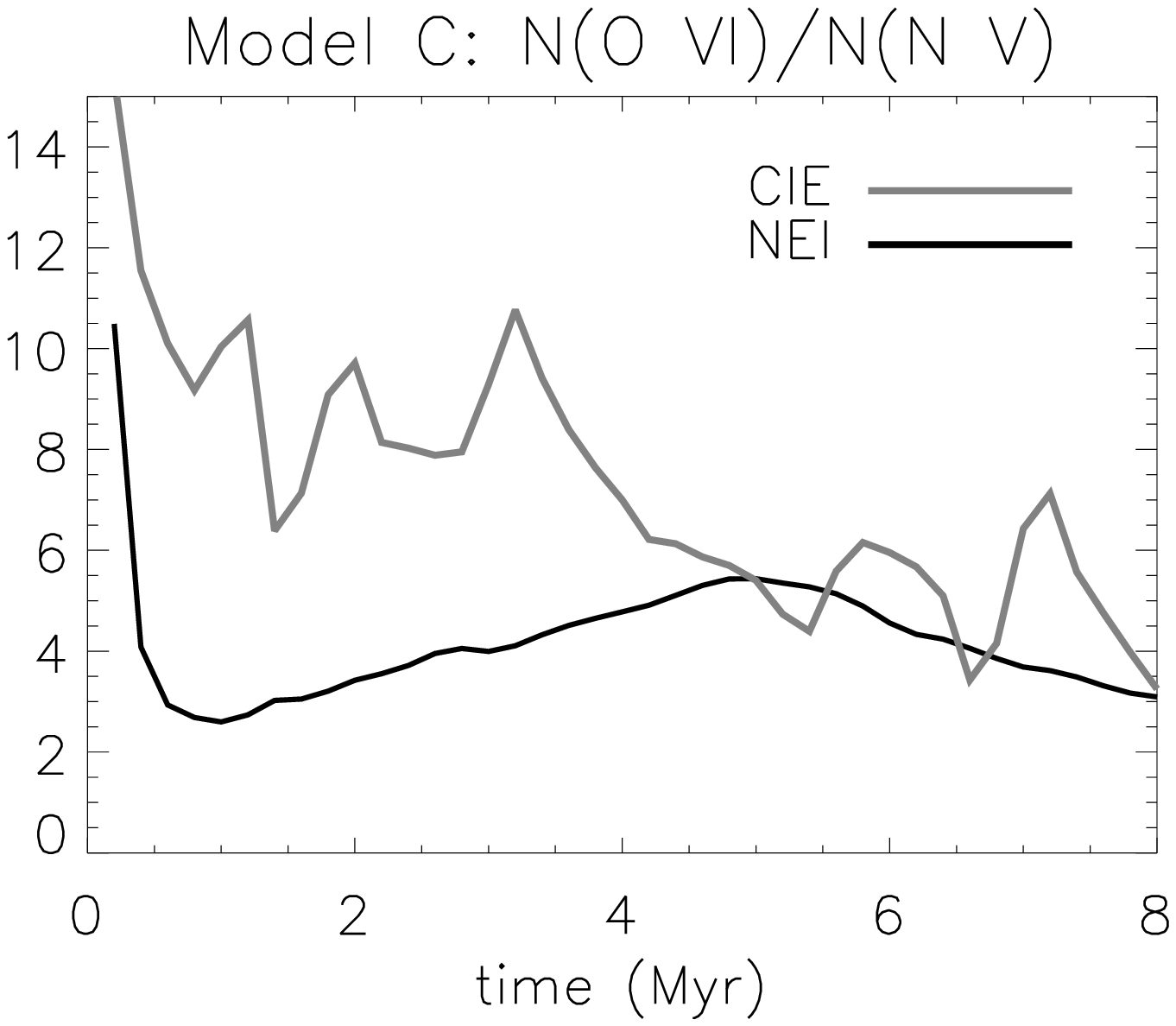}

\caption{Average high--stage ion column densities 
  for all vertical sightlines in 
  Model~A (left column), Model~B (middle column), 
  and Model~C (right column) as a function of time. 
  Top panels: column densities of 
  \ion{C}{4} (short dashed lines), \ion{N}{5} (long dashed lines), 
  and \ion{O}{6} 
  (solid lines), where the dark lines are from NEI calculations and 
  the gray lines are from CIE calculations. 
  Second row of panels: ratios of NEI to CIE column densities 
  for each ion. Third to fifth rows: 
  ratios between the column densities of different ions; 
  here the dark lines are from the NEI calculations and the gray 
  lines are from the CIE calculations. 
\label{modelABC_fig}}
\end{figure*}

We calculate the average column densities of \ion{C}{4}, \ion{N}{5}, 
and \ion{O}{6} ions along vertical sightlines, i.e., parallel 
to the $y$-axis through the Model~A domain, and plot them at 1~Myr 
intervals in Figure \ref{modelABC_fig}. The plots present the column 
densities obtained using both the NEI and CIE algorithms, showing 
that the NEI algorithms yield much higher column densities than do 
the CIE algorithms. This will be examined in more detail 
in \S \ref{neicie_S}. 
Calculating the column densities from the ion 
content of the entire domain is too CPU intensive, especially 
for CIE calculations. Instead, we
find them by averaging the column densities of a large number of 
evenly spaced, infinitely narrow sightlines (256 sightlines
for NEI and 86 sightlines for CIE) spread across the domain. 
For the CIE column densities, we confirm that 86 sightlines 
are sufficient by comparing their average with that found from 
256 sightlines for selected cases. Note that at any given time, 
the column density found along any given line of sight differs 
from those found along other lines of sight and thus differs 
from the average column density. The degree of variation is 
indicated, for example, in Figure \ref{all_sl_30myr_fig}, 
which shows column densities along each of 256 sightlines 
through the grid at $t=30$ Myr. Typical variation ranges from 
as little as standard deviation/mean = $16\%$ for CIE \ion{O}{6} 
to as much as $52\%$ for NEI \ion{C}{4}. 
The mean, median, standard deviation, and minimum 
and maximum values of the column density are summarized 
in Table \ref{all_sl_30myr_T}.

\subsubsection{Evolution of Column Densities} \label{evol_colm_dens_S}

A common question asked while comparing simulations with observations 
or the results of analytic calculations is ``which time frame should 
be used?'' In order to understand the answer, we discuss the time 
evolution of the high ion column densities here.

The top--left panel of Figure \ref{modelABC_fig} shows that 
the column densities 
of high stage ions are approximately $0$ at the beginning of the
simulation when all of the gas in the domain is either too cool or too 
hot to be rich in these ions. Once mixing begins, the high stage 
ion column densities, for both the NEI and CIE 
variants of Model A increase rapidly. Table \ref{increase_rate_T} 
presents the amount by which the \ion{C}{4}, \ion{N}{5}, and 
\ion{O}{6} column densities increase between $t=0$ and $t=10$ Myr. 
Although the growth rates vary, the column densities of all 
three ions in both the NEI and CIE variants of Model A 
continue to grow until $t\approx$ 20 to 30 Myr, when the depth 
of the mixed zone peaks for the first time. 
Between $t\approx30$ Myr and 80 Myr, 
the column densities of all three ions fluctuate. These fluctuations 
are also seen in the depth of the mixed zone 
in Figure \ref{modelA_dens_temp_fig}. 
However, the NEI \ion{C}{4}, in contrast 
with the NEI \ion{N}{5} and \ion{O}{6}, climbs to a new plateau at 
$t\approx50$ Myr.

The average \ion{C}{4} column density found using NEI calculations 
continues to grow until late in the simulation because some of 
NEI \ion{C}{4} is in the somewhat dense, radiatively cooled region 
below the interface (see Figure \ref{modelA_dens_temp_fig}) and 
this region grows, although sporadically, during the simulation time. 
The gas density in the radiatively cooled region is larger than that 
in the warmer, actively mixing region below it. Thus, even though 
the fraction of carbon atoms in the \ion{C}{4} stage (in NEI 
simulations) is smaller in the cooled gas than in the warm gas 
(see Figure \ref{modelA_dens_temp_fig}), the number of carbon 
atoms in the cooled gas is significant. 
To demonstrate this point, 
we estimate the average column density of each high stage ion 
in the radiately cooled region 
and in the actively mixing region at $t=80$ Myr, 
when the cooled region is most 
distinctive. We approximate the cooled region as that 
between $y=-60$ and $y=0$ pc 
and the actively mixing region as that 
between $y=-200$ and $y=-60$ pc. 
The column densities are averaged over 256 
sightlines for both the NEI and CIE calculations. 
The average column densities of each species in the cooled and 
actively mixing region are presented 
in Table \ref{cooled_mixing_T}. It is shown that 
in NEI, the average \ion{C}{4} column density 
in the cooled region is $47\%$ as large as 
that in the mixing region and that the average volume density
($\sim 1.94\times10^{-8}~\mbox{cm}^{-3}$) of
\ion{C}{4} ions in the cooled region is even slightly higher than that 
($\sim 1.79\times10^{-8}~\mbox{cm}^{-3}$) in the mixing region. 

\begin{deluxetable*}{ccccccccc}[pt]

\tablewidth{0pt}
\tabletypesize{\footnotesize}
\tablecaption{Column Densities along Vertical Sightlines at t=30 Myr 
in Model A \label{all_sl_30myr_T} }

\tablecolumns{9}

\tablehead{
\colhead{} & 
\multicolumn{4}{c}{NEI} & 
\multicolumn{4}{c}{CIE} \\
\colhead{Ions} &
\multicolumn{4}{c}{-----------------------------------------------------} & 
\multicolumn{4}{c}{-----------------------------------------------------} \\
\colhead{} & 
\colhead{~~mean} & \colhead{median} & 
\colhead{$\sigma$ \tablenotemark{a}} &
\colhead{[min, max]} &
\colhead{~~mean} & \colhead{median} & 
\colhead{$\sigma$ \tablenotemark{a}} &
\colhead{[min, max]} \\
\colhead{} &
\multicolumn{4}{c}{($10^{12}~\mbox{cm}^{-2}$)} & 
\multicolumn{4}{c}{($10^{12}~\mbox{cm}^{-2}$)} 
}

\startdata
C IV & 7.69 & 7.36 & 3.31 & [2.73, 16.17] &
1.56 & 1.54 & 0.81 & [0.13, 3.87] \\
N V & 1.49 & 1.43 & 0.35 & [0.92, 2.41] & 
0.58 & 0.54 & 0.24 & [0.17, 1.32] \\
O VI & 9.33 & 9.29 & 1.47 & [6.08, 12.19] & 
4.18 & 4.10 & 0.80 & [2.33, 6.98] 
\enddata

\tablenotetext{a}{standard deviation}
\end{deluxetable*}

The cooled region is poorer in \ion{N}{5} and \ion{O}{6} than in 
\ion{C}{4}. Thus, the NEI column densities of \ion{N}{5} and \ion{O}{6} 
in the cooled region are not as significant as that of \ion{C}{4}. 
They are $19\%$ and $17\%$ of those in the mixing 
region, respectively. The \ion{C}{4}, \ion{N}{5}, and \ion{O}{6} 
in the radiatively cooled zone principally derive from ions 
that were previously hotter and are now in the process of recombining 
with free electrons. Since the CIE calculations ignore the 
plasma's history as they consider only the current temperature, 
they predict very few high stage ions in the radiatively cooled zone. 
In CIE, the column densities 
of all three ions in the cooled region are less than $5\%$ of 
those in the mixing region.

\subsubsection{NEI vs. CIE}\label{neicie_S}

As soon as mixing begins in our simulations, the NEI calculations 
predict greater numbers of high stage ions than the CIE calculations. 
(Because we assume CIE at the initial timestep, the NEI/CIE ratios 
at $t=0$ Myr are unity for all ions.) This is indicated by the 
second row of panels in Figure \ref{modelABC_fig}, and in particular, 
the left--most panel in that row, which plots the ratios of column 
densities predicted by the NEI calculations to the column densities 
predicted by the CIE calculations for Model A. 

\begin{figure}[t]
\centering
\includegraphics[scale=0.35]{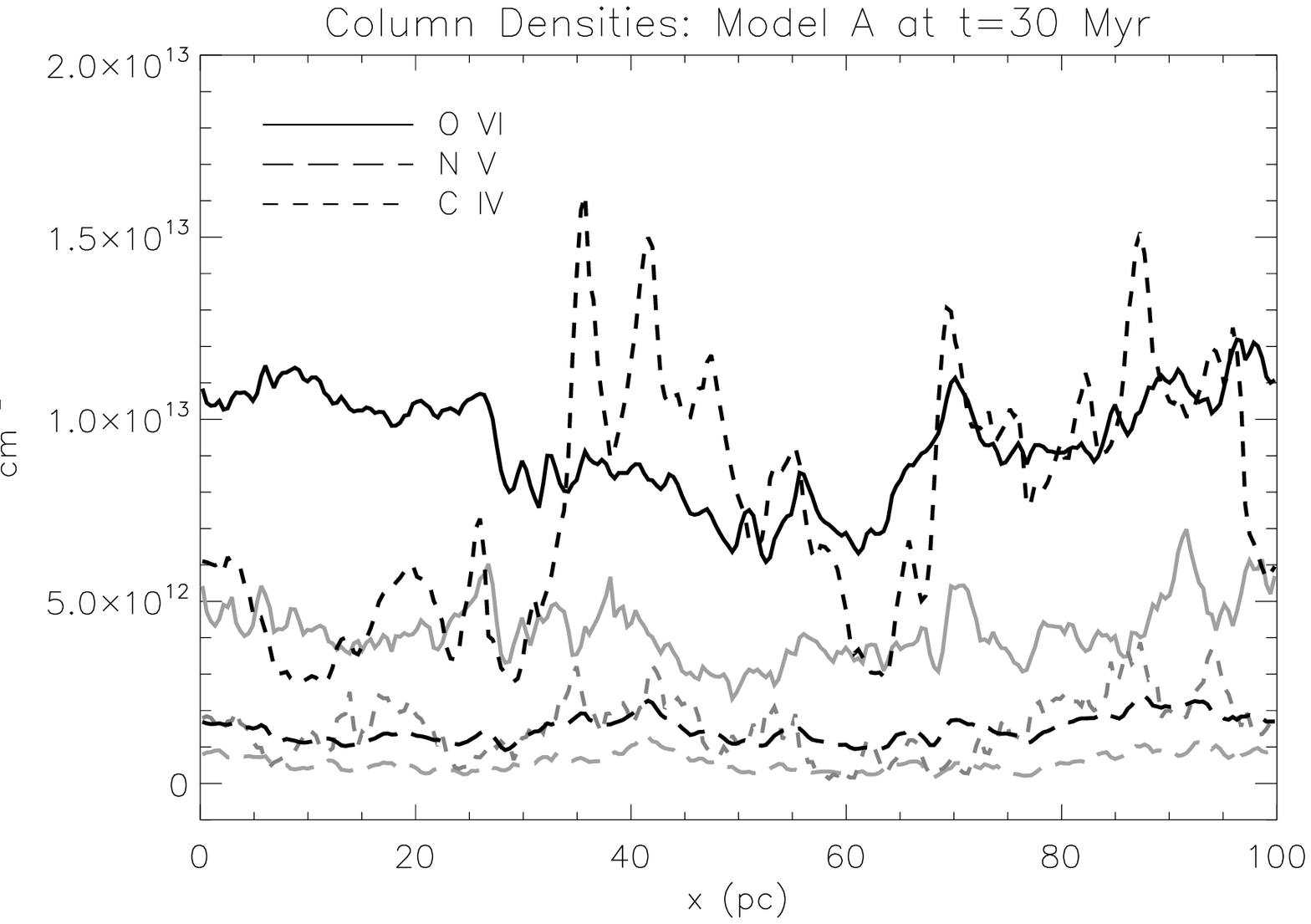}
\caption{Column densities of \ion{C}{4} (short dashed), 
  \ion{N}{5} (long dashed), 
  and \ion{O}{6} (solid) ions along 256 vertical sightlines through 
  Model~A's domain at $t=30$ Myr plotted as a function of position 
  on the $x$-axis. Dark and gray lines are from NEI 
  and CIE calculations, respectively. 
\label{all_sl_30myr_fig}}
\end{figure}

Mixing, which proceeds faster than ionization or recombination, 
forces the gas out of equilibrium. For the first 10 Myr, most of 
the mixed gas is hotter than the CIE temperatures for \ion{C}{4} 
and \ion{N}{5}, thus the CIE calculations predict low ion fractions 
for them. However, larger ion fractions 
and thus column densities of \ion{C}{4} 
and \ion{N}{5} are found in the NEI calculations. These are due to 
previously--cool atoms that have lost a couple of electrons after mixing 
with hotter gas, but have not yet ionized to their equilibrium
levels. Real observations of such gas should find wide absorption 
and emission features from this hot, `underionized' gas. In addition, 
the profiles should also be widened by Doppler broadening in the 
turbulent velocity field. 

\begin{deluxetable*}{cccccc}[t]

\tablewidth{0pt}
\tabletypesize{\footnotesize}
\tablecaption{Column Density Increase at Early Times 
\label{increase_rate_T}}
\tablecolumns{6}

\tablehead{
\colhead{} & \colhead{} & 
\colhead{~~time period~~} &
\multicolumn{3}{c}{column density increase ($10^{12}~\mbox{cm}^{-2}$)} \\
\colhead{Model} & \colhead{NEI/CIE} & 
\colhead{~~[$t_{min}$, $t_{max}$]~~} &
\multicolumn{3}{c}{--------------------------------------------------}\\
\colhead{} & \colhead{} & \colhead{(Myr)} &
\colhead{~~~~C IV} & \colhead{~~~~~~~N V} & \colhead{O VI} 
}

\startdata
A & NEI & [0, 10] & ~~~ 4.0  & ~~~~~~~ 0.56   & 2.9  \\
A & CIE & [0, 10] & ~~~ 0.49 & ~~~~~~~ 0.13   & 1.1 \\
B & NEI & [0, 10] & ~~~ 4.0  & ~~~~~~~ 0.59   & 3.4  \\
B & CIE & [0, 10] & ~~~ 0.76 & ~~~~~~~ 0.17   & 1.2 \\
C & NEI & [0, 2]  & ~~~ 0.89 & ~~~~~~~ 0.19   & 0.62 \\
C & CIE & [0, 2]  & ~~~ 0.17 & ~~~~~~~ 0.041  & 0.38 \\
D & NEI & [0, 10] & ~~~ 1.1  & ~~~~~~~ 0.18   & 1.1  \\
E & NEI & [0, 10] & ~~~ 4.4  & ~~~~~~~ 0.68   & 4.4  \\
F & NEI & [0, 10] & ~~~ 1.4  & ~~~~~~~ 0.2    & 1.2 
\enddata

\end{deluxetable*}

\begin{deluxetable*}{cccccccc}

\tablewidth{0pt}
\tabletypesize{\footnotesize}
\tablecaption{Volume and Column Densities in Cooled and Mixing Layers 
  of Model~A
  \label{cooled_mixing_T}}
\tablecolumns{8}

\tablehead{
\colhead{} & \colhead{} & 
\multicolumn{2}{c}{C IV} &
\multicolumn{2}{c}{N V} &
\multicolumn{2}{c}{O VI} \\
\colhead{} & \colhead{} &
\multicolumn{2}{c}{---------------------------------} &
\multicolumn{2}{c}{---------------------------------} &
\multicolumn{2}{c}{---------------------------------} \\
\colhead{} & \colhead{} & 
\colhead{NEI} & \colhead{CIE} &
\colhead{NEI} & \colhead{CIE} &
\colhead{NEI} & \colhead{CIE} 
}

\startdata
Cooled \tablenotemark{a} & $n$ \tablenotemark{b} ($\mbox{cm}^{-3}$) & 
$1.94\times10^{-8}$ & $2.69\times10^{-10}$ & $1.22\times10^{-9}$ &
$2.50\times10^{-11}$ & $4.36\times10^{-9}$ & $2.11\times10^{-12}$ \\
layer & $N$ \tablenotemark{c} ($\mbox{cm}^{-2}$) & 
$3.60\times10^{12}$ & $4.98\times10^{10}$ & $2.26\times10^{11}$ &
$4.63\times10^{9}$ & $8.07\times10^{11}$ & $3.91\times10^{8}$ \\
Mixing \tablenotemark{d} & $n$ \tablenotemark{b} ($\mbox{cm}^{-3}$) & 
$1.79\times10^{-8}$ & $5.73\times10^{-9}$ & $2.80\times10^{-9}$ &
$9.86\times10^{-10}$ & $1.07\times10^{-8}$ & $4.76\times10^{-9}$ \\
layer & $N$ \tablenotemark{c} ($\mbox{cm}^{-2}$) & 
$7.74\times10^{12}$ & $2.48\times10^{12}$ & $1.21\times10^{12}$ &
$4.26\times10^{11}$ & $4.62\times10^{12}$ & $2.06\times10^{12}$
\enddata

\tablenotetext{a}{$y \in [-60, 0]$ pc}
\tablenotetext{b}{volume density of C IV, N V, and O VI}
\tablenotetext{c}{column density obtained by multiplying volume
  density with the length of layer, 60 pc for cooled layer and 140 pc 
for mixing layer}
\tablenotetext{d}{$y \in [-200, -60]$ pc}

\end{deluxetable*}

\begin{deluxetable*}{ccccccccccccc}[b]

\tablewidth{0pt}
\tabletypesize{\footnotesize}
\tablecaption{Ratios of Column Densities Calculated Using NEI to Those 
  Calculated Using CIE
  \label{nei_cie_ABC_T} }
\tablecolumns{13}

\tablehead{
\colhead{} & 
\multicolumn{4}{c}{C IV} &
\multicolumn{4}{c}{N V} &
\multicolumn{4}{c}{O VI} \\ 
\colhead{Model} & 
\multicolumn{4}{c}{--------------------------------------------------} &
\multicolumn{4}{c}{--------------------------------------------------} &
\multicolumn{4}{c}{--------------------------------------------------} \\
\colhead{} & 
\colhead{mean} & \colhead{median} &
\colhead{$\sigma$ \tablenotemark{a}} & 
\colhead{[min, max]} & \colhead{mean} & \colhead{median} & 
\colhead{$\sigma$ \tablenotemark{a}} & 
\colhead{[min, max]} & \colhead{mean} & \colhead{median} & 
\colhead{$\sigma$ \tablenotemark{a}} & 
\colhead{[min, max]} 
}
  
\startdata
A \tablenotemark{b} & 4.64 & 4.62 & 0.44 & [3.82, 5.67] & 
2.91 & 2.88 & 0.34 & [2.24, 3.82] & 
2.26 & 2.25 & 0.20  & [1.83, 2.81] \\
B \tablenotemark{b} & 4.30 & 4.27 & 0.54 & [3.30, 5.53] & 
3.00 & 2.89 & 0.42 & [2.28, 4.03] & 
2.29 & 2.32 & 0.25 & [1.69, 2.83] \\
C \tablenotemark{c} & 3.80 & 3.68 & 0.42 & [3.21, 4.65] & 
2.73 & 2.66 & 0.34 & [2.27, 3.50] & 
2.09 & 2.03 & 0.28 & [1.73, 2.68] 
\enddata

\tablenotetext{a}{standard deviation}
\tablenotetext{b}{averaged over $t \in [20, 80]$ Myr}
\tablenotetext{c}{averaged over $t \in [6, 8]$ Myr}

\end{deluxetable*}

At their worst, the CIE calculations underpredict the \ion{C}{4} 
and \ion{N}{5} column densities (relative to the NEI predictions) 
by factors of $\sim8$ and $\sim4$, respectively, but this only 
occurs during the first $\sim12$ Myr of the simulation. Subsequently, 
as the mixed zone develops a wider temperature profile, a radiatively 
cooled zone develops, and more ionization and recombination occurs, 
the CIE and NEI predictions begin to track each other with average 
ratios of 4.6, 2.9, and 2.3 for \ion{C}{4}, \ion{N}{5}, and
\ion{O}{6}, respectively (see Table \ref{nei_cie_ABC_T}).

\begin{figure}[hb]

\centering

\includegraphics[scale=0.5]{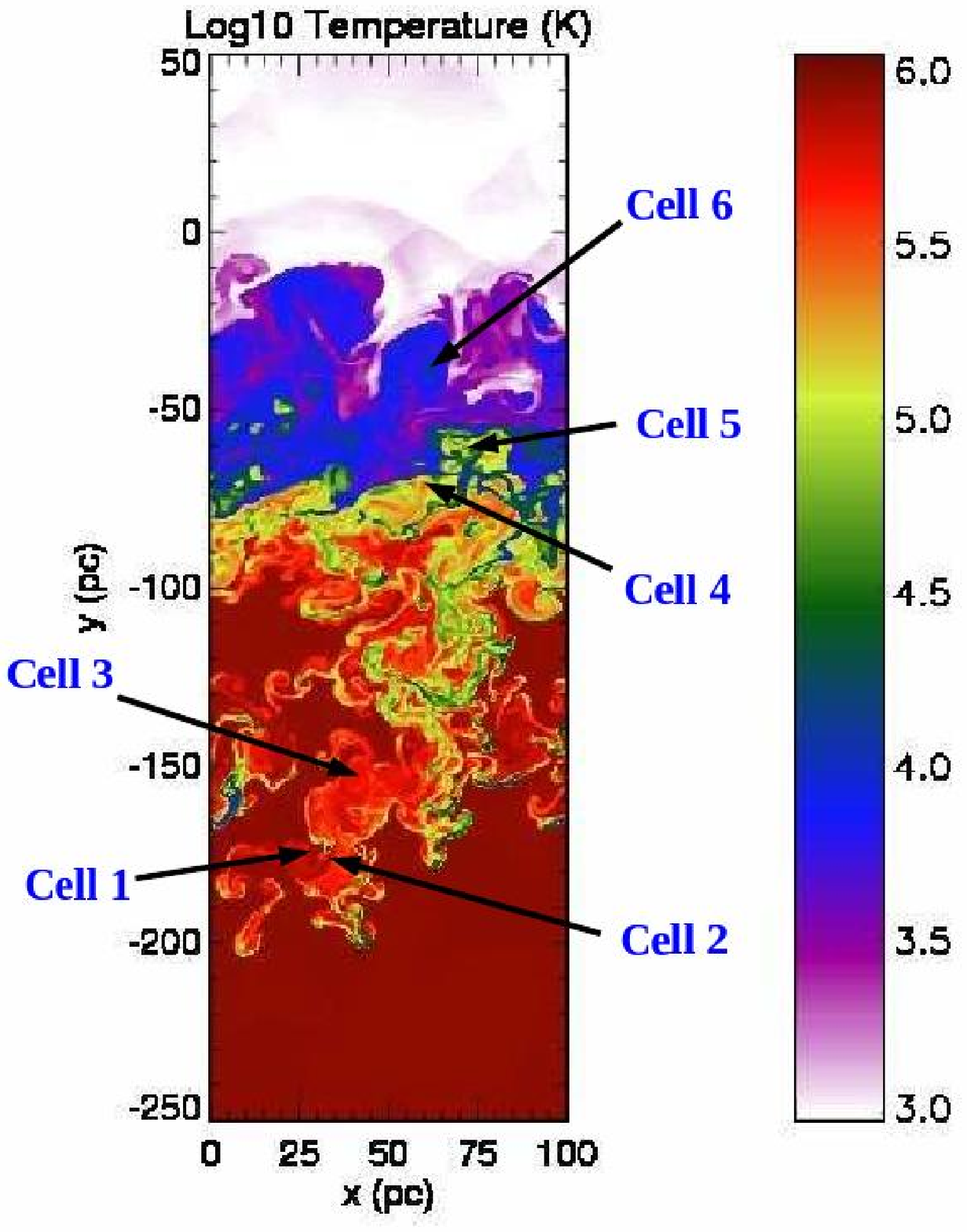}

\caption{Cell locations for ionization level calculations for carbon, 
  nitrogen, and oxygen. In this temperature map for Model~A 
  at $t=80$ Myr, six cells are identified. The information for 
  each cell is given in Table \ref{cell_neicie_T} and the ionization 
  level calculations are shown in Figure \ref{modelA_ioniz_fig}. 
  \label{cell_location_fig}}

\end{figure}

The effects of mixing, radiative cooling, ionization, and
recombination can be seen in individual cells. Ideally, we could 
examine a single cell at various stages in its time evolution, 
from the moment after it first experienced mixing until long after 
it radiatively cooled. While this is not practical, it is possible 
to examine a variety of cells at a single moment in time. 
Figure \ref{cell_location_fig} shows an expanded image of the domain 
and points out 6 cell locations. 
Cells 1--3 belong to the lower region of the computational domain, 
contain newly mixed gas on the boundary of the mixing zone, and 
are hotter. Most of the material in these cells came from 
the hot reservoir and minimal radiative cooling has occurred in them. 
Cells 4--5 are deeper in the interior of the mixing zone. Roughly 
half of their material came from the hot reservoir and half from 
the cool gas. These cells have radiated away significant fractions of 
their thermal energy and are now roughly 1/2 the temperature of the 
hot reservoir. Cell 6 is nearest to the cool gas. Among all of our 
sample cells, cell 6 has mixed for the largest period of time, 
entrained the largest fraction of cool gas, and lost the largest 
fraction of its thermal energy to radiation. It has lost $\ge85\%$ 
of its thermal energy to radiation and with a temperature of 
$\sim15,000~\mbox{K}$, exemplifies the radiatively cooled zone 
at the `base' of the mixed zone.


Table \ref{cell_neicie_T} lists the locations (column 2) 
and hydrogen number densities (column 3) of 
these cells. The FLASH code allows us to trace the mass fraction 
of the initial cool and hot gas at each cell over time. 
The mass fraction of cell material 
that originally came from the hot reservoir 
($f_{hot}$, column 4) and the mass fraction 
from the cool reservoir ($f_{cool}$, column 5)
are also listed in Table \ref{cell_neicie_T}. 
From these mass fractions, 
we calculate the temperature that the gas would have had if 
radiative cooling had not occurred,    
$\bar{T}=T_{hot} \times f_{hot}+T_{cool} \times f_{cool}$,  where
$T_{hot}=1.0 \times 10^6~\mbox{K}$ and 
$T_{cool}=1.0 \times 10^3~\mbox{K}$. This temperature is tabulated 
in (column 6) in Table \ref{cell_neicie_T}. The measured temperature 
at each cell from the simulation ($T$, column 7) is the temperature 
that the mixed gas reaches after the radiative cooling is in effect. 
Comparing $\bar{T}$ (column 6) and $T$ (column 7) shows 
how significantly radiative cooling has lowered 
the temperature of the mixed gas.

As the temperature drops (progressing from cell 1 to larger numbered 
cells) due to radiative cooling and advection of cooler gas, the 
fraction of very high ions predicted by CIE calculations decreases 
much faster than those predicted by NEI calculations. Because the 
NEI recombination rate lags the cooling rate so severely, we can find
\ion{C}{4} in the $15,000~\mbox{K}$ gas in cell 6, next to the cool 
gas. This is consistent with the sudden increase of 
column density of \ion{C}{4} 
at later times shown in the top--left panel of Figure
\ref{modelABC_fig} (\S \ref{evol_colm_dens_S}).

Similarly, when cells advect cool gas, they gain low ions, which 
raise their low ion fractions. These ions are slow to ionize, 
causing the fractions of NEI \ion{C}{4}, \ion{N}{5}, and \ion{O}{6} 
in cell 1 to be significantly greater than those predicted by CIE 
calculations. 
In cells 4 and 5, the NEI predictions for once, twice, and thrice 
ionized carbon, nitrogen, and oxygen are more similar to those of 
CIE calculations, but this is not solely due to recombinations. 
Instead, it occurs because radiative cooling has lowered the gas 
temperature enough for the CIE fractions of low ions to converge 
with the NEI ion fractions. Such gas is not actually in ionization 
equilibrium, as is apparent from the fact that the fractions of 
high ions greatly exceed those predicted by CIE calculations.

\begin{deluxetable*}{ccccccccc}[t]

\tablewidth{0pt}
\tabletypesize{\footnotesize}
\tablecaption{Cells for Ionization Level Calculations \label{cell_neicie_T}}
\tablecolumns{9}

\tablehead{
\colhead{} & \colhead{coordinates} & \colhead{$n_H$ \tablenotemark{a}} &
\colhead{hot gas} & \colhead{cool gas} &
\colhead{$\bar{T}$ \tablenotemark{b}} & 
\colhead{T \tablenotemark{c}} \\ 
\colhead{cell} & \colhead{(x,y) pc} & \colhead{($10^{-4}~\mbox{cm}^{-3}$)} &
\colhead{mass fraction} & \colhead{mass fraction} & 
\colhead{(K)} & \colhead{(K)}
}

\startdata
1 & (24.9, -176.0) & 0.87 & 0.76 & 0.24 & 760,000 &
653,000 \\
2 & (34.3, -181.4) & 1.72 & 0.65 & 0.35 & 650,000 &
462,000 \\
3 & (40.7, -148.6) & 1.83 & 0.60 & 0.40 & 600,000 &
371,000 \\
4 & (59.5, -71.8) & 2.51 & 0.44 & 0.56 & 441,000 & 
251,000 \\
5 & (78.0, -59.5) & 3.65 & 0.36 & 0.64 & 361,000 &
152,000 \\
6 & (61.9, -40.9) & 40.9 & 0.11 & 0.89 & 110,890 &
15,000 
\enddata

\tablenotetext{a}{Hydrogen number density}
\tablenotetext{b}{Mixed temperature if there had been no radiative 
  cooling. It is calculated from 
  $\bar{T}=T_{hot} \times f_{hot}+T_{cool} \times f_{cool}$ where
  $T_{hot}=1.0\times10^6$ K, $T_{cool}=1.0\times10^3$ K, and $f_{hot}$ and
  $f_{cool}$ are mass fractions of hot and cool gas, respectively}
\tablenotetext{c}{Temperature measured at the cell from the simulation}
\end{deluxetable*}

\begin{figure*}[t]

\centering

\includegraphics[scale=0.08]{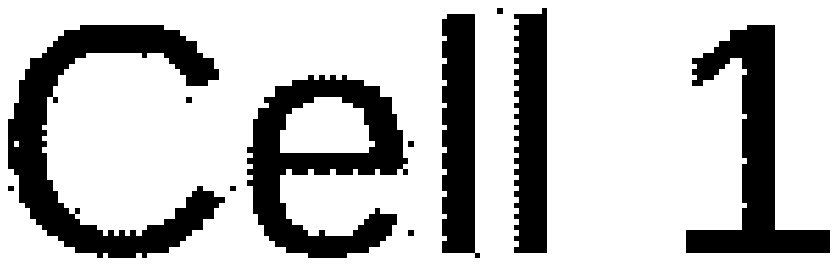}
\includegraphics[scale=0.18]{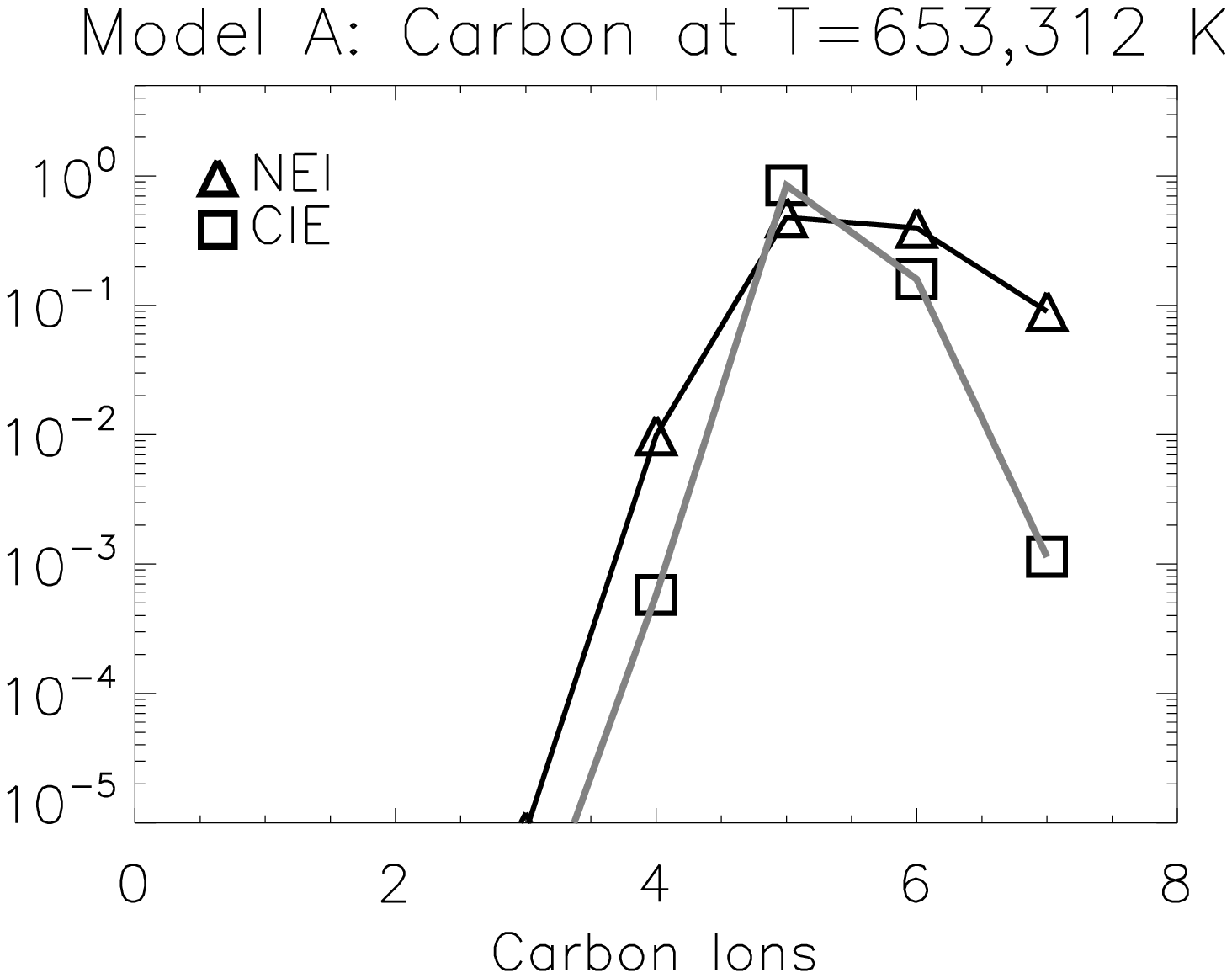}
\includegraphics[scale=0.18]{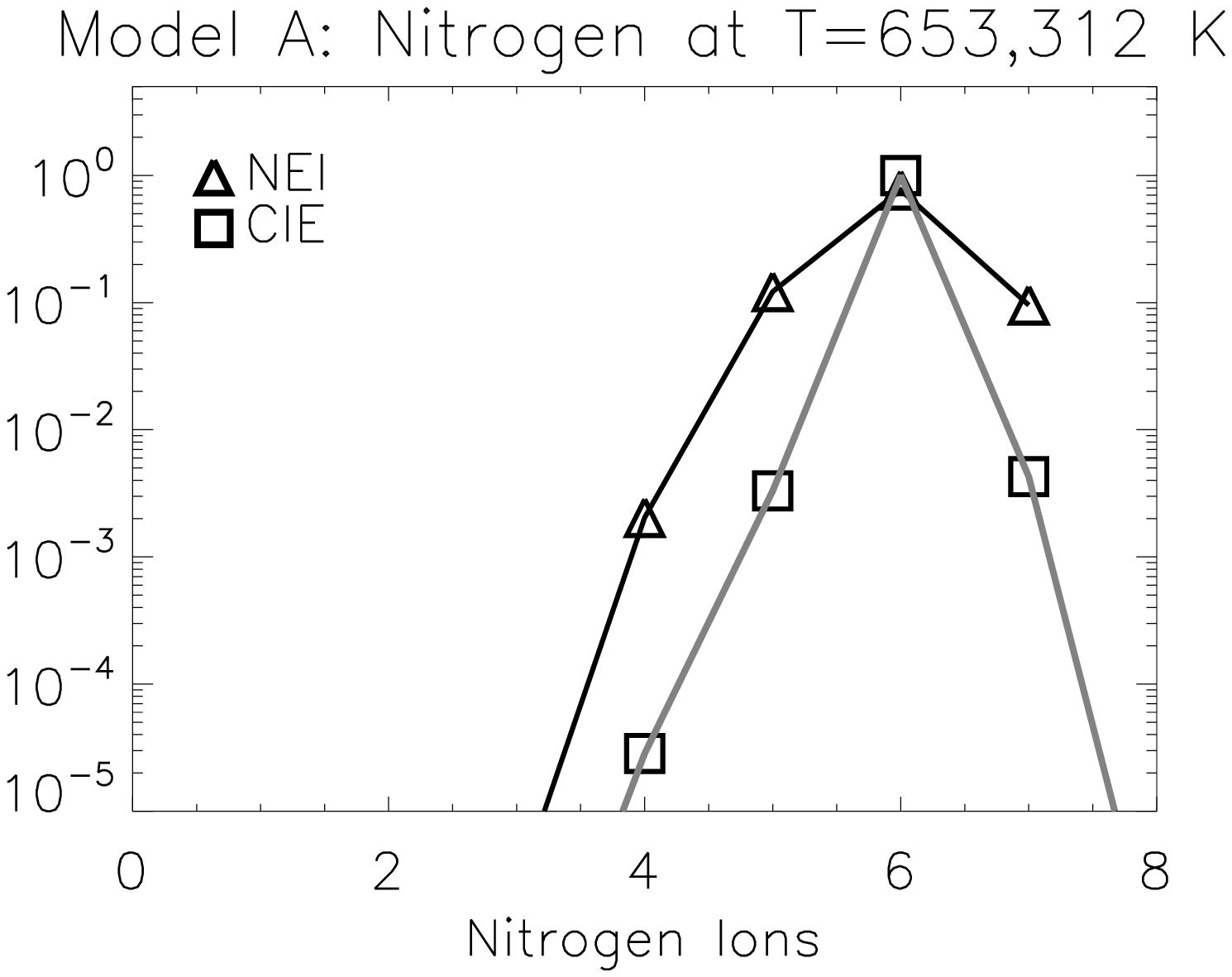}
\includegraphics[scale=0.18]{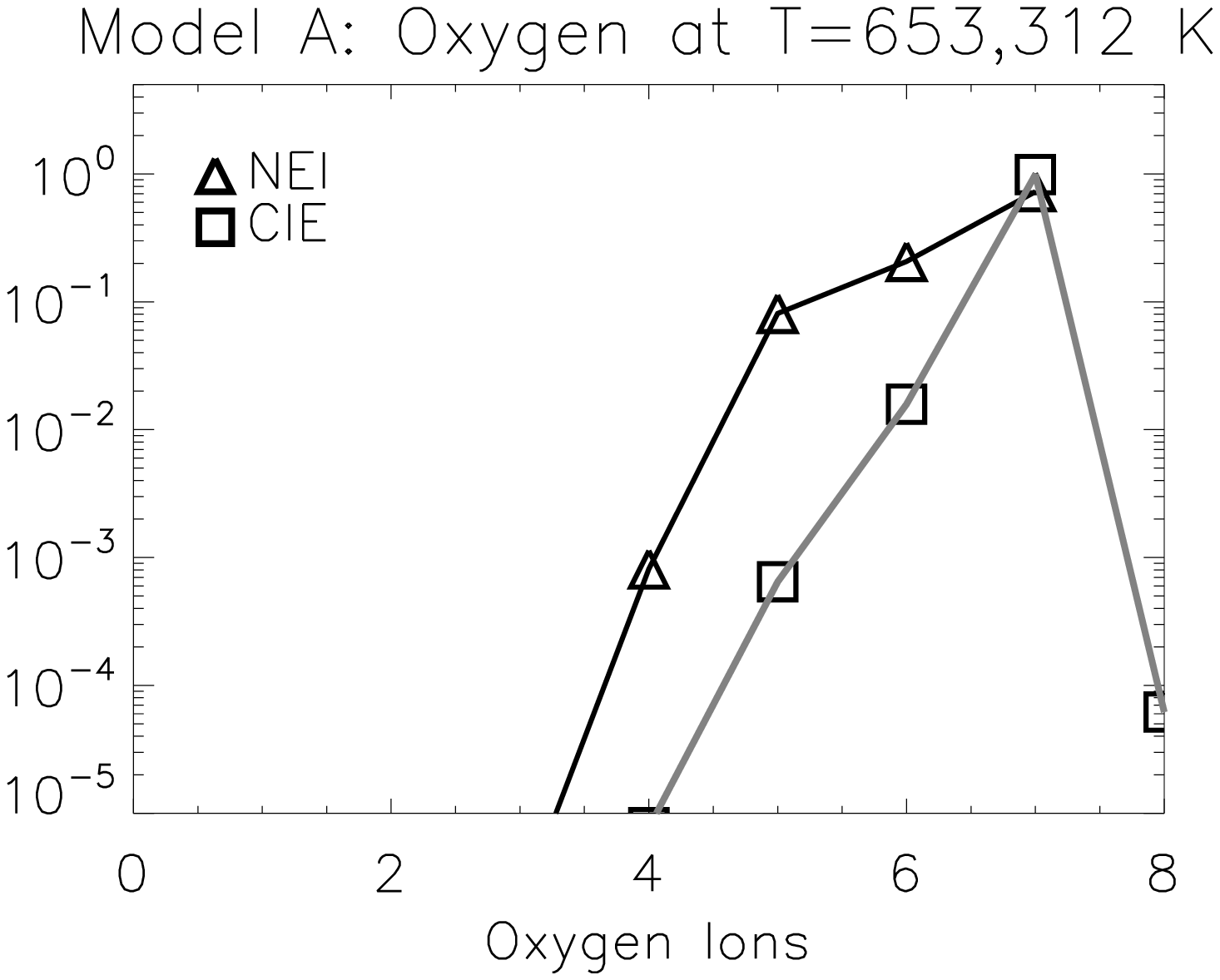} \\
\includegraphics[scale=0.08]{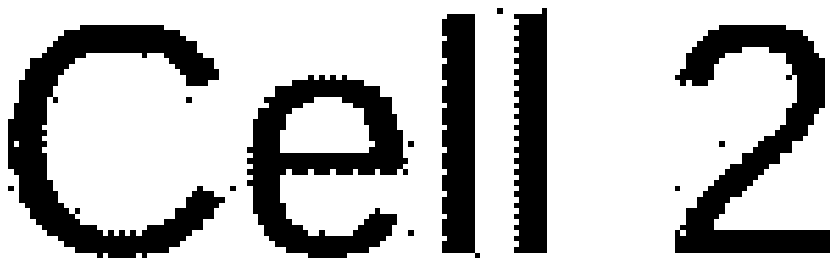}
\includegraphics[scale=0.18]{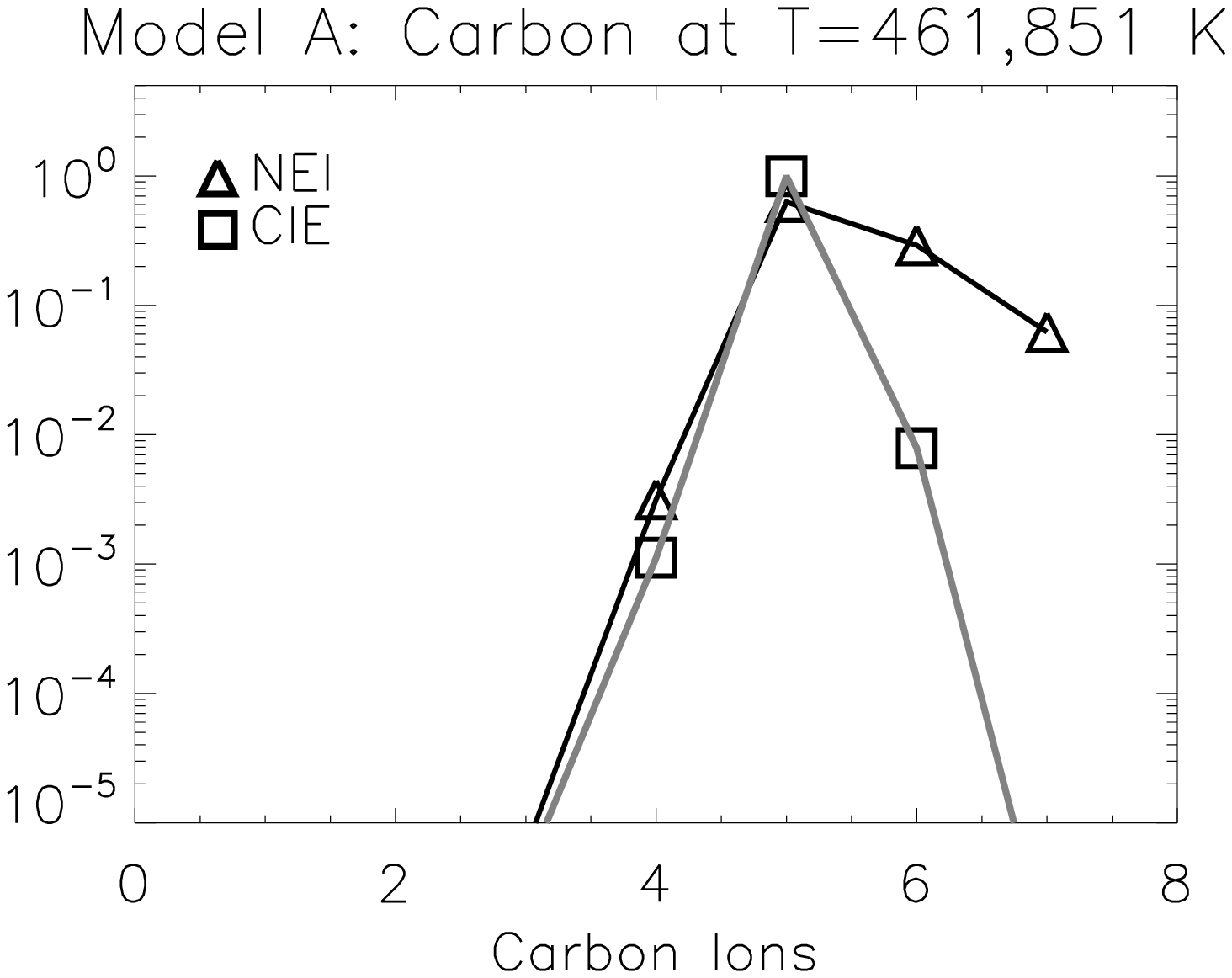}
\includegraphics[scale=0.18]{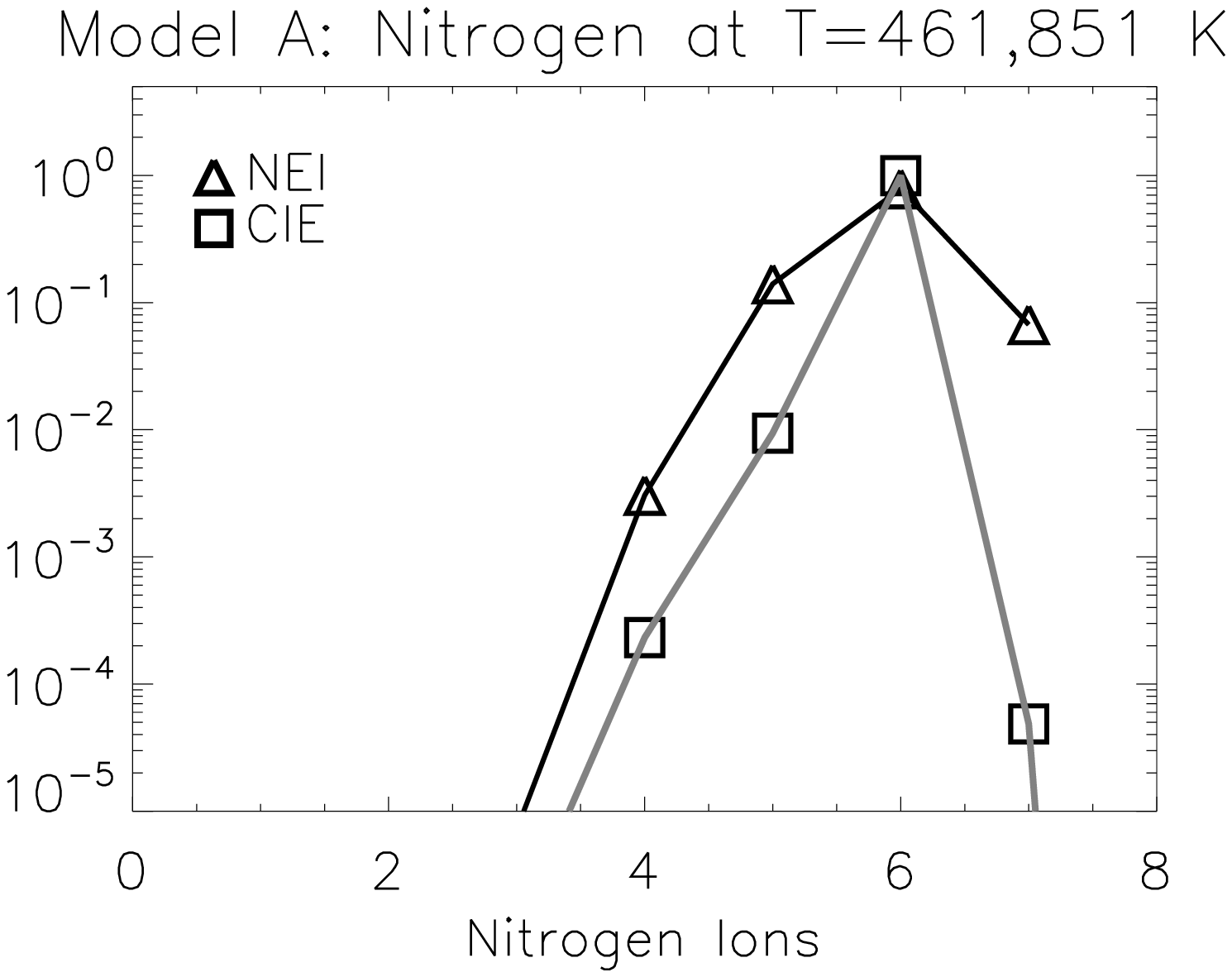}
\includegraphics[scale=0.18]{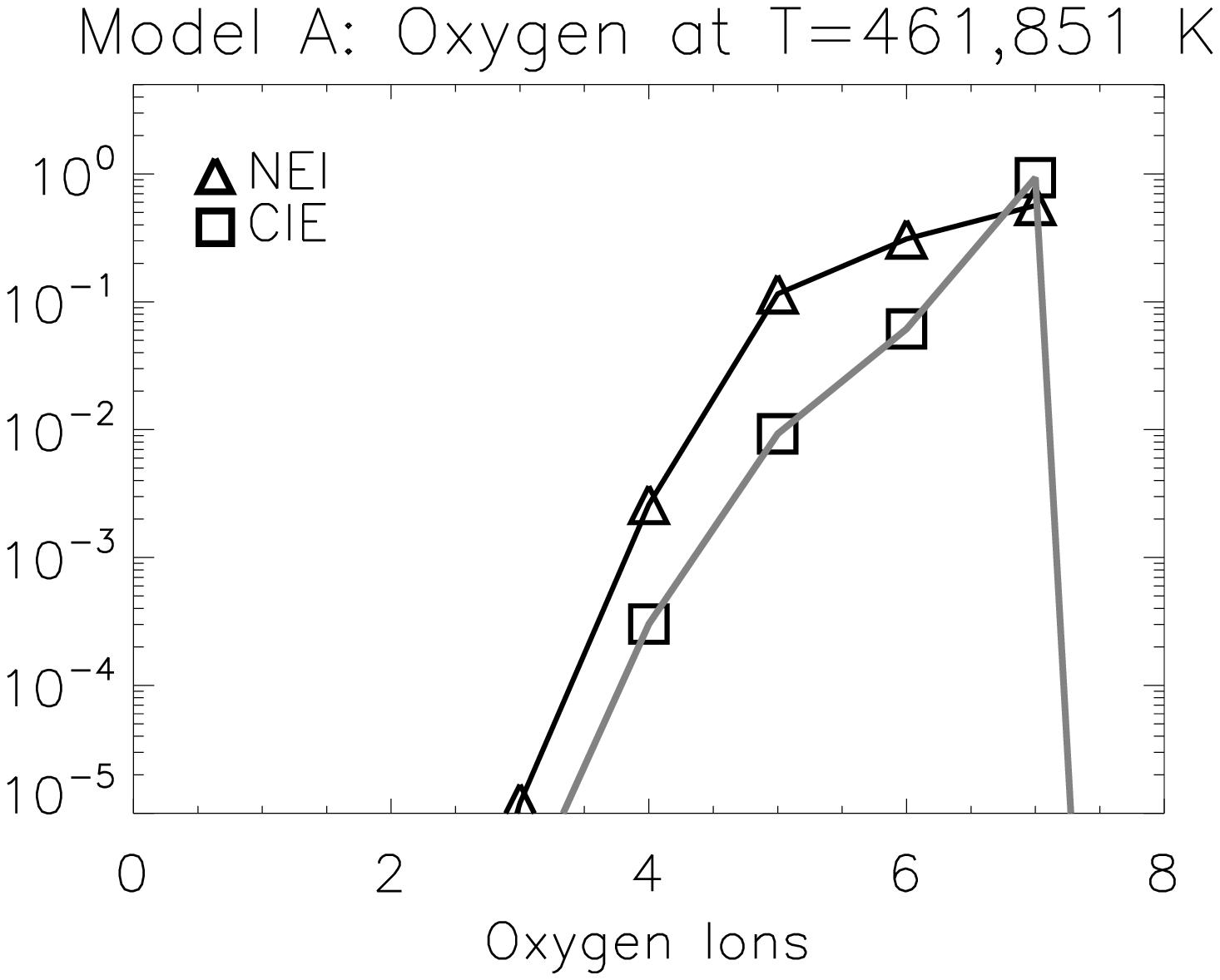} \\
\includegraphics[scale=0.08]{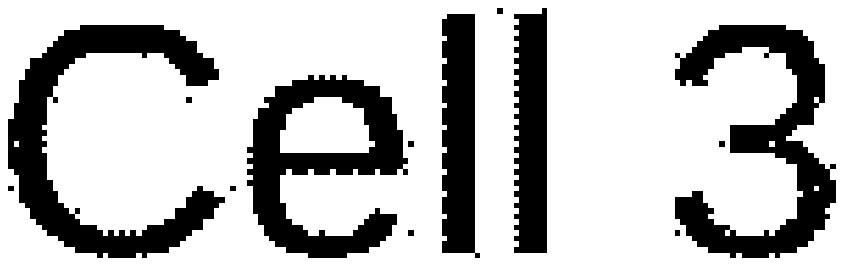}
\includegraphics[scale=0.18]{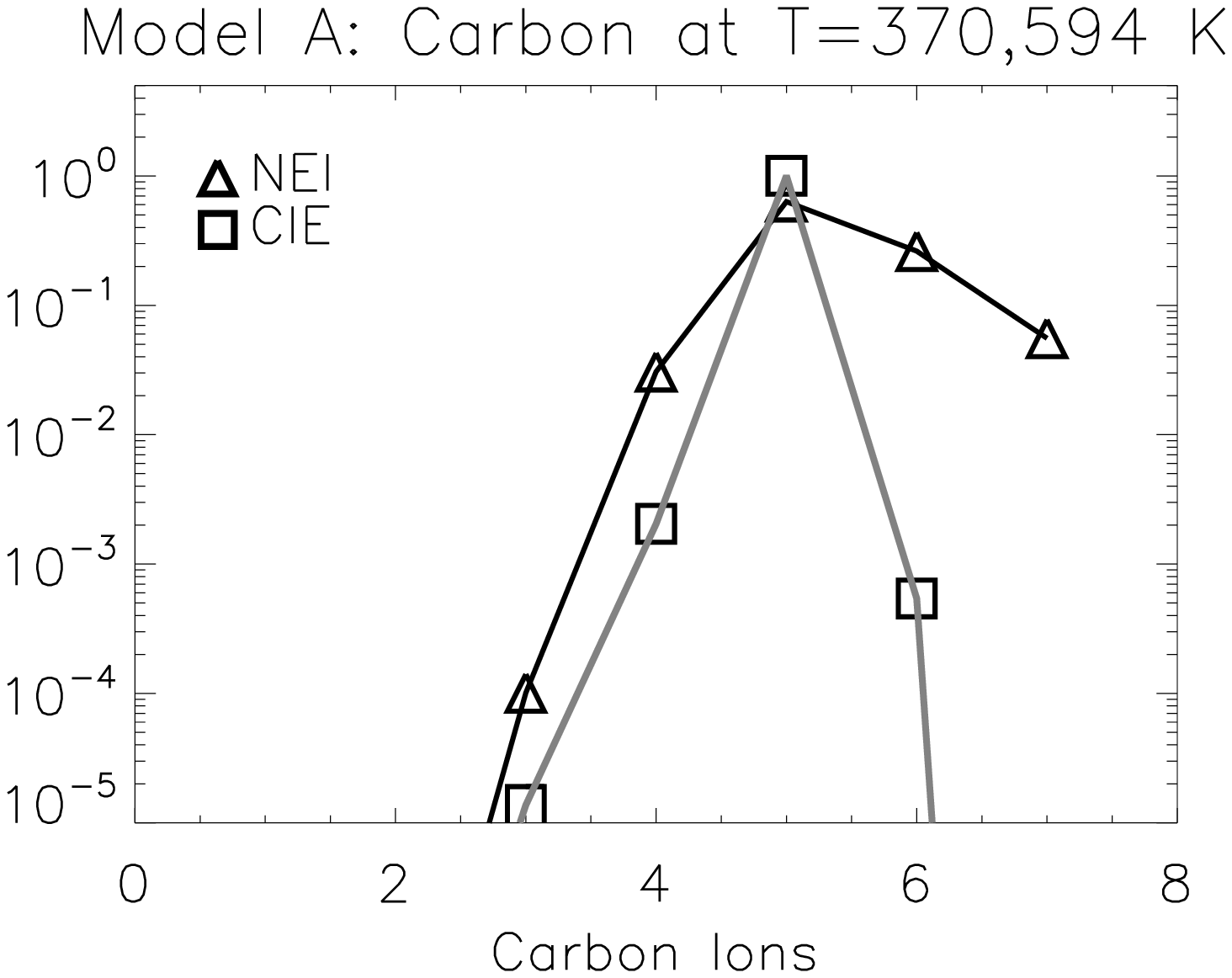}
\includegraphics[scale=0.18]{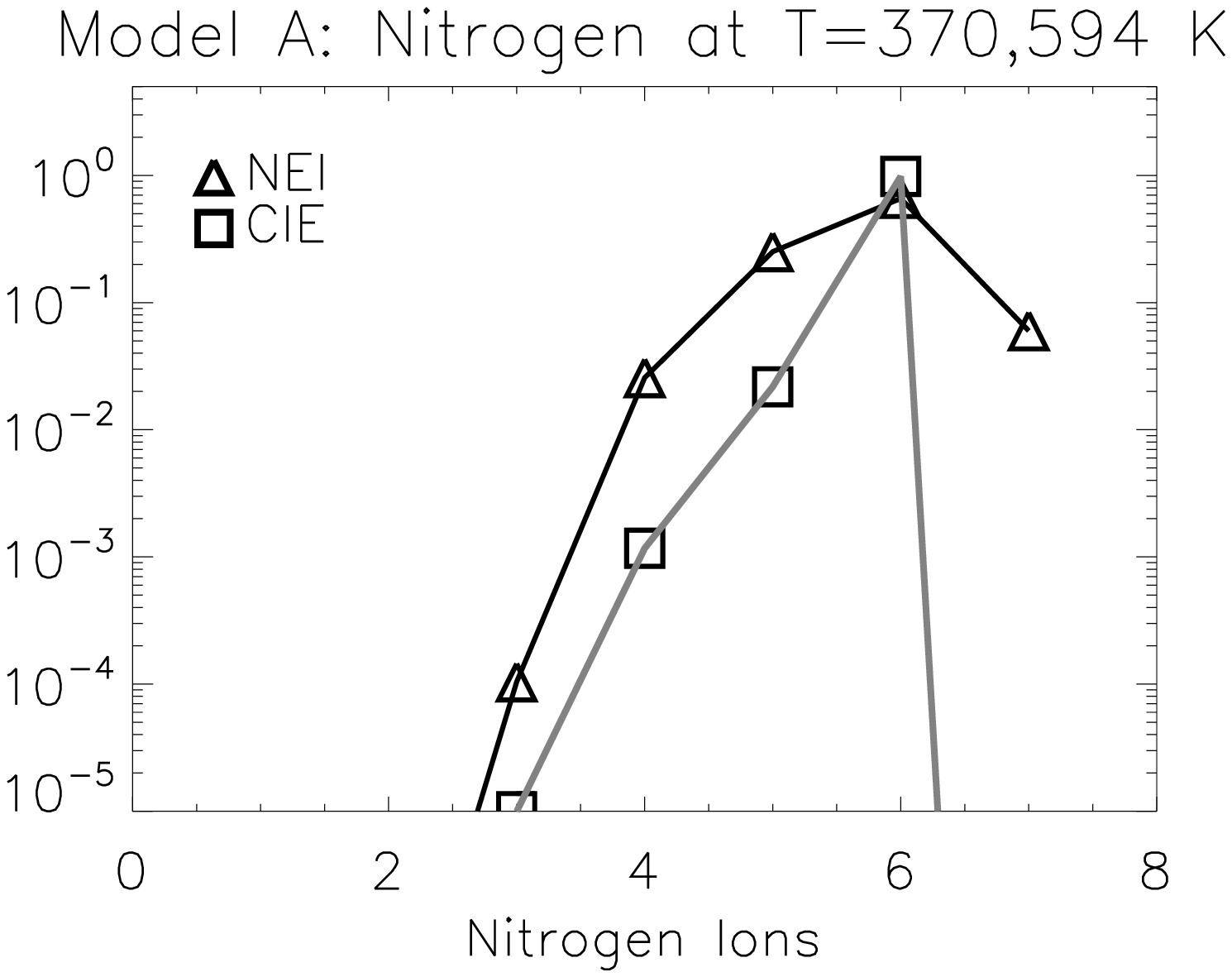}
\includegraphics[scale=0.18]{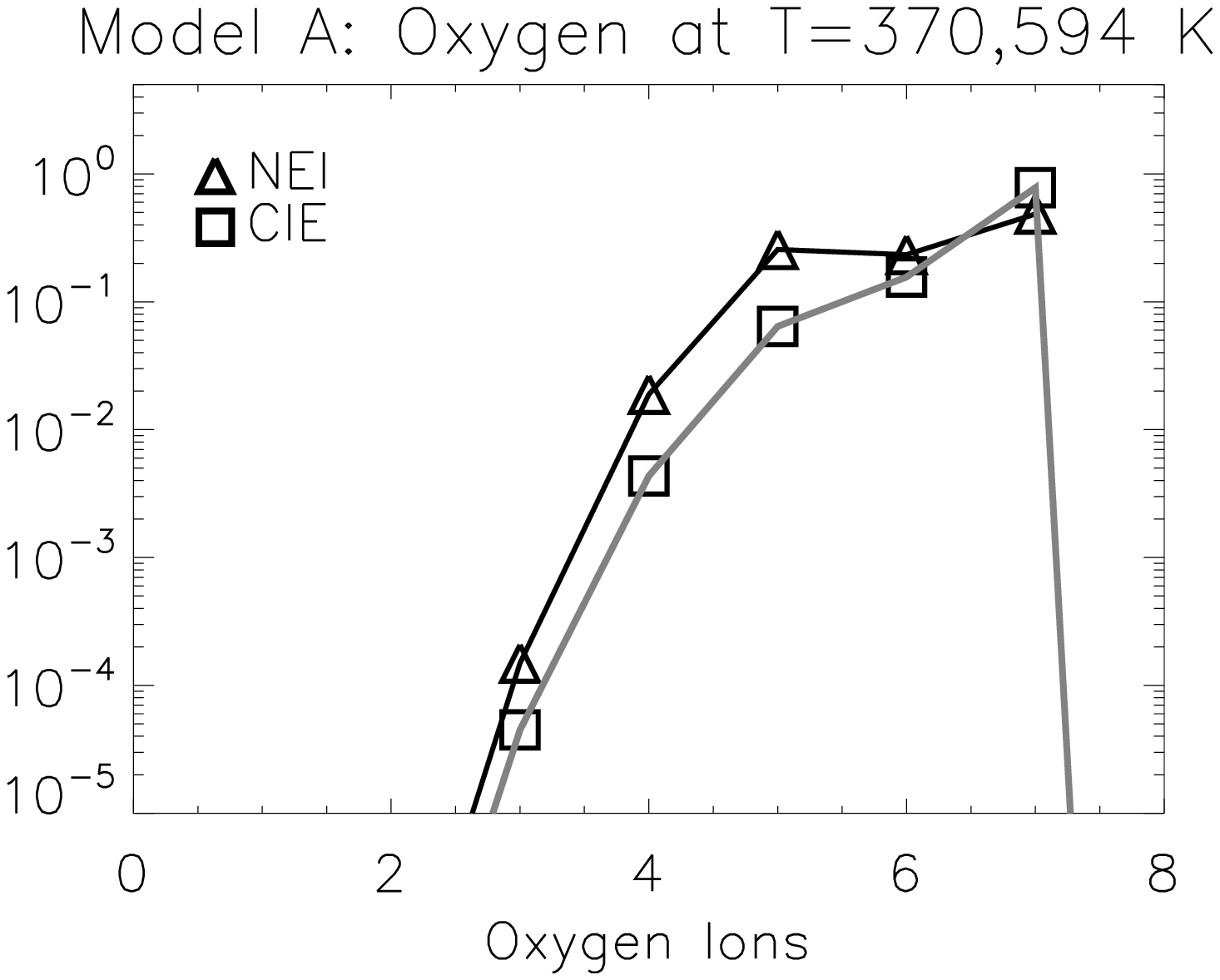} \\
\includegraphics[scale=0.08]{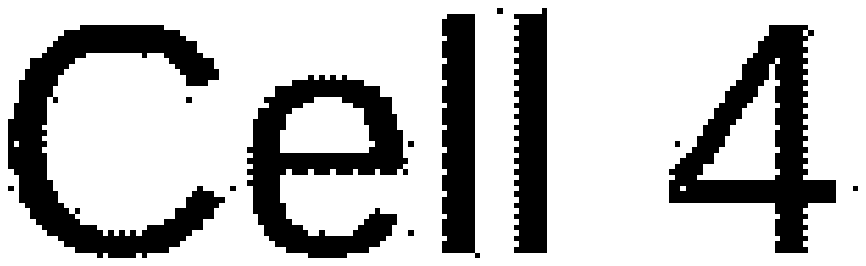}
\includegraphics[scale=0.18]{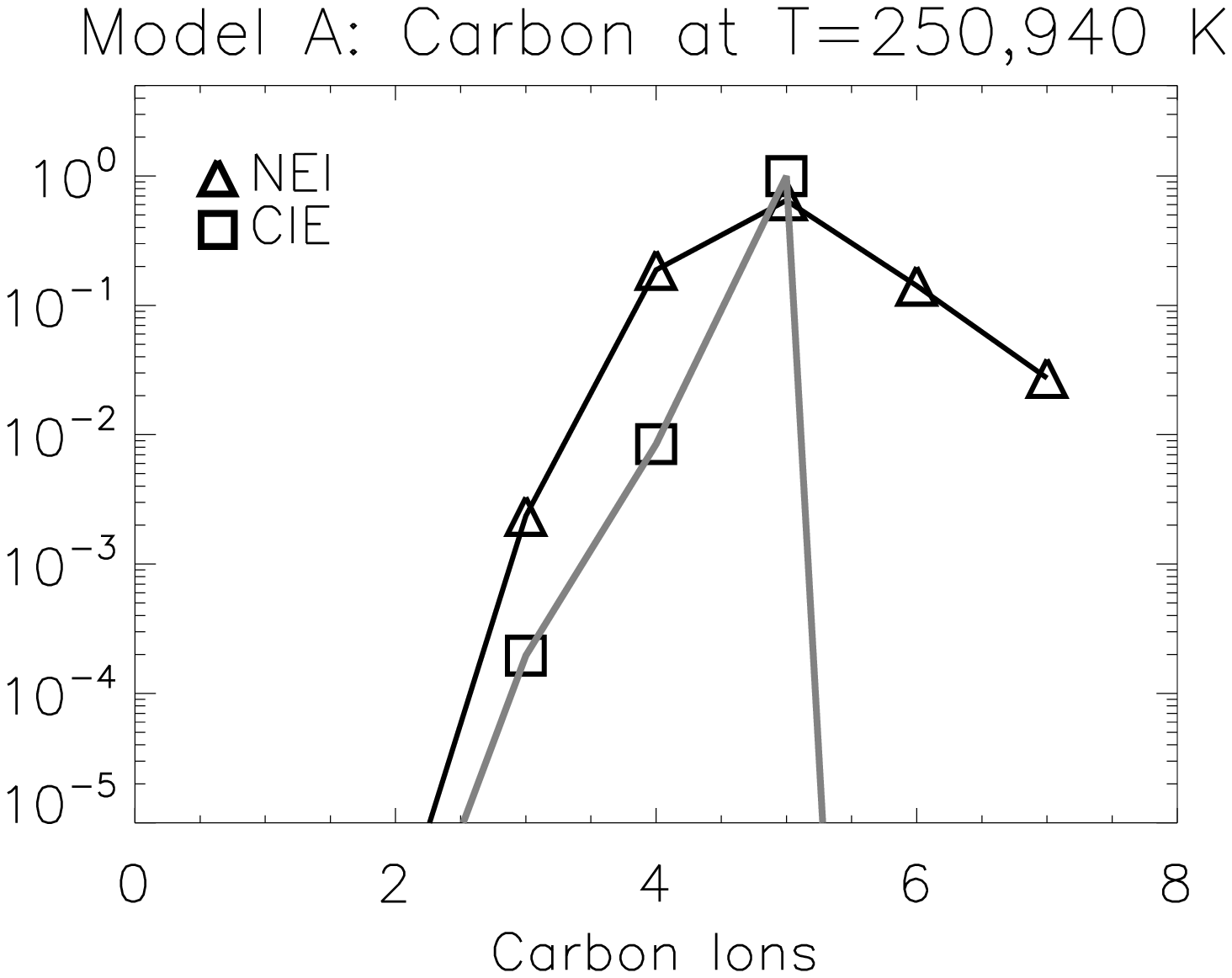}
\includegraphics[scale=0.18]{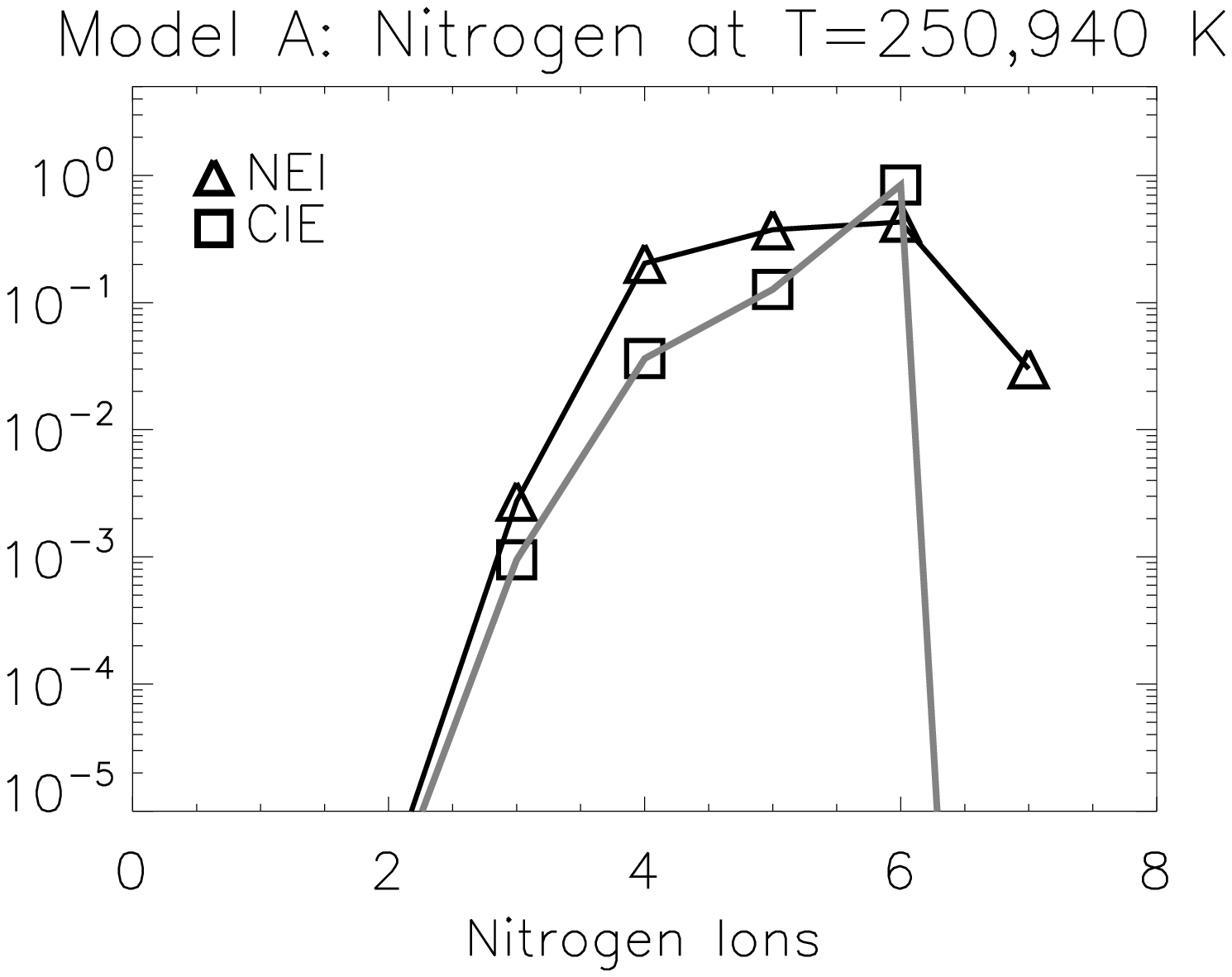}
\includegraphics[scale=0.18]{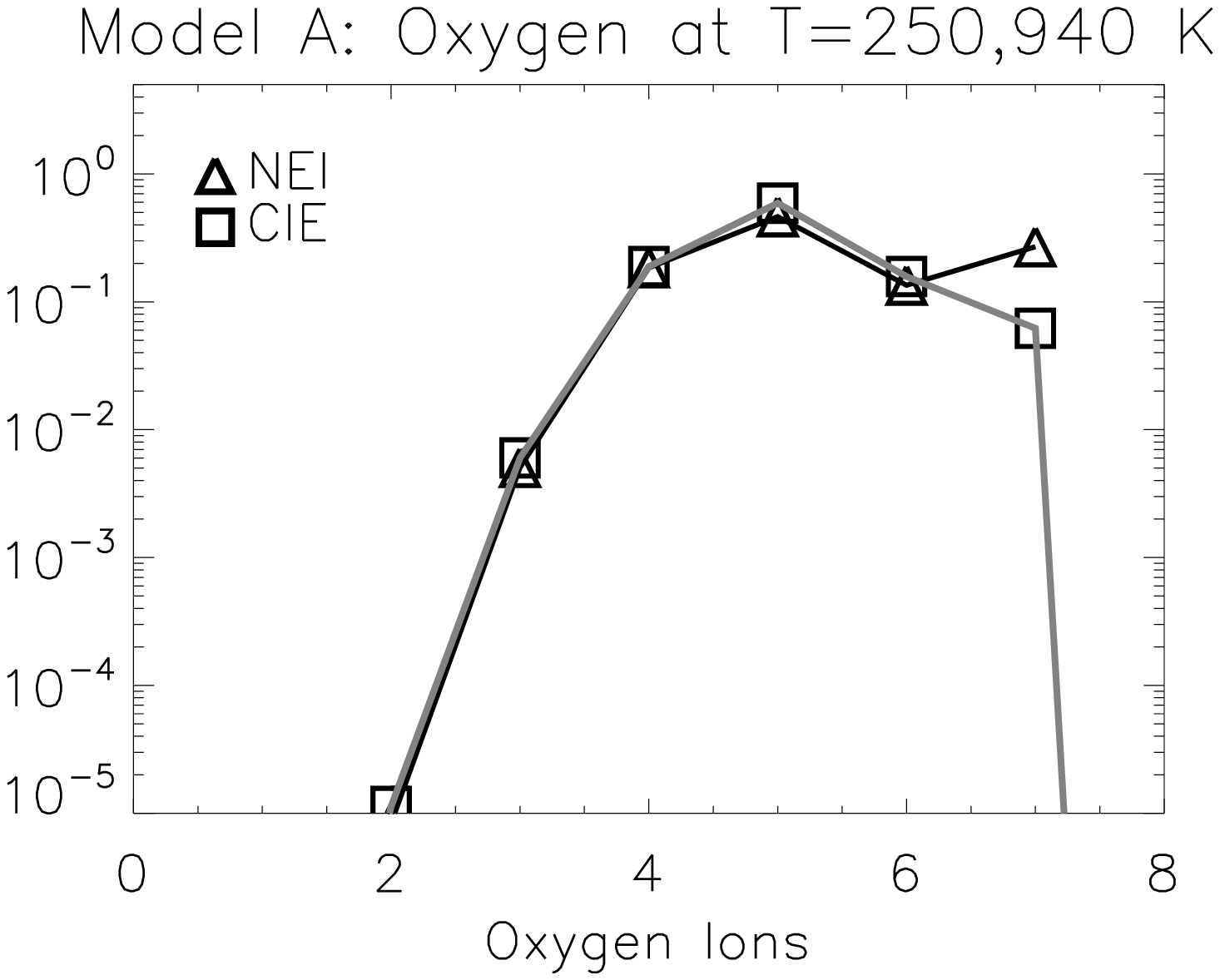} \\
\includegraphics[scale=0.08]{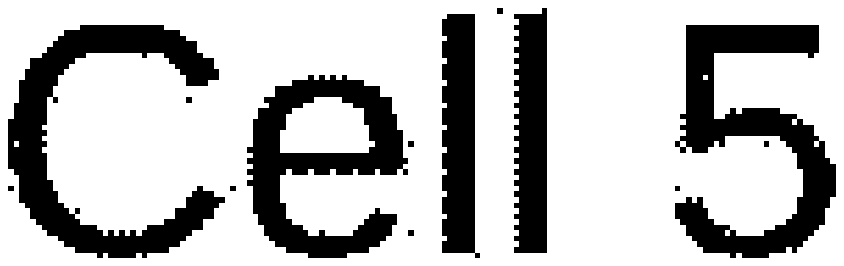}
\includegraphics[scale=0.18]{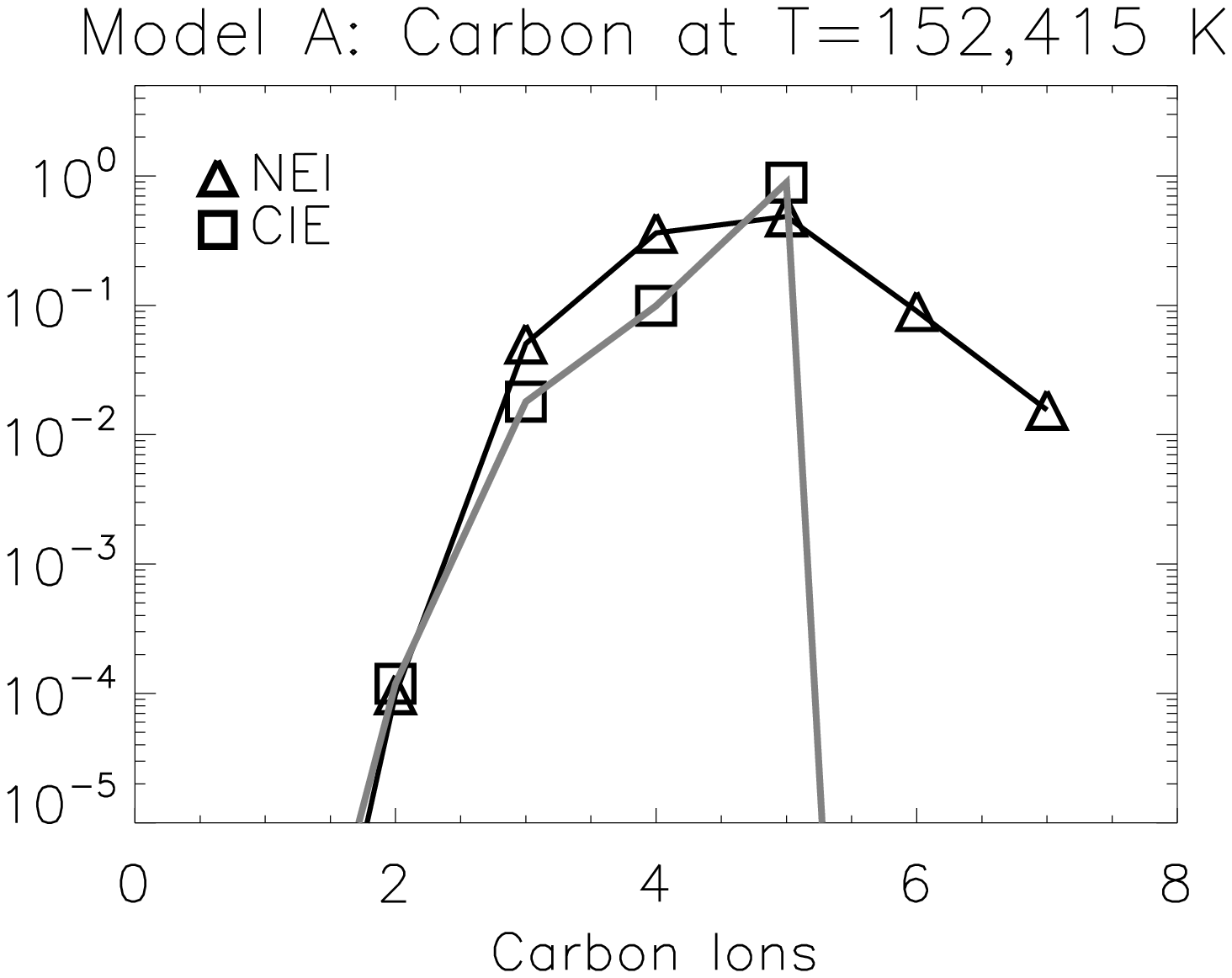}
\includegraphics[scale=0.18]{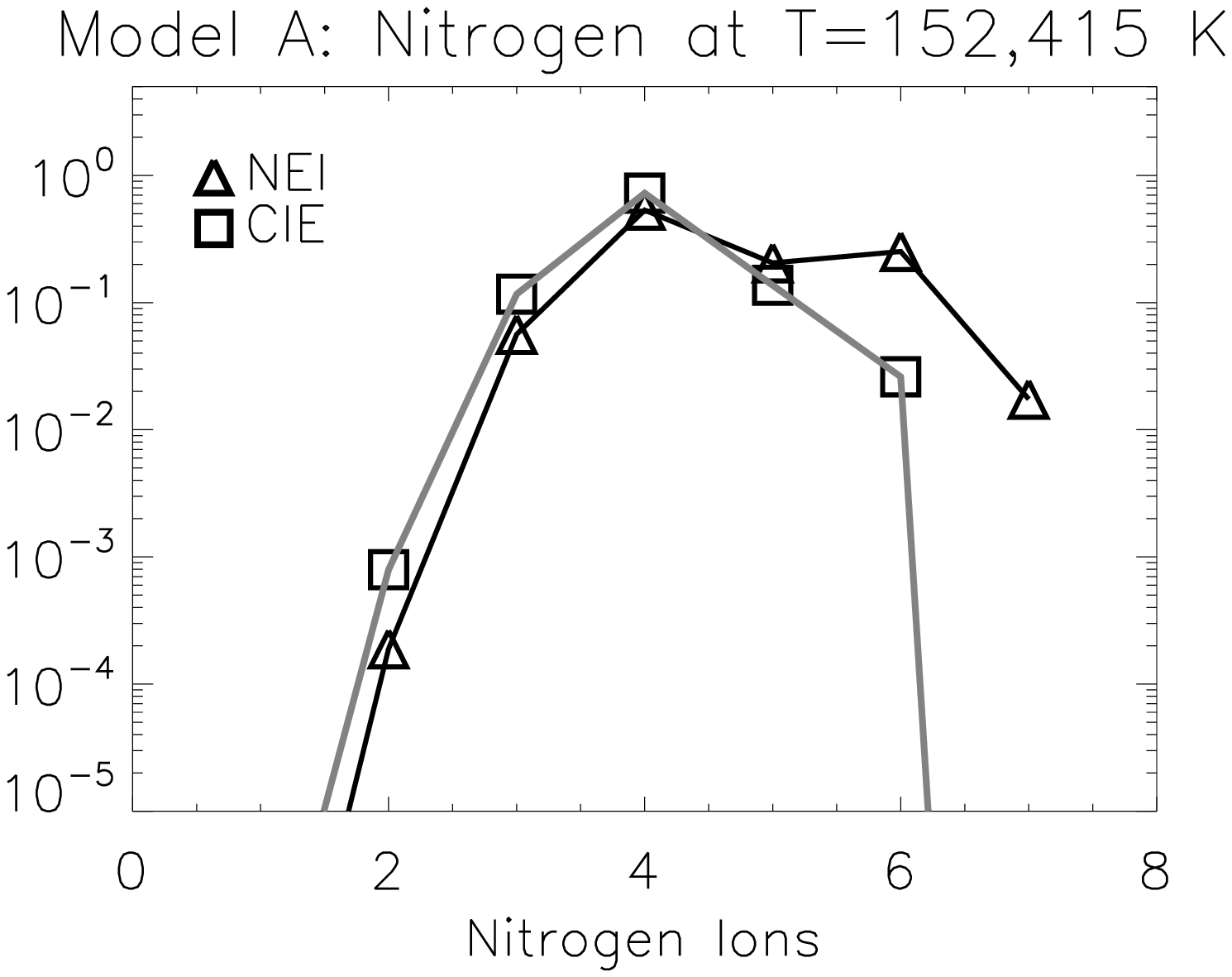}
\includegraphics[scale=0.18]{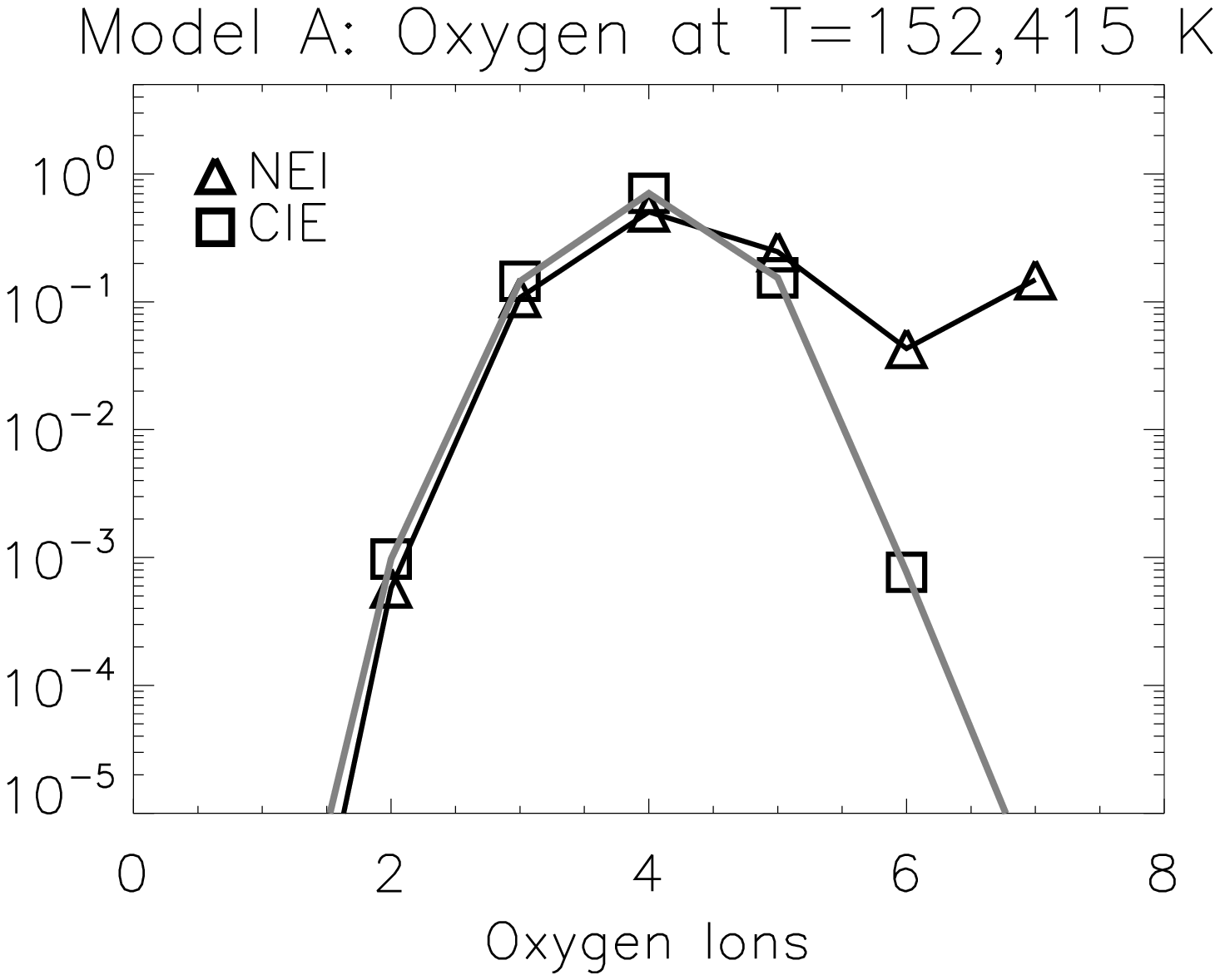} \\
\includegraphics[scale=0.08]{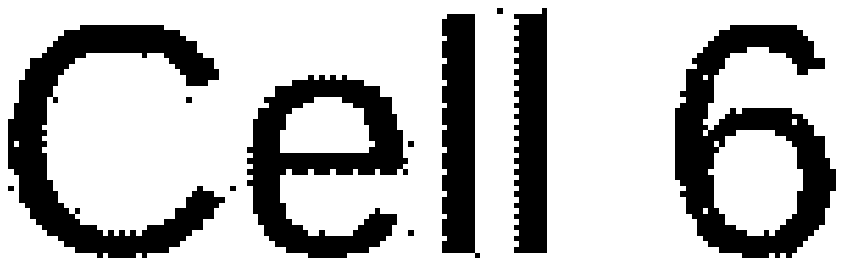}
\includegraphics[scale=0.18]{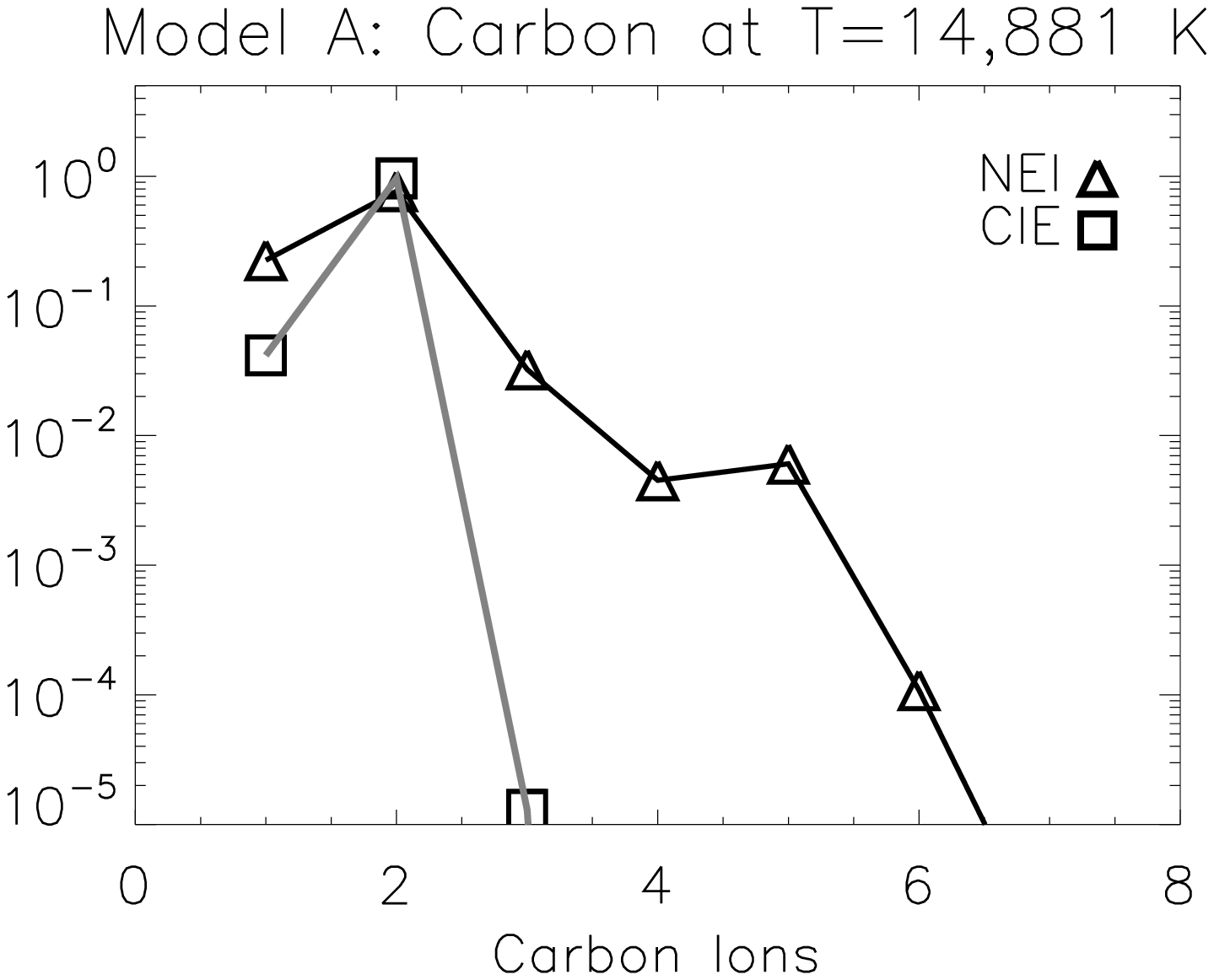}
\includegraphics[scale=0.18]{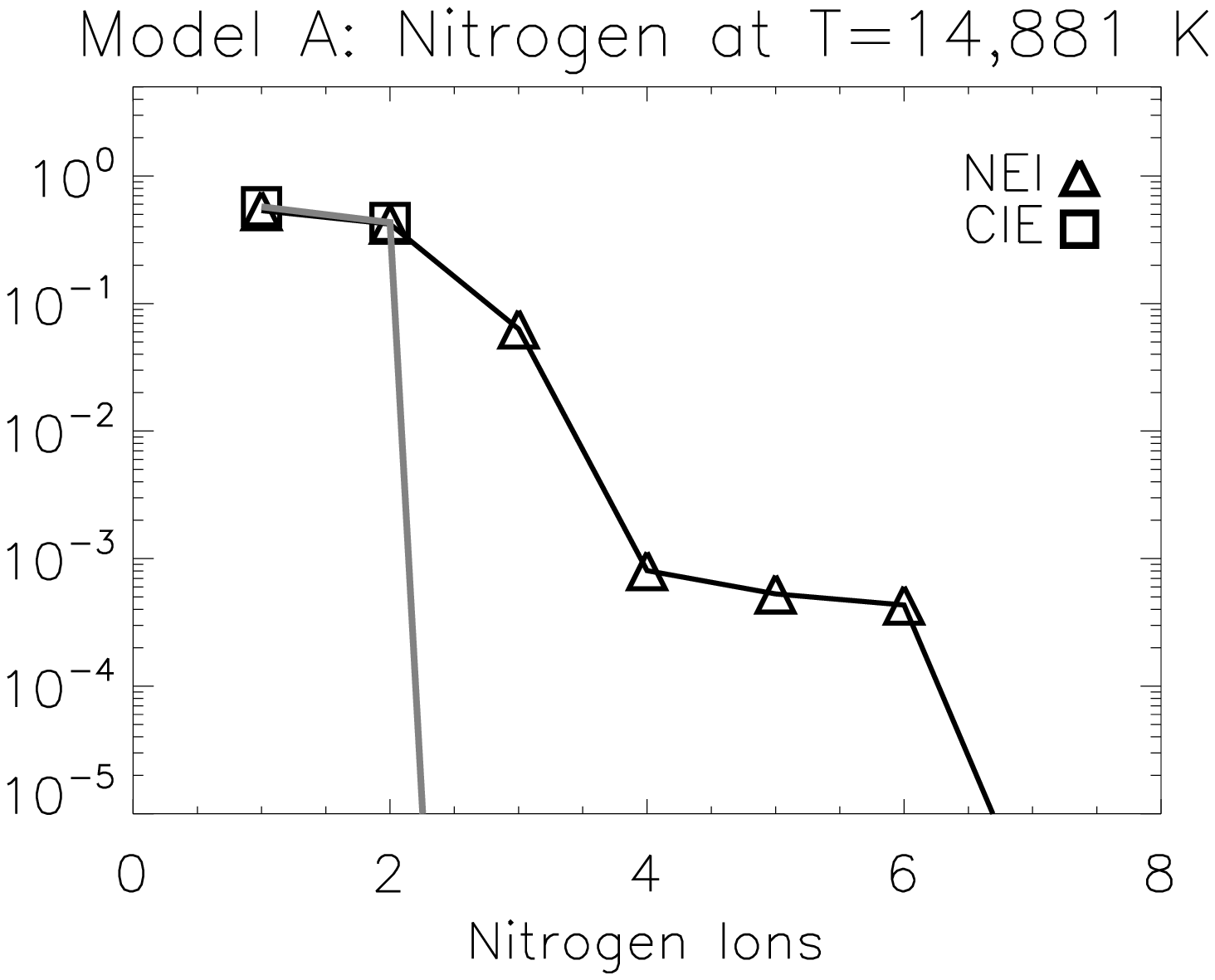}
\includegraphics[scale=0.18]{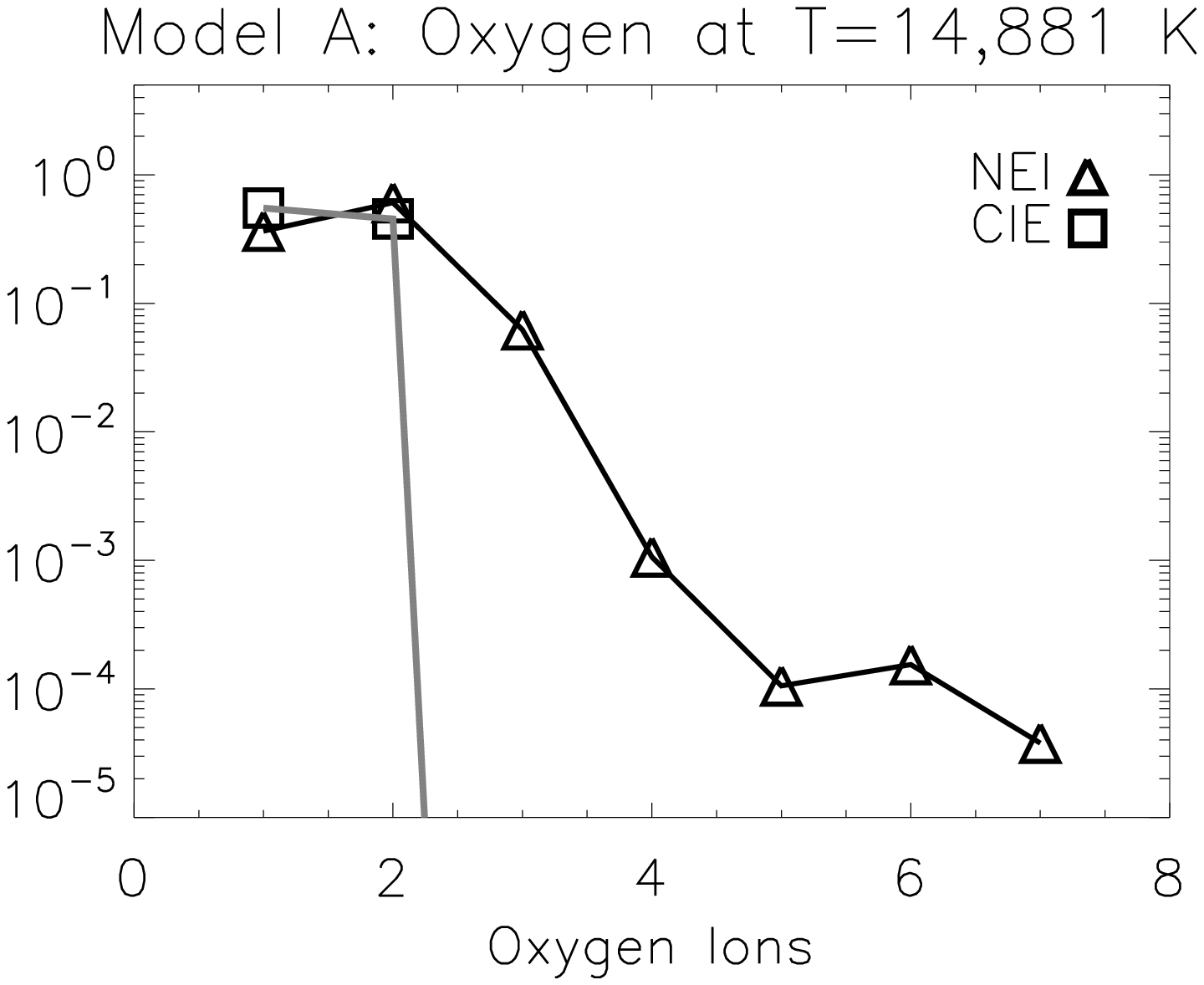}

\caption{Model A at $t=80$ Myr: fraction of atoms in various
  ionization levels versus ionization level for three elements, 
  carbon, nitrogen, and oxygen. The ionization levels are indicated 
  by 1, 2, 3, etc on the abscissa, where, for example, 1 on a carbon
  plot refers to \ion{C}{1}. Values from NEI calculations are
  indicated by triangles, while values from CIE calculations are
  indicated by squares. The NEI ion fractions of \ion{N}{8}, 
  \ion{O}{8}, and \ion{O}{9} are not included in the plot although 
  they were traced in our NEI calculations. Note that the CIE ion 
  fractions at low temperatures are as small as the numerical 
  precision in our CIE calculation code. 
  The top row of panels pertains to cell 1 in
  Figure \ref{cell_location_fig}, while the subsequent panels pertain
  to cells 2 through 6 in Figure \ref{cell_location_fig}. Information
  about each zone is given in Table \ref{cell_neicie_T}. 
  \label{modelA_ioniz_fig}}

\end{figure*}

The hot reservoir is in or near collisional ionization equilibrium 
and thus mixing such gas into a cell raises the fraction of very 
highly ionized atoms in the cell. For this reason, cell 1 contains 
large numbers of hydrogen--like, helium--like, and fully stripped 
carbon, as indicated by the top--left panel of 
Figure \ref{modelA_ioniz_fig}. 
(Figure \ref{modelA_ioniz_fig} shows the ion fractions, calculated 
from both the NEI and CIE algorithms, except for the NEI ion 
fractions of \ion{N}{8}, \ion{O}{8}, and \ion{O}{9}, which are not 
included in the plot although they were traced 
in our NEI calculations.) The greater prevalence of very high ions 
in the NEI case than in the CIE case is a sign of delayed 
recombination in NEI calculations; the recombination rate is slower 
than the mixing rate.

The ionization behavior in our NEI simulations is consistent with
previous studies with NEI calculations. 
\citet{Balletetal1986AA} studied the 
evaporation of a spherical gas cloud and 
found that in NEI calculations, ionization to He--like stages 
is delayed. In their calculations, 
the gas is heated via conduction, while in our simulations, 
the gas is heated due to mixing. 
\citet{BoehringerHartquist1987MNRAS} added radiative
cooling to the calculations of \citet{Balletetal1986AA} and found
a similar delayed ionization trend in the conductive interface of the
evaporating cloud. Radiative cooling plays a more important role 
in the production of high ions 
when hot ($\ge$ a few times $10^6$ K) gas cools radiatively. 
Recently, \citet{GnatSternberg2007ApJS} numerically calculated 
the ionization states and corresponding radiative cooling
rates for the elements H, He, C, N, O, Ne, Mg, Si, S, and Fe. 
They included recent atomic data and investigated 
the effect of various metallicities. Their NEI calculations confirm 
that recombination is delayed relative to the radiative cooling rate. 
Similar recombination lags have been shown in earlier NEI 
calculations, which are summarized in \citet{GnatSternberg2007ApJS}. 
Previous CIE calculations in the context of radiatively cooled gas are
also reviewed in \citet{GnatSternberg2007ApJS}.

\subsubsection{Ratios between Column Densities of Different Ions}

\begin{figure*}

\centering

\includegraphics[scale=0.25]{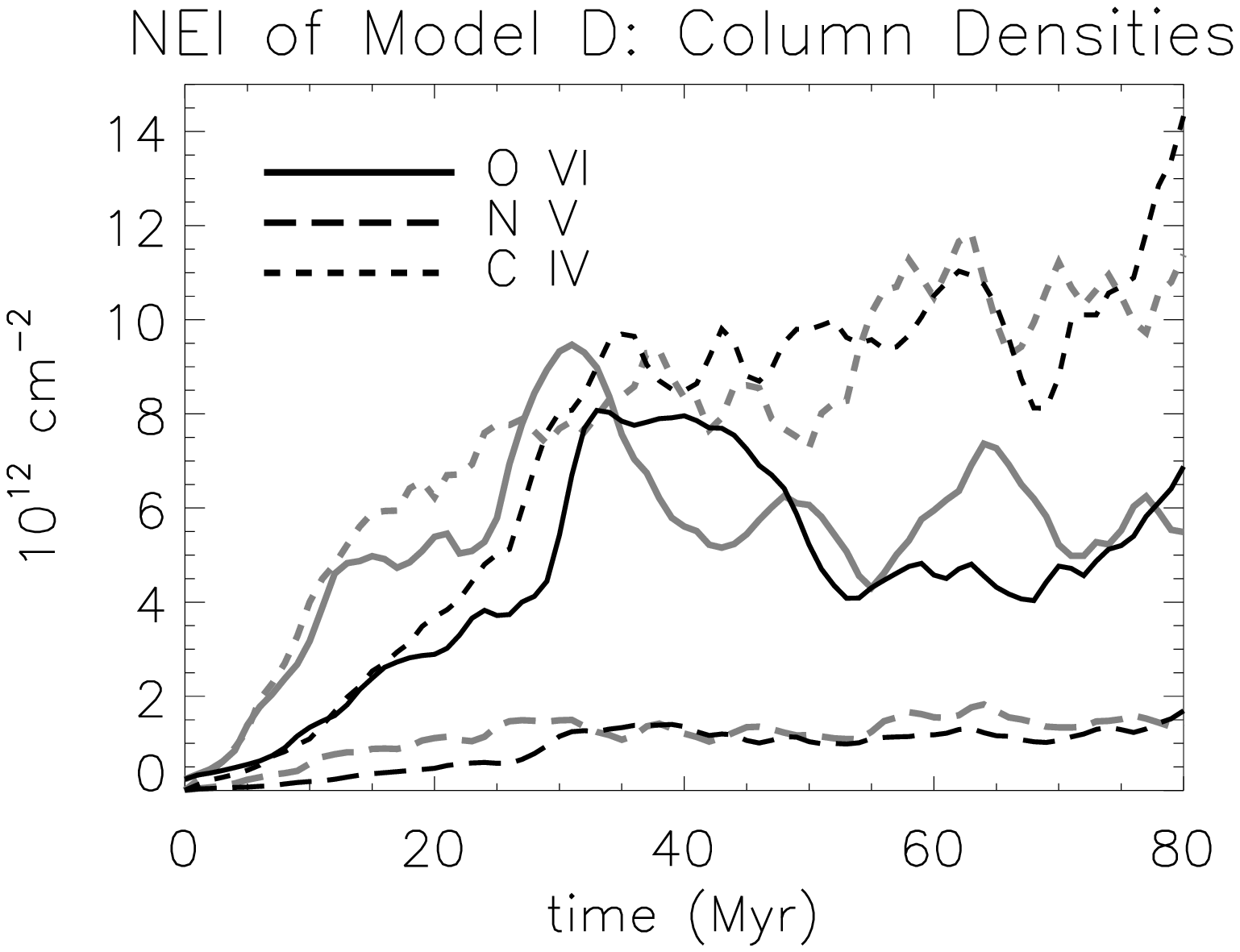}
\includegraphics[scale=0.25]{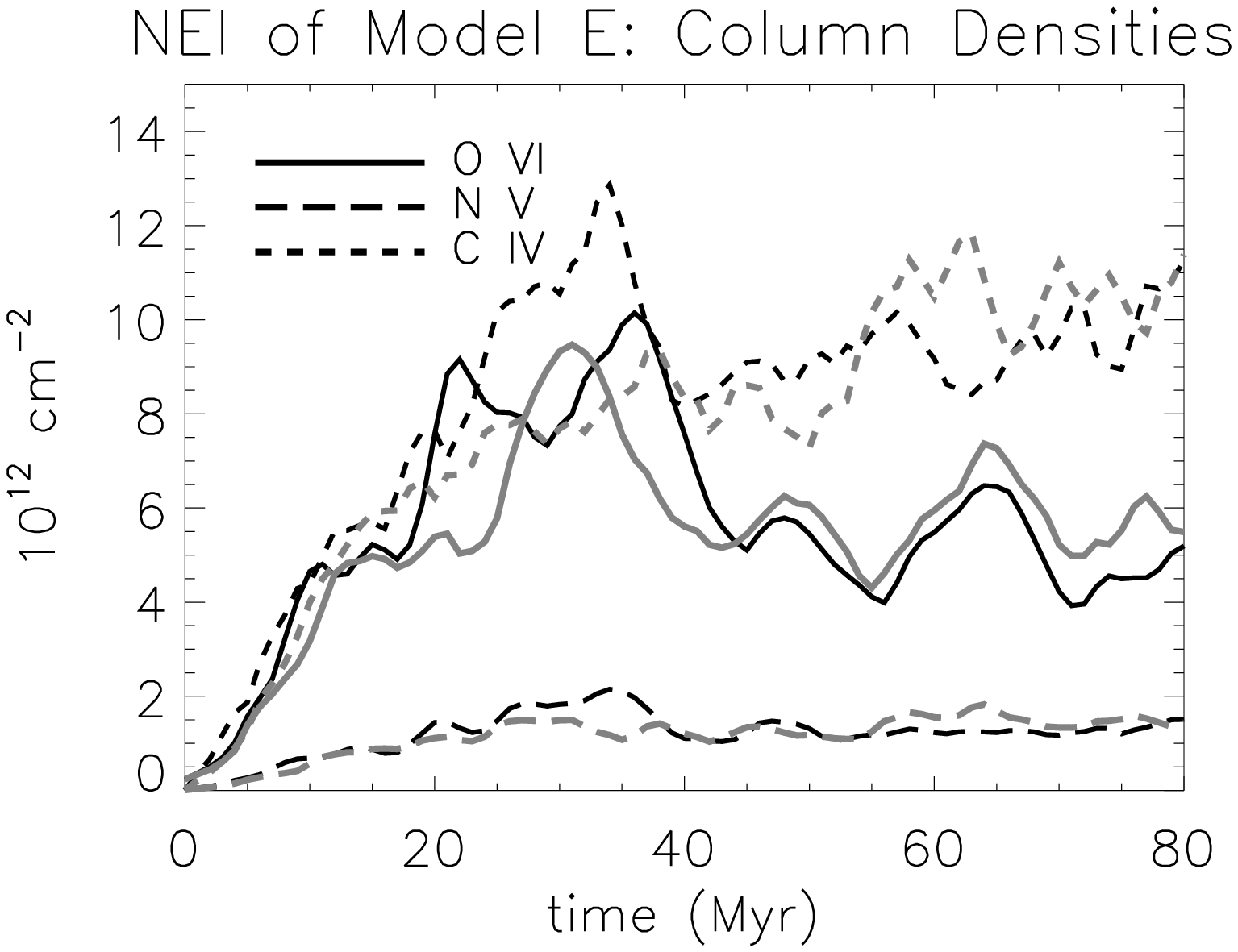}
\includegraphics[scale=0.25]{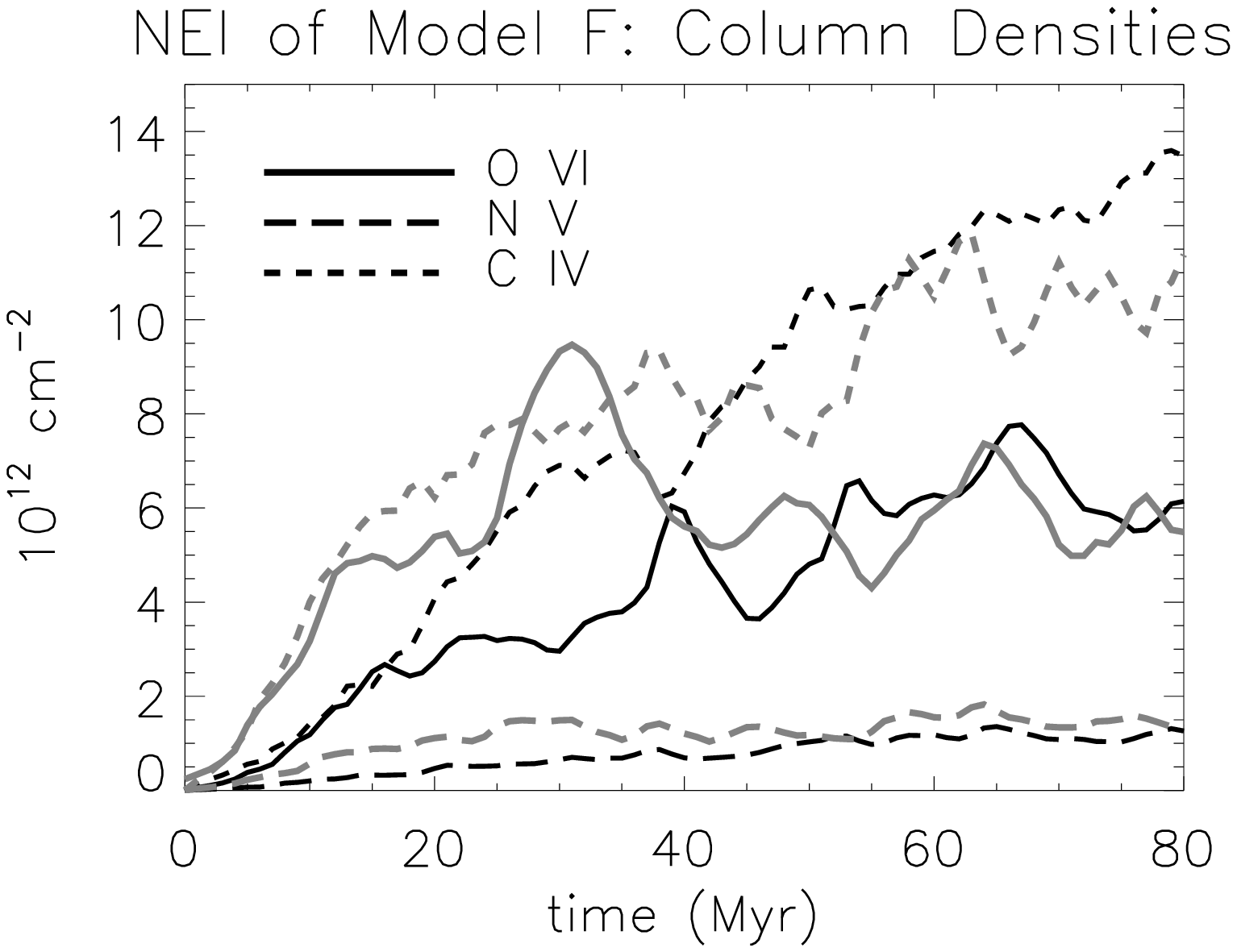} \\
\includegraphics[scale=0.25]{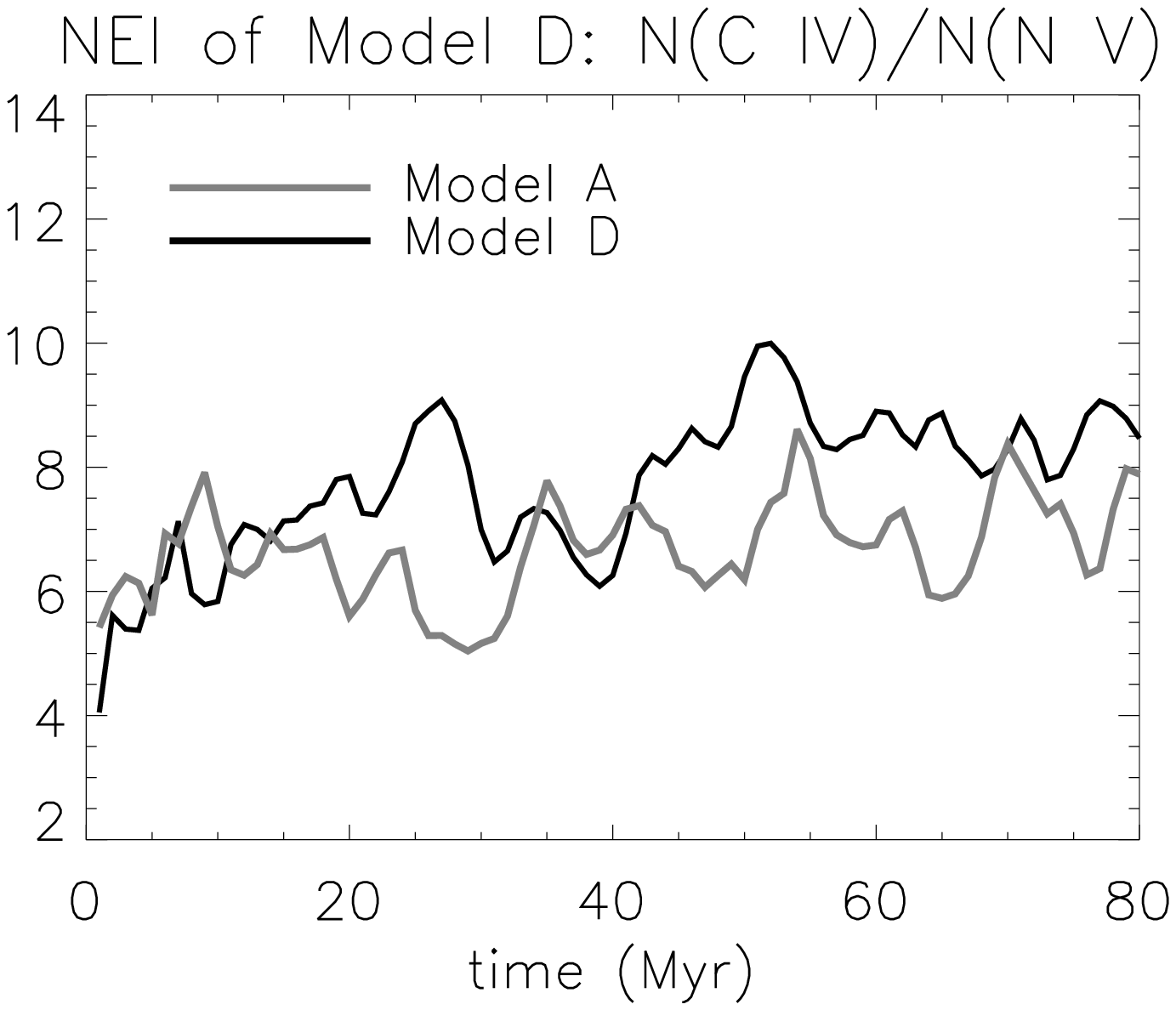}
\includegraphics[scale=0.25]{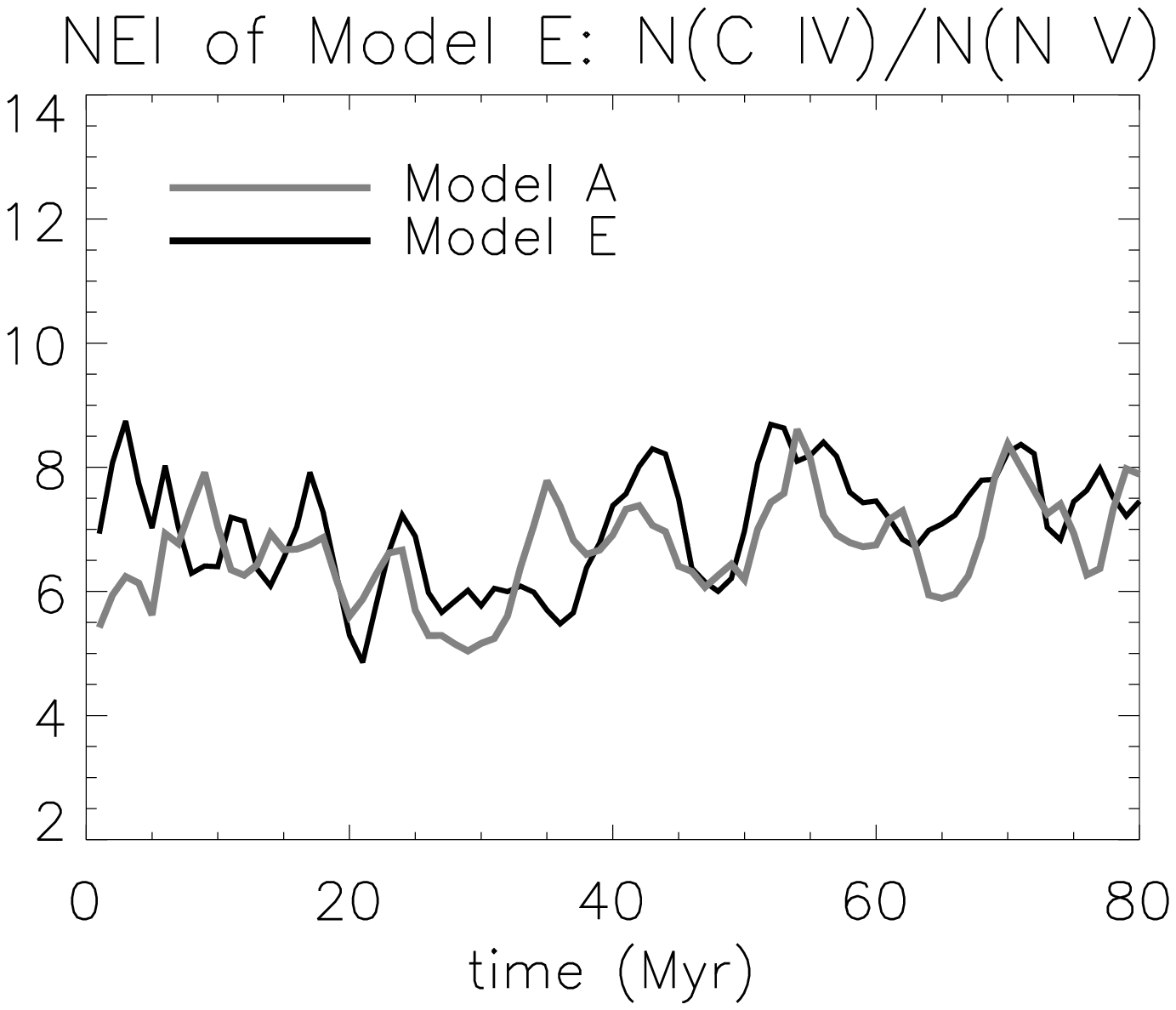}
\includegraphics[scale=0.25]{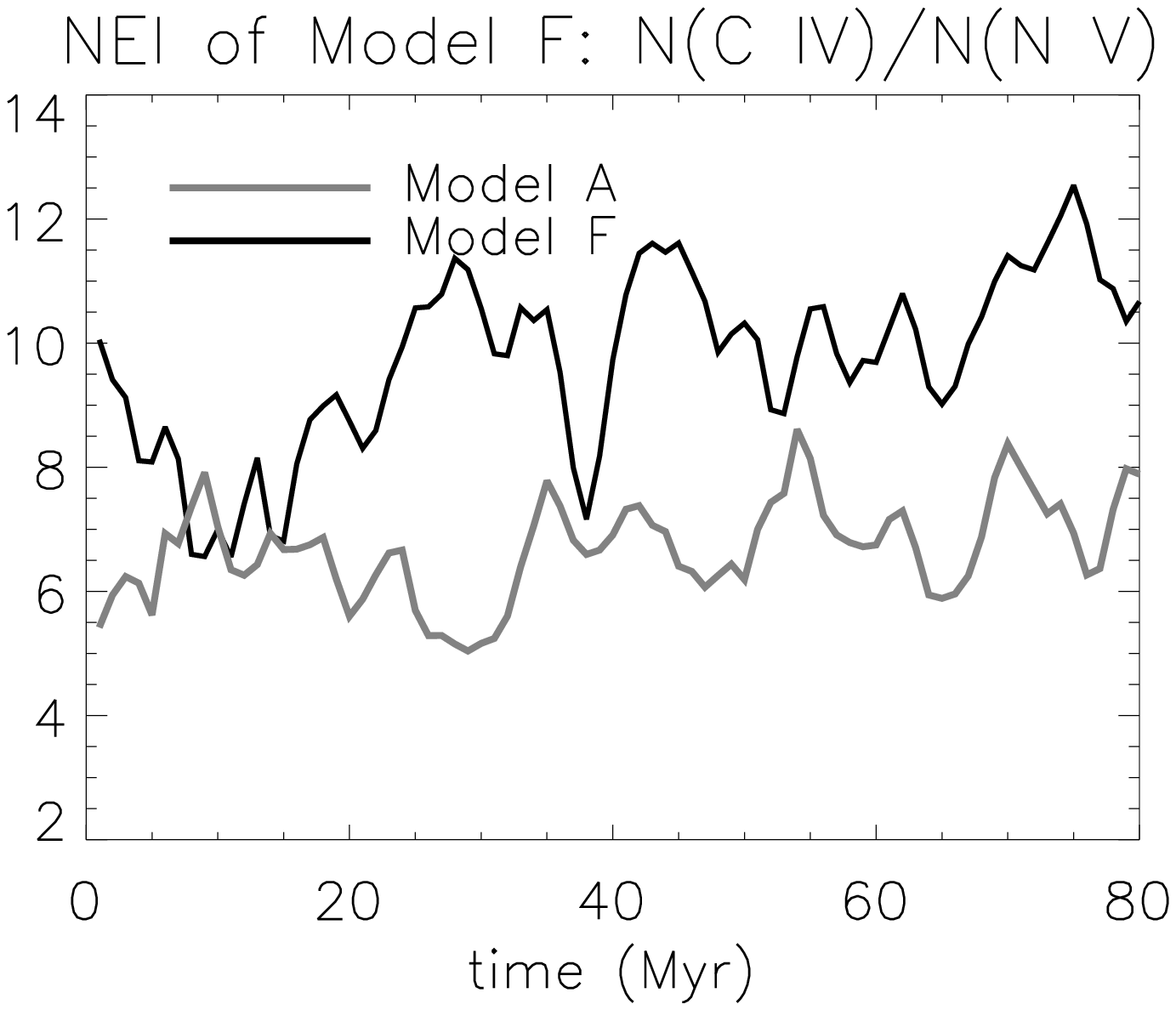} \\
\includegraphics[scale=0.25]{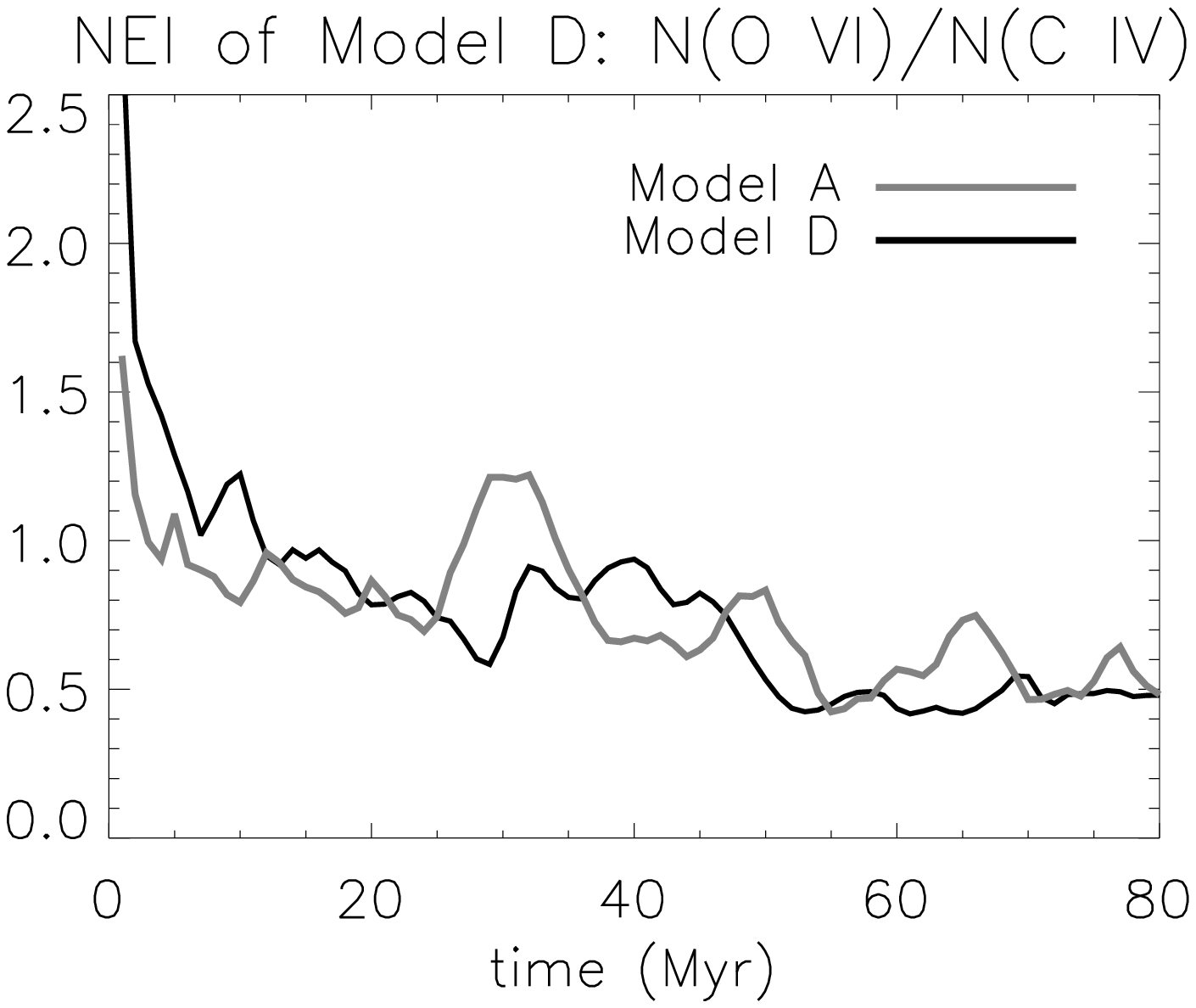}
\includegraphics[scale=0.25]{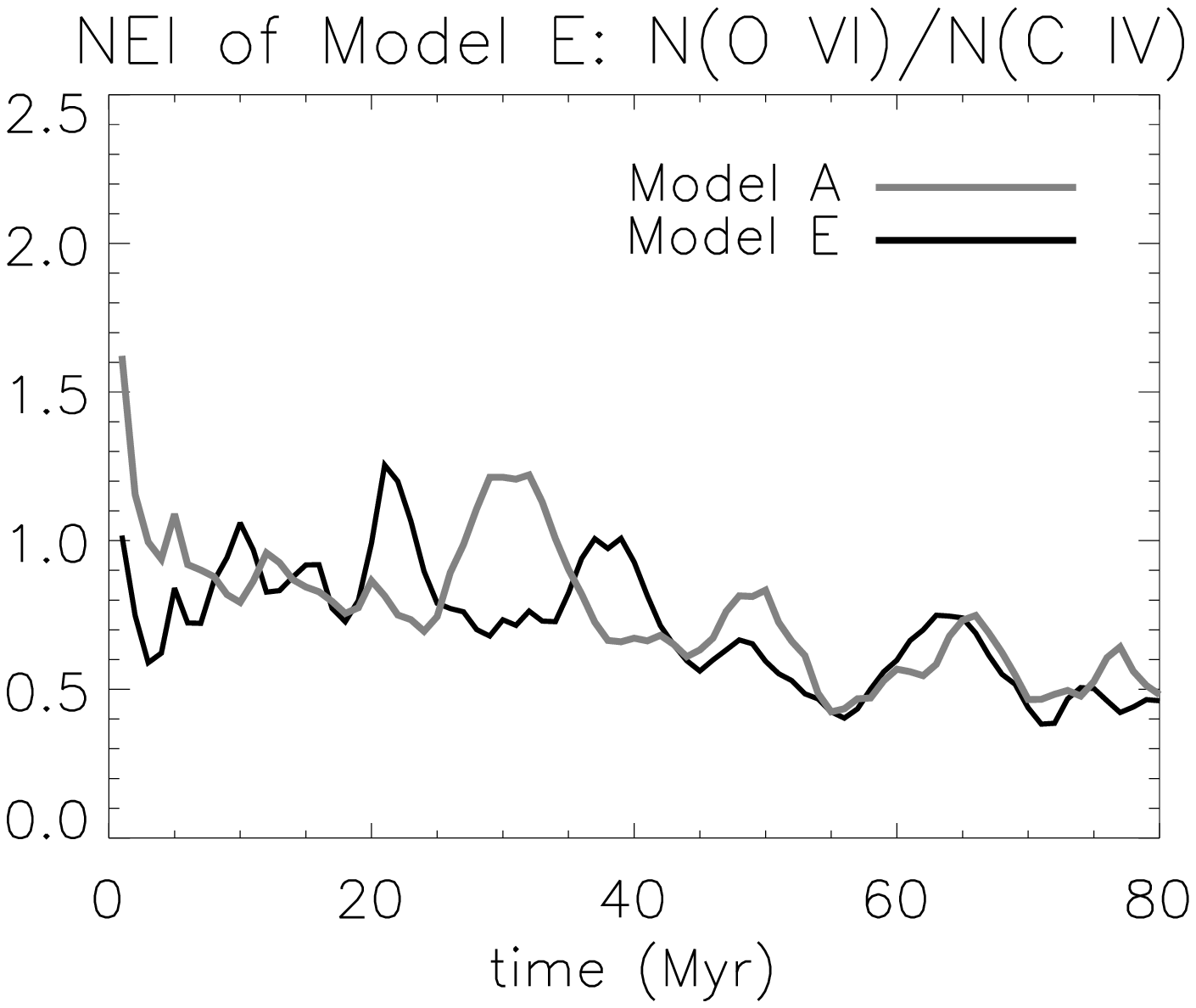}
\includegraphics[scale=0.25]{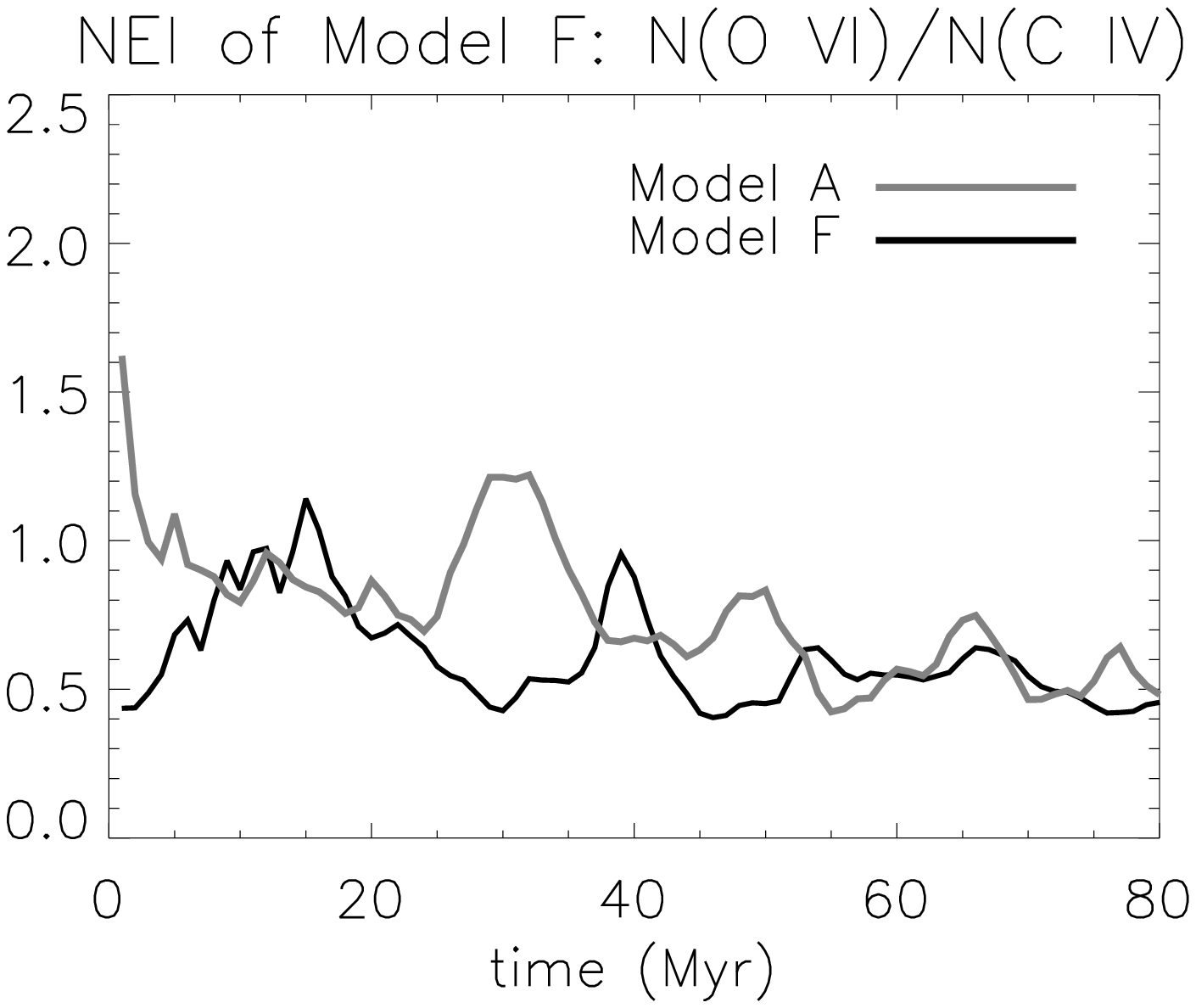} \\
\includegraphics[scale=0.25]{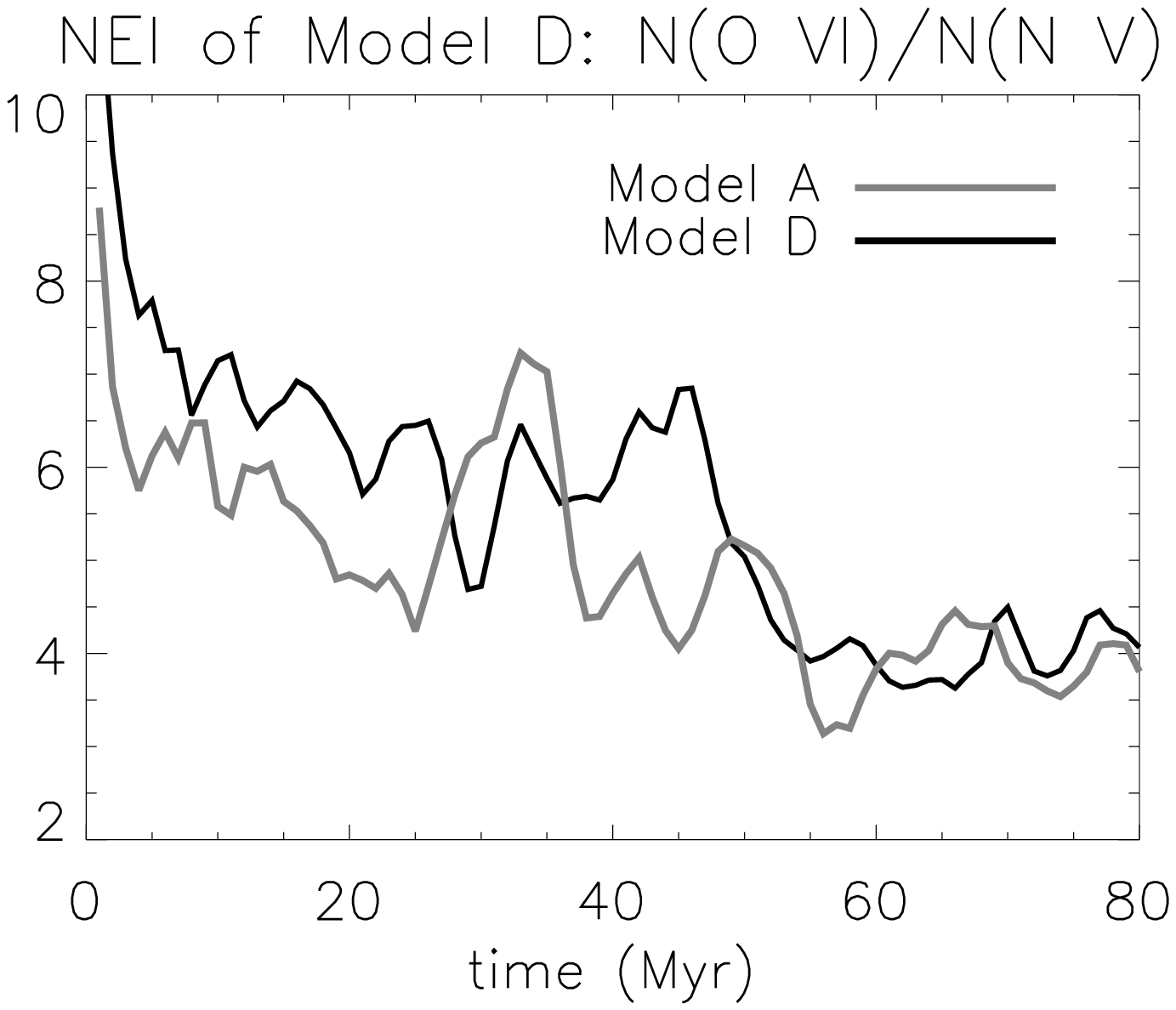}
\includegraphics[scale=0.25]{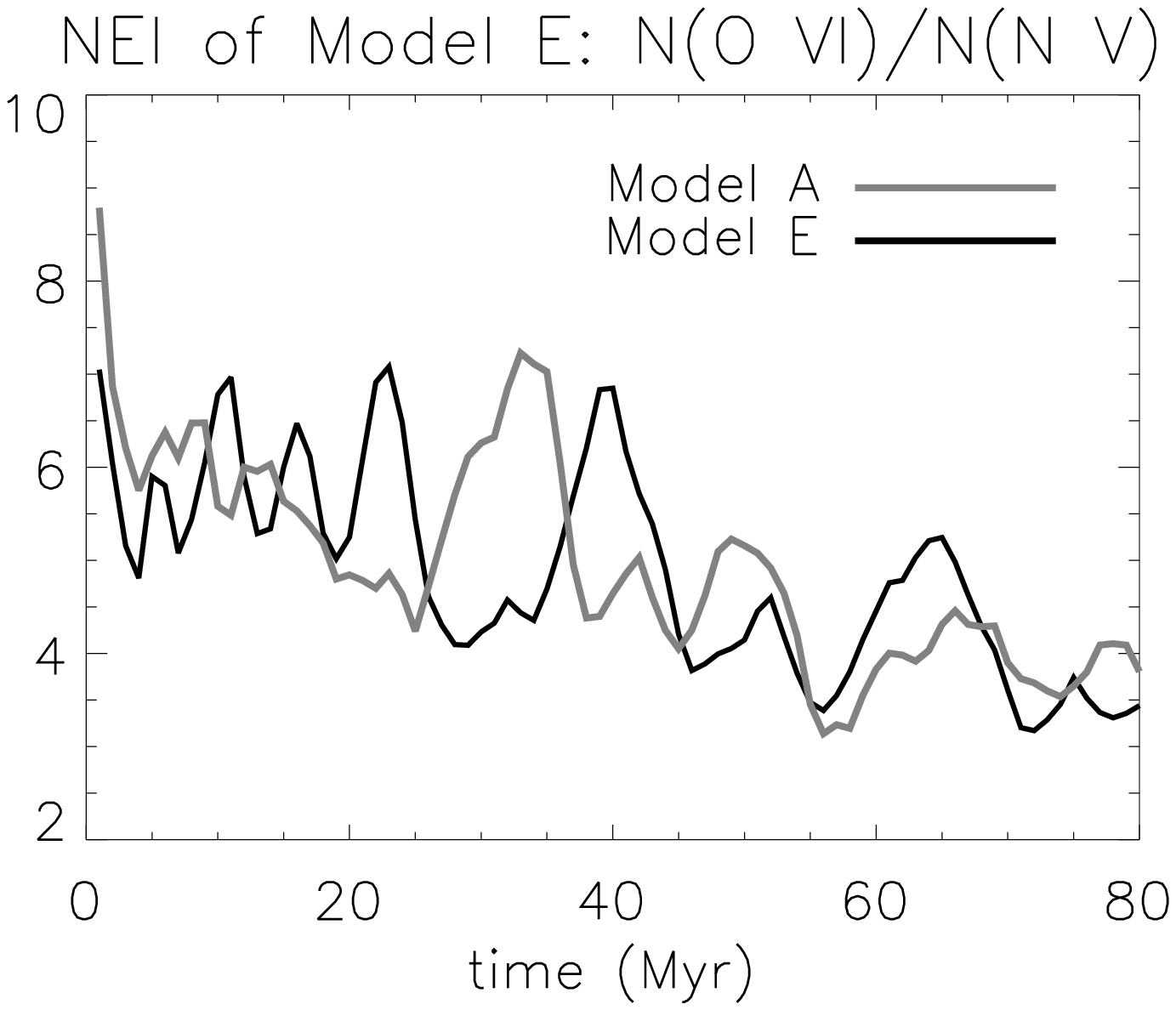}
\includegraphics[scale=0.25]{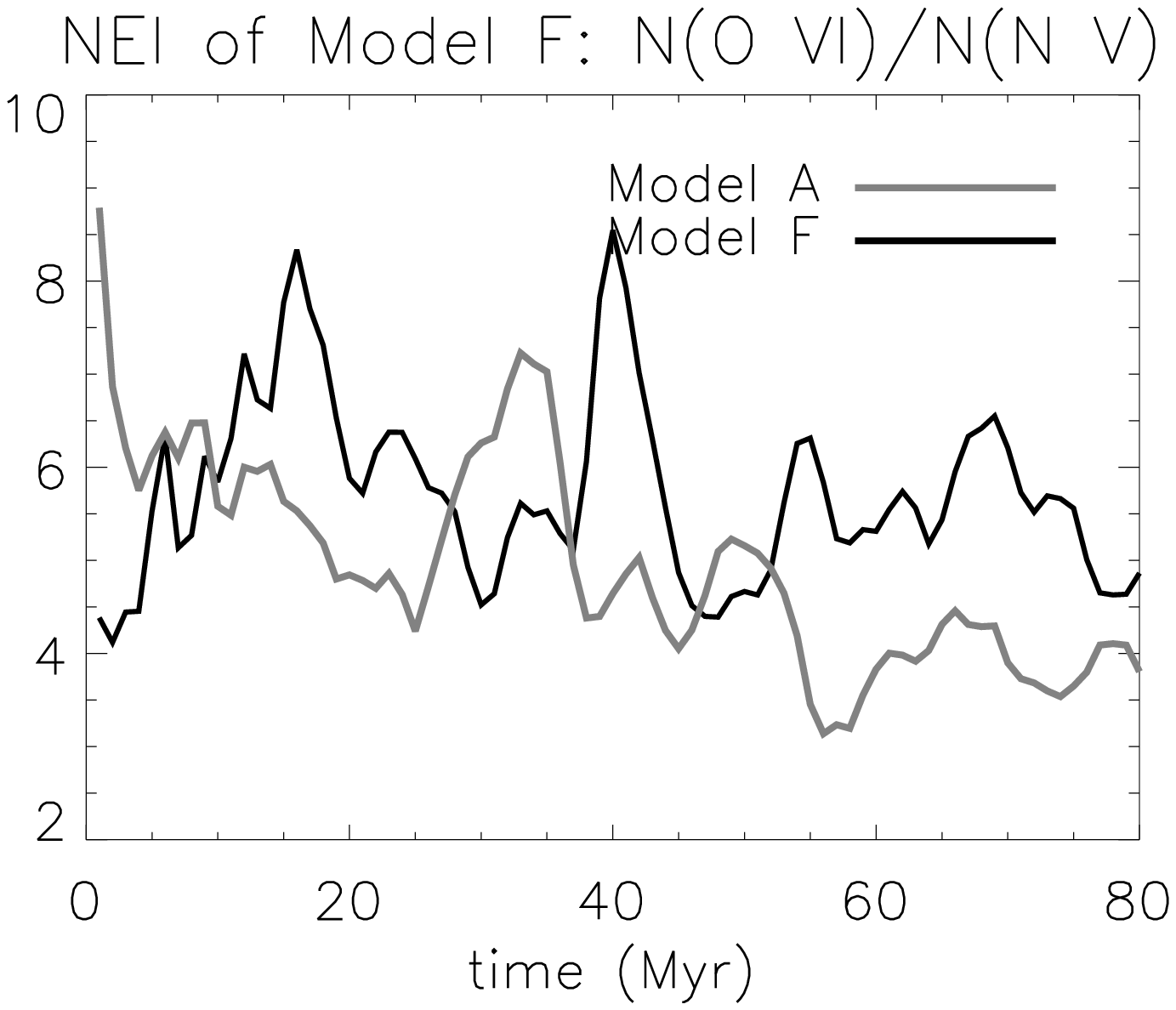}

\caption{Column densities and ratios of column densities for Model~D (left
  column), Model E (middle column), and Model F (right column) 
  from NEI calculations compared with similar results for Model~A. 
  Top panels: column densities; gray lines are the Model~A NEI results 
  and dark lines are the NEI results of Models~D, E, and F in the left, 
  middle, and right panel, respectively. Second, third, and fourth 
  panels: column density ratios of \ion{C}{4}/\ion{N}{5}, 
  \ion{O}{6}/\ion{C}{4}, and \ion{O}{6}/\ion{N}{5}, respectively. 
\label{modelDEF_fig}}
\end{figure*}

The third--left to fifth--left panels in Figure \ref{modelABC_fig} 
show ratios between the column densities of different ions, 
N(\ion{C}{4})/N(\ion{N}{5}), N(\ion{O}{6})/N(\ion{C}{4}), and 
N(\ion{O}{6})/N(\ion{N}{5}), respectively, calculated both from NEI 
and CIE for Model A. The ratios follow a similar trend as the column density 
evolution (top--left panel) and NEI/CIE ratios for each ion
(second--left panel) such that they start to stabilize around 
$t=20$ Myr. Estimated mean, median, standard deviation, 
and minimum and maximum
values of the column density ratios during the time 
period of $t=20$ to $t=80~\mbox{Myr}$
for Model A are given in Table \ref{colm_dens_ratios_T}.

The ratios of various ions' column densities are the most frequently 
used diagnostics for evaluating observations because 
models that are based upon differing 
physical processes predict different 
values. In the following sections, we will estimate the column density 
ratios between different ions for all of our model simulations and 
summarize them in Table \ref{colm_dens_ratios_T}. 
Comparisons between our results and other models, including the 
turbulent mixing calculations of 
\citet{Esquiveletal2006ApJ} and \citet{SlavinShullBegelman1993ApJ} 
will be given in \S \ref{discussion_S}.

\subsection{Model B: Higher Resolution than Model A} \label{ModelB_S}

In Model~B, the smallest cells are half the height and half the 
width of the smallest cells in Model~A. All of the other 
parameters are the same as in Model~A. The high ion column 
densities and ratios predicted for the Model~B simulation 
are shown in the middle column of Figure \ref{modelABC_fig}. 
The panels are placed in the same order as used for 
Model A (left column). A comparison between the left and middle column 
shows that the results of our high resolution simulation (Model~B) 
are very similar to those of the reference simulation (Model~A). 
The top--left and top--middle panels show that the evolution 
of column densities in Model B follows the same trends as Model~A such 
that all column densities, both in NEI
and CIE, continue to increase until 
$t\approx20~\mbox{to}~30~\mbox{Myr}$ 
and then fluctuate 
between $t=20$ and $t=80$~Myr. Similarly, the column density 
ratios between NEI and CIE (second--left and second--middle panels) 
stabilize after $t=20$~Myr for both models. 
As in Model~A, the NEI calculations predict more high ions than 
do the CIE calculations. 
Between $t=20$ and $t=80~\mbox{Myr}$, the mean/standard deviation 
of the ratios of column densities calculated using NEI to 
the column densities calculated using CIE 
are $4.30$/$0.54$, 
$3.00$/$0.42$, and $2.29$/$0.25$ for \ion{C}{4}, \ion{N}{5}, and
\ion{O}{6}, respectively (Table \ref{nei_cie_ABC_T}). 
These values in Model B are close to those 
calculated in Model A (\S \ref{evol_colm_dens_S}). From 
the information in the 
third--middle to fifth--middle panels, we tabulate the mean, median, 
standard deviation, and minimum and maximum values of 
N(\ion{C}{4})/N(\ion{N}{5}), N(\ion{O}{6})/N(\ion{C}{4}), and 
N(\ion{O}{6})/N(\ion{N}{5}), both for the NEI and CIE predictions. 
The values are given in Table \ref{colm_dens_ratios_T} and confirm 
the similarity between Model~B and Model~A.

Even though the overall results of Model B are very similar to those 
of Model A, we do find that resolution affects the temperature 
of the mixed gas and therefore the column densities at early times. 
Between $t=0$ and $t=10$~Myr, 
the column densities of each ion in Model~B 
(top--middle panel in Figure \ref{modelABC_fig}) increase by 
$4.0\times10^{12}$ ($7.6\times10^{11}$), 
$5.9\times10^{11}$ ($1.7\times10^{11}$), and 
$3.4\times10^{12}$ ($1.2\times10^{12}$) $\mbox{cm}^{-2}$
for \ion{C}{4}, \ion{N}{5}, and \ion{O}{6}, 
respectively, in NEI (CIE) (see Table \ref{increase_rate_T}). 
Although these values for Model B are close 
to those from Model~A (\S \ref{evol_colm_dens_S}) in both 
NEI and CIE, the increase in the column densities 
in Model~B is slightly larger than that in Model~A for 
all ions, which indicates that hot and cool gas mix slightly faster 
in the high resolution simulation between $t=0$ and $t=10$~Myr. 
Because \ion{C}{4} in CIE traces the mixed gas around $10^5$ K, 
the gas around this temperature is most affected. It is produced much 
faster in the high resolution simulation.

\subsection{Model C: 1/10 Scale Simulation}

In Model C, we run the simulation for 1/10th the time period 
(8~Myr) in a 1/10th scale computational 
domain ($10\times30$ pc) than in Model A. The purpose of performing 
this model is to check whether
the results of Model A (our reference simulation) are still valid 
on a smaller scale and to study whether the difference 
between NEI and CIE begins at early times in the smaller domain. 
This is an important issue because we include radiative cooling 
which is not scale--invariant. 
If radiative cooling were not included, the smaller spatial scale
simulation would be identical to the larger scale simulation 
because the spatial 
coordinates and time are proportional to each other in the 
hydrodynamics equations without any dissipative terms, 
making the calculations scale--invariant. 
For those researchers who are interested in the early evolution 
of the mixing zone, Model~C can be seen as a close--up, with 
greater spatial resolution and smaller time periods.

The column densities and ratios predicted for Model~C 
are shown in the right column of 
Figure \ref{modelABC_fig}, in which the 
plots are placed in the same order (from top to bottom) as for 
Models A and B (left and middle column in Figure \ref{modelABC_fig}). 
At early times, the column densities of each ion in Model~C 
(top--right panel) evolve in a similar manner 
as those in Models~A and B, such that they continue to increase 
until $t\approx3$~Myr, i.e., $\sim1/10$ of the onset periods 
in Models A and B. The column density rises 
for each ion both in NEI and CIE 
for Model~C are given in Table \ref{increase_rate_T}. 
These increases are close to those in Models~A and B 
over 10~Myr, showing that the early evolution of the mixing layer 
does not depend on the scale as long as the radiative cooling rate is 
not faster than the mixing rate.

There are ways in which Model~C behaves more like Model~A during 
its first 8~Myr than like a time--scaled version of Model~A. 
In Model~A, the depth of the mixed zone and the column densities 
of Li--like ions increase steadily for the first 8~Myr of the 
simulation. Similarly, they increase in Model~C during this 
time period. But, if Model~A and C acted like time--scaled version 
of each other, then the depth of the mixed zone and the column 
densities of Li--like ions would stall at 2~Myr 
($=20/80 \times 8~\mbox{Myr}$) in Model~C, which they do not. 
Model~C's mixing 
zone grows so steadily that it overflows the computational domain 
by 8~Myr. Model~C, therefore, does not run long enough to develop 
a \ion{C}{4}--rich, radiatively cooled ($T<2\times10^4~\mbox{K}$) 
layer similar to that which appears in Model~A around 10~Myr. 
As a result, most of the existing \ion{C}{4} in Model~C resides 
in the mixed layer of actively cooling gas ($T\ge10^5~\mbox{K}$), 
the same region where \ion{O}{6} resides. The same can be said 
of Model~A at early times.

\begin{deluxetable*}{ccccccccccccc}

\tablewidth{0pt}
\tabletypesize{\scriptsize}
\tablecaption{Column Density Ratios between Different Ions\label{colm_dens_ratios_T} }
\tablecolumns{13}

\tablehead{
\colhead{Model~~} & 
\multicolumn{4}{c}{N(C IV) / N(N V)} &
\multicolumn{4}{c}{N(O VI) / N(C IV)} &
\multicolumn{4}{c}{N(O VI) / N(N V)} \\ 
\colhead{} & 
\multicolumn{4}{c}{--------------------------------------------------} &
\multicolumn{4}{c}{--------------------------------------------------} &
\multicolumn{4}{c}{--------------------------------------------------} \\
\colhead{NEI / CIE~~} & 
\colhead{~mean} & \colhead{median} &
\colhead{$\sigma$ \tablenotemark{a}} & 
\colhead{[min, max]} & \colhead{~mean} & \colhead{median} &  
\colhead{$\sigma$ \tablenotemark{a}} & 
\colhead{[min, max]} & \colhead{~mean} & \colhead{median} &
\colhead{$\sigma$ \tablenotemark{a}} & 
\colhead{[min, max]} 
}
  
\startdata
A \tablenotemark{b} ~ NEI & 6.72 & 6.75 & 0.84 & [5.04, 8.61] & 
0.70 & 0.66 & 0.21 & [0.42, 1.22] & 
4.61 & 4.38 & 0.96 & [3.14, 7.23] \\
~~~~~~ CIE & 4.27 & 4.07 & 0.94 & [2.39, 7.10] & 
1.46 & 1.38 & 0.48 & [0.65, 2.96] & 
5.88 & 5.72 & 1.00 & [4.26, 8.57] \\ 
B \tablenotemark{b} ~ NEI & 6.77 & 7.05 & 1.10 & [4.20, 8.25] & 
0.68 & 0.66 & 0.21 & [0.41, 1.23] & 
4.41 & 4.30 & 0.84 & [3.18, 6.76] \\
~~~~~~ CIE & 4.76 & 4.73 & 1.01 & [3.01, 6.92] & 
1.28 & 1.28 & 0.43 & [0.63, 2.32] & 
5.77 & 5.71 & 1.15 & [3.68, 8.69] \\ 
C \tablenotemark{c} ~ NEI & 6.42 & 6.43 & 0.34 & [5.90, 6.98] & 
0.59 & 0.55 & 0.08 & [0.51, 0.72] & 
3.76 & 3.69 & 0.47 & [3.09, 4.56] \\
~~~~~~ CIE & 4.68 & 4.57 & 0.92 & [3.49, 6.32] & 
1.08 & 1.13 & 0.19 & [0.71, 1.31] & 
5.04 & 5.10 & 1.19 & [3.25, 7.13] \\
D \tablenotemark{d} ~ NEI & 8.17 & 8.33 & 0.94 & [6.08, 10.00] & 
0.61 & 0.49 & 0.18 & [0.42, 0.94] & 
4.81 & 4.36 & 1.03 & [3.63, 6.85] \\ 
E \tablenotemark{b} ~ NEI & 7.02 & 7.17 & 0.96 & [4.86, 8.69] & 
0.67 & 0.65 & 0.20 & [0.38, 1.25] & 
4.56 & 4.33 & 1.02 & [3.17, 7.08] \\ 
F \tablenotemark{b} ~ NEI & 10.25 & 10.37 & 1.06 & [7.16, 12.55] & 
0.55 & 0.54 & 0.11 & [0.40, 0.96] & 
5.61 & 5.56 & 0.84 & [4.39, 8.55]
\enddata
\tablenotetext{a}{standard deviation}
\tablenotetext{b}{averaged over $t \in [20, 80]$ Myr}
\tablenotetext{c}{averaged over $t \in [6, 8]$ Myr}
\tablenotetext{d}{averaged over $t \in [30, 80]$ Myr}

\end{deluxetable*}

\subsection{Models D, E, and F: Modified Initial Configurations}

Models D, E, and F are presented in order to address 
the following questions regarding 
the physical conditions of the mixing layer. 
Do different physical conditions affect the growth of mixing 
layers and do different physical conditions affect the 
column density ratios such that the observed ratios can be used 
as diagnostics of mixing layers? Answering these questions, 
we find that modifying the initial amplitude of the disturbance 
between the hot and cool gases does not change the characteristics 
of mixing layer (see \S \ref{ModelE_S} regarding Model E). 
However, adjusting the speed difference between 
the hot and cool layers (see \S \ref{ModelD_S} regarding Model D) 
and the temperature of the hot gas 
(see \S \ref{ModelF_S} regarding Model F) 
affect the column density ratios. 
Because NEI provides more realistic diagnostics than CIE, 
only the NEI results of these models 
are compared with those of Model A, the reference simulation. 
Figure \ref{modelDEF_fig} shows the results of Model~D, E, and F 
in the left, middle, and right column, respectively, together 
with the NEI results of Model A for comparison.

\subsubsection{Model D: Slow Speed}\label{ModelD_S}

In Model~D, the hot and cool gases slide past each other 
with a smaller speed ($50~\mbox{km}~\mbox{s}^{-1}$) than in Model A 
($100~\mbox{km}~\mbox{s}^{-1}$). Mixing occurs more slowly in 
Model~D than in Model~A because mixing is caused by shear 
instabilities, which grow with the speed difference between 
the two bodies. As a result, the early--time high ion column 
densities build up more slowly (see the top--left panel 
of Figure \ref{modelDEF_fig} and Table \ref{increase_rate_T}). 
Although slightly delayed, the column densities of Li--like 
ions in Model~D do approach those of Model~A by the middle 
of the simulation time.

The column densities from Model~D continue to increase 
until $t\approx30$ Myr when they begin to fluctuate. 
We estimate the column density ratios between different 
ions for Model~D between $t=30$ and $t=80$ Myr and list them 
in Table \ref{colm_dens_ratios_T}. Compared with Model~A during 
a similar time frame, Model~D has a higher average 
N(\ion{C}{4})/N(\ion{N}{5}) ratio but a lower average 
 N(\ion{O}{6})/N(\ion{C}{4}) ratio, while the 
N(\ion{O}{6})/N(\ion{N}{5}) ratios are similar. 
The noticeable variations in the column density ratios due to 
only $50~\mbox{km}~\mbox{s}^{-1}$ of velocity difference 
suggest that the ratios may serve as diagnostics of velocity 
when the velocity difference between the hot and cool gas 
is large (see \S \ref{diagnostics_S}).

\subsubsection{Model E: Large Initial Amplitude}\label{ModelE_S}

The sine wave shape of the interface between the hot and cool 
gases is larger in Model~E than in Model~A, allowing us to 
test the effect of the interface's curvature. We find that 
Model~E's characteristics, including depth of mixing region, 
high ion column densities, and column density ratios, are 
similar to those of Model~A (see Figure \ref{modelDEF_fig} 
and Table \ref{increase_rate_T}). 
These results show that the perturbation amplitudes 
in the initial interface between the hot and cool gas do not 
significantly affect the physical properties 
of the turbulent mixing layer as long as mixing is efficient.

\subsubsection{Model F: Hot Gas with Higher Temperature}
\label{ModelF_S}

In Model~F, the hot gas has a higher temperature 
($3\times10^6~\mbox{K}$) 
than in Model~A. Because we require the thermal pressure of 
the initial hot layer to equal that of the cool gas layer, 
the hot gas in Model~F has a smaller density 
($\onethird \times 10^{-4}$ H atoms $\mbox{cm}^{-3}$) 
than that in Model~A.

The turbulent mixing layer in Model~F evolves differently 
than that in Model~A; the high ion column densities 
increase at a slower pace 
and continue to rise throughout the simulation period 
(see Table \ref{increase_rate_T} and 
Figure \ref{modelDEF_fig}). Furthermore the newly mixed 
gas is hotter and the radiatively cooled gas 
($T\le 2\times10^4~\mbox{K}$) is shallower in Model~F 
than in Model~A because it takes longer for the hotter 
mixed gas in Model~F to cool. We find that the high ion 
population is low in the hotter, newly mixed gas in Model~F 
than in Model~A. This effect seems to yield larger 
N(\ion{C}{4})/N(\ion{N}{5}) and 
N(\ion{O}{6})/N(\ion{N}{5}) ratios than in Model~A 
(see Figure \ref{modelDEF_fig} and Table \ref{colm_dens_ratios_T}). 
Model~F's N(\ion{C}{4})/N(\ion{N}{5}) ratio is greater than that 
of Model~A for nearly the entire simulation. Thus, this ratio, 
especially if used in conjunction with 
the N(\ion{O}{6})/N(\ion{N}{5}) ratio may serve as a diagnostic 
of the hot gas temperature when interpreting observations of mixing 
gas.

\subsubsection{Diagnostics for Observations} \label{diagnostics_S}

The column densities on sightlines that run perpendicular to the 
turbulent mixing layers are on the order of $10^{13}~\mbox{cm}^{-2}$ 
for \ion{C}{4}, $10^{12}~\mbox{cm}^{-2}$ for \ion{N}{5}, and 
$5\times10^{12}~\mbox{cm}^{-2}$ for \ion{O}{6} in our model 
simulations (except Model~C). In real observations, 
the sightlines probably pass through the layers at sharper angles 
and intersect multiple mixing layers or multiple portions of 
curved mixing layers that surround individual clouds. Each of these 
effects increases the column densities by a multiplicative factor. 
When comparing with an observation, the model column density ratios 
are more useful than the model column densities because the 
multiplicative factor is unknown. In \S \ref{othermodels_S}, we 
compare the ion ratios from our simulations with actual observations. 
Here, we discuss the use of column density ratios as diagnostics 
of the physical conditions producing the turbulent mixing layers.

The results of Models D and F reveal that the 
ratios between the column densities of 
different ions can be used as diagnostics for 
the shear speed and hot gas temperature. 
If the N(\ion{C}{4})/N(\ion{N}{5}) and 
N(\ion{C}{4})/N(\ion{O}{6}) ratios along a sightline 
through a specific cloud 
are larger than those along sightlines through other clouds 
while the N(\ion{O}{6})/N(\ion{N}{5}) ratio is similar along 
all of these sightlines, then it is likely that the mixing 
layer on the first cloud forms with a slower sheer speed 
(Model~D). Another possible diagnostic comes 
from Model~F. If observations along one sight line have higher 
N(\ion{C}{4})/N(\ion{N}{5}) and N(\ion{O}{6})/N(\ion{N}{5}) ratios 
but lower N(\ion{O}{6})/N(\ion{C}{4}) ratios 
compared with observations of other clouds, 
it is possible to assume that the first mixing layer 
formed in hotter ambient gas than the others.

Generally, more \ion{C}{4} is produced in the NEI calculations 
of the models than in the CIE calculations, especially 
in the radiatively cooled regions. This high population of \ion{C}{4} 
affects the column density ratios between different ions. However, 
\ion{C}{4} can also be produced by photoionization and 
is produced more easily than other high ions if there are 
nearby photoionizing sources because the photoionization threshold 
of \ion{C}{4} is only $48~\mbox{eV}$. Therefore, care must be taken 
when evaluating observational results.

\section{Discussion} \label{discussion_S}

The column densities of high ions located in the Galactic halo 
are observable from the absorption lines in the spectra of halo 
stars and extragalactic objects. Here we compare observed high 
ion column densities and their ratios with those estimated from 
our model simulations. We also compare our results with predictions 
from other turbulent mixing models and from other sorts of models 
for high stage ions. 
However, there are some uncertainties 
in estimating the column densities 
from model calculations. Therefore, we discuss these uncertainties 
before addressing how much they affect the results of model 
calculations and the comparisons with observations.

\subsection{Uncertainties in Model Calculations} \label{uncertainty_S}

High ion column density predictions are strongly dependent on the 
assumed metallicity of the gas experiencing the modeled physical 
processes. However, the metallicity 
in the Galactic halo is not well constrained and the uncertainty 
increases from the Milky Way to the intergalactic medium and external 
galaxies. Therefore, the column density calculations vary according to 
the metallicity used in the model. 

\begin{deluxetable}{cccccccc}[b]

\tablewidth{0pt}
\tabletypesize{\footnotesize}
\tablecaption{Abundances Normalized to \citet{Allen1973asqu.book}
\label{metallicity_T} }
\tablecolumns{8}

\tablehead{
\colhead{Reference} & \colhead{C \tablenotemark{a}} & 
\colhead{N \tablenotemark{a}} &
\colhead{O \tablenotemark{a}} & \colhead{C/N} & 
\colhead{O/C} & \colhead{O/N}
}

\startdata
1 & 1.34 & 1.00 & 1.12 & 1.34 & 0.83 & 1.12 \\
2 & 1.48 & 1.07 & 1.23 & 1.38 & 0.83 & 1.15 \\
3 & 1.10 & 1.23 & 1.29 & 0.89 & 1.17 & 1.05 \\
4 & 1.07 & 1.02 & 1.12 & 1.05 & 1.05 & 1.10 \\
5 & 0.74 & 0.93 & 0.74 & 0.80 & 1.00 & 0.80 \\
6 & 0.054 & 0.013 & 0.12 & 4.15 & 2.22 & 9.23 \\
7 & 0.74 & 0.66 & 0.69 & 1.12 & 0.93 & 1.05 
\enddata

\tablerefs{(1) \citet{AndersEbihara1982GeCoA} metallicities 
  used in \citet{BoehringerHartquist1987MNRAS} 
  and in \citet{SlavinCox1993ApJ} 
  (2) \citet{Grevesse1984PhST} used in 
  \citet{Borkowskietal1990ApJ,SlavinShullBegelman1993ApJ}
  (3) \citet{AndersGrevesse1989GeCoA} used in 
  \citet{SutherlandDopita1993ApJS,Shelton1998ApJ,
    Esquiveletal2006ApJ}
  (4) \citet{Grevesse1996ASPC} used in
  \citet{IndebetouwShull2004aApJ}.
  (5) \citet{Foxetal2004ApJ}: adopted solar metallicity 
  from \citet{Holweger2001AIPC,Allendeetal2002ApJL,
    Allendeetal2001ApJL}
  (6) \citet{Foxetal2004ApJ}: Complex C metallicity estimated along 
  the PG 1259+593 sightline
  (7) \citet{Asplundetal2005ASPC}: solar photospheric metallicity
}

\tablenotetext{a}{carbon, nitrogen, and oxygen abundance ratio 
to that in \citet{Allen1973asqu.book} for each reference. 
In \citet{Allen1973asqu.book}, carbon, nitrogen, 
and oxygen abundance is $3.31\times10^{-4}$, $9.12\times10^{-5}$, 
and $6.61\times10^{-4}$ per hydrogen atom, respectively.}

\end{deluxetable}

We find that most of previous model calculations 
used solar metallicity, which are sometimes better constrained 
by meteorite measurements than other methods, 
although the abundances of 
carbon, nitrogen, and oxygen still have large uncertainties, 
even in this method. 
Because metallicity measurements have been revised repeatedly over 
the years, different model calculations used 
slightly different metallicities, 
especially for carbon, nitrogen, and oxygen. 
For example, the turbulent mixing calculations of 
\citet{SlavinShullBegelman1993ApJ} used the solar photospheric 
metallicities of \citet{Grevesse1984PhST} and assumed that the 
cool gas has depleted metallicities (such that $50\%$ of carbon, 
nitrogen, and oxygen are depleted into dust grains) while all dust 
grains are destroyed in the initial hot gas. In contrast, 
the CIE calculations \citep{Benjaminetal2001ApJL} 
in the numerical study of turbulent mixing layers of 
\citet{Esquiveletal2006ApJ} 
used the updated solar photospheric metallicity data from
\citet{AndersGrevesse1989GeCoA} which have different abundances 
for carbon, nitrogen, and oxygen from \citet{Grevesse1984PhST} 
(see Table \ref{metallicity_T} for the difference between these two 
metallicities).

For consistency between our NEI and CIE calculations, we use 
the cosmic abundances of \citet{Allen1973asqu.book} because 
they are the default abundances both in the FLASH NEI module and 
in the HEASARC version of the Raymond and 
Smith code, which we use for our CIE calculations. 
Note that the abundances of carbon, nitrogen, and oxygen 
from most of the references (before the year of 2000) 
are higher than those from \citet{Allen1973asqu.book} although 
recent measurements \citep{Asplundetal2005ASPC,Allendeetal2002ApJL,
Allendeetal2001ApJL,Holweger2001AIPC} measured lower abundances 
of these elements. Table \ref{metallicity_T} 
shows the ratios of solar metallicities from various references to the 
abundance from \citet{Allen1973asqu.book} for carbon, nitrogen, and 
oxygen. The ratios for fractions between different atoms such as 
carbon/nitrogen, oxygen/carbon, and oxygen/nitrogen are also 
presented. They can be used to convert our column density 
ratios to those for different metallicities. 
For example, if the solar metallicities from 
\citet{AndersGrevesse1989GeCoA} were desired instead of those from 
\citet{Allen1973asqu.book}, then our column density ratio predictions 
would need to be multiplied by $0.89$, $1.17$, and $1.05$ for 
N(\ion{C}{4})/N(\ion{N}{5}), N(\ion{O}{6})/N(\ion{C}{4}), and 
N(\ion{O}{6})/N(\ion{N}{5}), respectively. 
Note that this conversion for different abundances is approximate 
for the complete effects of abundances on the high ions 
(and their ratios) 
because abundances of metal ions 
also affect the cooling rates thus 
influencing the dynamical evolution of the gas 
containing the high ions. In optically thin gas, large 
abundances of metal ions increase the cooling rates because the 
dominant cooling process is resonant line emission from these 
metal ions. As mentioned in \S \ref{method_S}, more complete 
future studies are required to address the full effects 
of abundances of all relevant ions on the cooling rates.

Photoionization also significantly affects 
the column densities of high ions 
and the column density ratios between different ions 
although we do not consider its effect in our simulations. 
Because the photoionization energy of \ion{C}{4} is much lower 
than those of \ion{N}{5} and \ion{O}{6}, it is likely that 
photoionization preferentially enhance the 
\ion{C}{4} column densities. In this case, 
as already mentioned in \S \ref{diagnostics_S}, 
it is not easy to distinguish observationally between collisionally 
ionized \ion{C}{4} in the radiatively cooled part of the mixing 
layer and photoionized \ion{C}{4}. 
Note that \citet{SlavinShullBegelman1993ApJ} 
considered the effect of photoionization for the 
case where photons radiated from the mixed layer irradiate 
the cool gas. 

\begin{figure*}
\centering

\includegraphics[scale=0.8]{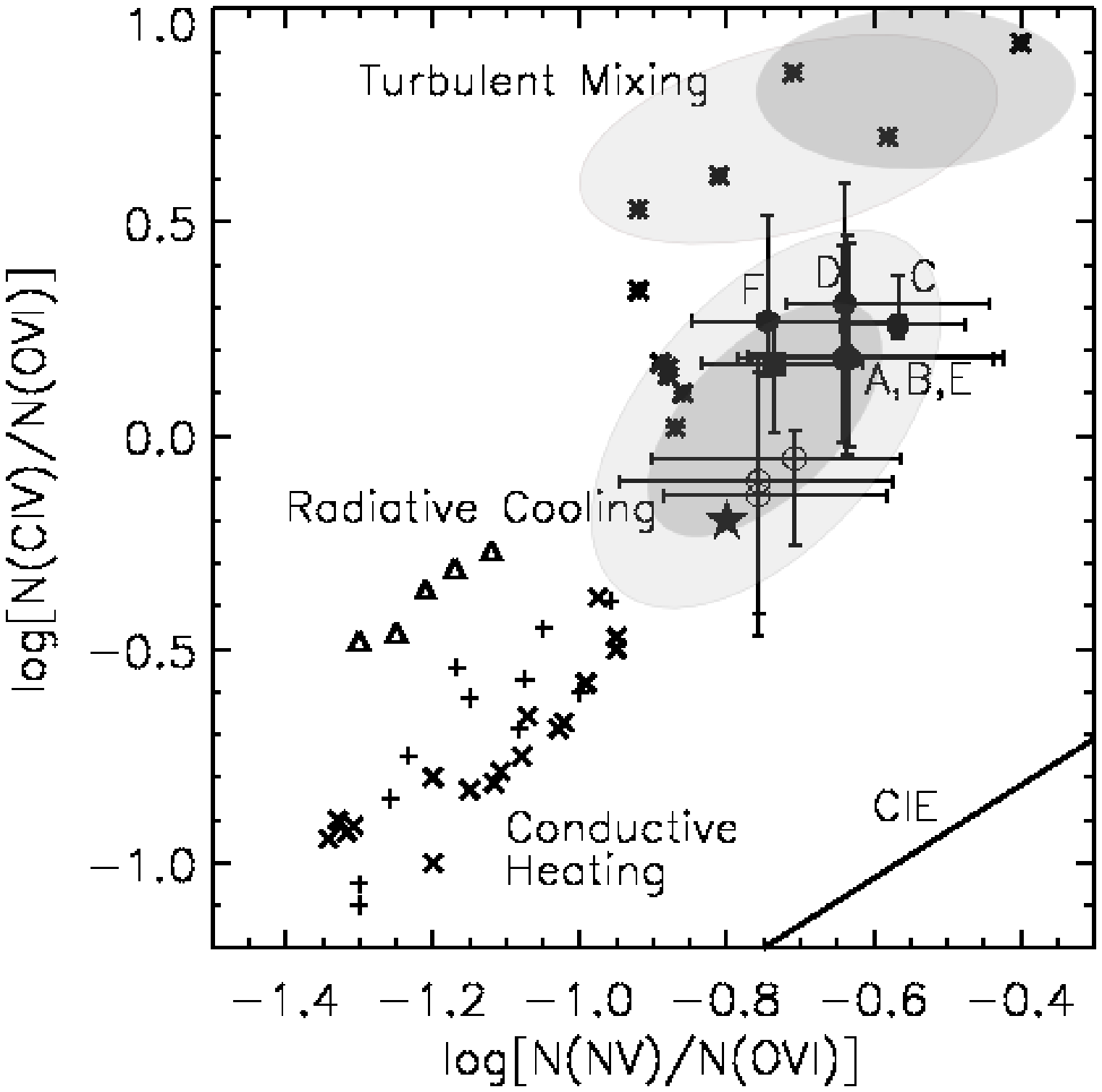}

\caption{Column density ratios between different ions in 
  log[N(N~V)/N(O~VI)]--log[N(C~IV)/N(O~VI)] 
  space. Re-plotted from Fig. 1 of \citet{IndebetouwShull2004aApJ}, 
  with additions. 
  Models: radiative cooling of Galactic fountain gas ({\it triangles};
  \citet{ShapiroBenjamin1993sfgi.conf,BenjaminShapiro1993uxrs.conf}), 
  turbulent mixing layers ({\it asterisks}; 
  \citet{SlavinShullBegelman1993ApJ}), conductive heating and 
  evaporation of spherical and planar clouds ({\it crosses}; 
  \citet{BoehringerHartquist1987MNRAS} and \citet{Borkowskietal1990ApJ}, 
  respectively), cooling supernova remnant shells ({\it plus signs}; 
  \citet{SlavinCox1993ApJ,Shelton1998ApJ}), and CIE ({\it solid line}; 
  \citet{SutherlandDopita1993ApJS}). The suite of average values from 3-D 
  MHD turbulent mixing calculations of \citet{Esquiveletal2006ApJ} are 
  indicated by bright shaded regions (no magnetic field) and dark 
  shaded regions (with magnetic field). 
  Note that the shaded regions near the 
  center are the results of their models with radiative cooling, 
  while the shaded regions in the upper right 
  are those without radiative cooling. 
  The median values 
  in Table \ref{colm_dens_ratios_T} are plotted as filled (NEI) and 
  empty (CIE) circles and the averaged value of halo observations 
  along 34 sightlines \citep{IndebetouwShull2004bApJ} is indicated 
  by a filled star. The filled square is the result of the additional
  simulation mentioned in \S \ref{othermodels_S}. 
  \label{othermodels_fig}}
\end{figure*}

Besides the above two, there are two more uncertainties 
in dynamical models of turbulent mixing layers: 
time and sightline dependence. 
Column densities can be calculated 
along specific sightlines at specific times and can be directly 
compared with the observed ones. 
However, the computed column densities vary significantly with 
viewing angle and position, and in dynamical models also vary 
with time. For example, Figure 
\ref{all_sl_30myr_fig} and Table \ref{all_sl_30myr_T} show 
column density variations between sightlines at a specific time 
($t=30$ Myr), while the top panels in Figure \ref{modelABC_fig} 
and \ref{modelDEF_fig} show variations of sightline averaged 
column densities over time. In general, when these sources of 
variation are taken into account, the computed column densities 
from the dynamical models have wider ranges of predicted values 
than those from the static models.

It is necessary to address again the issue of our 
simulated turbulent mixing layers 
reaching steady states, particularly the question of 
whether the evolution of column densities and their ratios 
can be applied to interstellar and intergalactic clouds 
that are relatively small so travel the length of the 
cloud in less than $80~\mbox{Myr}$ (see \S \ref{ModelA_S}). 
In our model simulations, 
the column densities and their ratios evolve until 
$80~\mbox{Myr}$ (except in Model~C) which corresponds to 
a length of roughly 
$8~\mbox{kpc}$ for the 100~pc long mixing layers and a 
speed of $100~\mbox{km}~\mbox{s}^{-1}$. 
This size may be too large for the realistic size of the 
clouds in the Milky Way although it is a possible size for some 
of HVCs relatively far away in the halo such as Complex~C whose
distance and area are measured as roughly $10~\mbox{kpc}$ and 
$3 \times 15~\mbox{kpc}^2$, respectively \citep{Thometal2008ApJ}. 
In the case that the size of the cloud in the Milky Way is 
much smaller than $8~\mbox{kpc}$, care needs to be taken 
when using the column densities of high ions 
in our simulations because the column densities 
continue to increase during 
the early times of the simulation. However, the ratios between 
high ion column densities settle down quickly and do not vary 
significantly at later times implying that the ratios 
in Table \ref{colm_dens_ratios_T} are applicable 
at early times as well and for small clouds. In addition, 
the Model~C simulations which use a smaller domain 
and concentrate on the earlier times 
show that the ratios between high 
ion column densities settle down very quickly 
(around $t=2~\mbox{Myr}$, which is similar to 
the cooling time scale in \S \ref{ModelA_S}) 
and remain similar afterward, supporting 
the use of the values in Table \ref{colm_dens_ratios_T} for earlier 
times (and for smaller clouds).

Finally, we discuss the effects of some physical processes 
that we do not include in our simulations but could affect 
the results of our simulations. First, the results of 
our 2--D simulations for the NEI calculations of high 
ion column densities could be different from more realistic 
3--D simulations because it has been known that turbulent motion 
due to instabilities in 3--D is different from that in 2--D 
\citep{Hussain1984.conf,Bayly1986PhRvL,CraikCriminale1986RSPSA}. 
The magnetic field is 
known to suppress turbulence both in 2--D and 3--D 
\citep{Franketal1996ApJ,Jonesetal1997ApJ,Jeongetal2000ApJ,
Ryuetal2000ApJ}. 
These two effects may affect the column densities of 
high ions because the region for the high ions' existence 
is where the turbulent motion occurs. 
However, as the comparison of our 2--D hydrodynamic simulations 
with the 3--D MHD simulations of \citet{Esquiveletal2006ApJ} 
in the following section (\S \ref{othermodels_S}) shows, 
including the effects of the 3--D geometry and magnetic field 
would not significantly affect the ratios 
between high ion column densities.  
The turbulent mixing model would still 
distinguish itself from models with other physical processes 
regardless of these two effects (see Figure \ref{othermodels_fig}). 
(Note that the CIE and NEI calculations of our simulations 
predict similar ratios and that \citet{Esquiveletal2006ApJ} 
used only the CIE calculations for the high ion column densities.)

The effects of thermal conduction for the turbulent mixing layers 
were discussed in \citet{Esquiveletal2006ApJ}. Following their 
estimation, our model parameters lead to 
a similar Spitzer thermal diffusion 
coefficient $\kappa_{Sp} \sim 10^{24}~\mbox{cm}^2~\mbox{s}^{-1}$ with 
$T_{mixed}\sim10^5~\mbox{K}$ and $n_{mixed}\sim 10^{-2}~\mbox{cm}^{-3}$ 
but a larger turbulent diffusion coefficient $\kappa_{turb}\sim \onethird
v_{turb}L_{inj}\sim10^{27} ~\mbox{cm}^2~\mbox{s}^{-1}$ with $L_{inj}\sim
100~\mbox{pc}$ and $v_{turb} \sim 100 ~\mbox{km}~\mbox{s}^{-1}$, 
where $L_{inj}$ is the energy injection scale (approximated as 
the domain size) and $v_{turb}$ is the turbulent speed. In this 
estimation, we assume that the mixed gas has a typical temperature 
of $T\sim10^5~\mbox{K}$ and a density of $n\sim10^{-2}~\mbox{cm}^{-3}$ 
as in \citet{Esquiveletal2006ApJ}. 
As mentioned in \citet{Esquiveletal2006ApJ}, 
the larger turbulent diffusion 
coefficient than the Spitzer thermal diffusion coefficient implies 
that the heat transfer is dominated by turbulence rather than 
thermal diffusion which has a smaller scale. 
However, the small scale 
diffusion process (including numerical diffusion) 
could affect our simulations, especially at early times. 
(The effect of spatial resolution that corresponds to the numerical 
diffusion is discussed in \S \ref{ModelB_S}.) 
In our simulations as well as in the simulations of 
\citet{Esquiveletal2006ApJ}, the thermal diffusion coefficients at 
early times when the mixing zone is very shallow 
are larger than the above estimated values because 
the gas in the initial interface between hot and cool gas 
has a larger temperature 
gradient than the gas in the interfaces 
between mixed and hot 
gas or between mixed and cool gas at later times. 
Therefore, both in our and 
their hydrodynamic simulations, the heat transfer 
would be faster at early times with thermal conduction included. 
(Note that \citet{Esquiveletal2006ApJ} ran both 
hydrodynamic and MHD simulations but did not include thermal 
conduction.)

\subsection{Comparison with Observations and Other Models}
\label{othermodels_S}

The column density ratios between different ions for all of our 
model simulations are summarized in Table \ref{colm_dens_ratios_T}. 
Note that in our column density calculations, 
we use sightline geometries 
that are perpendicular to the initial interface 
between hot and cool gas (i.e., perpendicular to the initial 
velocity vector of the cool gas). 
This choice of sightlines is the same as 
\citet{SlavinShullBegelman1993ApJ} and \citet{Esquiveletal2006ApJ}. 
Along these sightlines, the measured velocities of each ion 
are so small that they can be directly compared 
with the halo observations ($\bar{v}\approx0$).

The column density ratios in Table \ref{colm_dens_ratios_T} are 
plotted in the 
log[N(\ion{N}{5})/N(\ion{O}{6})]--log[N(\ion{C}{4})/N(\ion{O}{6})]
space for the comparison with halo observations and other models 
(Figure \ref{othermodels_fig}). The averaged value from 
the halo observations along 34 sightlines in 
\citet{IndebetouwShull2004bApJ} is indicated by a filled 
star in Figure \ref{othermodels_fig}. 
The median values in Table \ref{colm_dens_ratios_T}
are plotted as filled (NEI) and empty (CIE) circles. Error bars 
indicate minimum and maximum values.

The median values from our model simulations are with a factor 
of $\sim5$ of the analytic turbulent mixing results of 
\citet{SlavinShullBegelman1993ApJ}. 
The 3--D hydrodynamic and MHD 
simulations of \citet{Esquiveletal2006ApJ} 
in which radiative cooling was allowed 
produce similar column density ratios as our simulations 
although they only used CIE calculations 
for the column density estimations (Figure \ref{othermodels_fig}). 
Note that the bright and dark shaded regions near the center 
of Figure \ref{othermodels_fig} represent their models 
that include radiative cooling, while the shaded regions 
in the upper right in Figure \ref{othermodels_fig} 
represent the results of their models without any radiative cooling. 
(The bright and dark shaded regions indicate their models without 
magnetic field and with magnetic field, respectively.) 
The turbulent mixing phenomenon, 
evaluated by both analytic and numerical 
means yields larger N(\ion{N}{5})/N(\ion{O}{6}) and 
N(\ion{C}{4})/N(\ion{O}{6}) ratios when compared with other 
phenomena such as radiative cooling, supernova remnants, and 
conductive heating (Figure \ref{othermodels_fig}).

The average ion ratios predicted by our models are similar to the 
average from halo observations, however, our models produce 
more \ion{C}{4} and \ion{N}{5}. From our model simulations, 
we find a trend that models with small velocity differences 
between the cool and hot gas (i.e. Model~D) produce 
more \ion{C}{4} and we find that models with higher hot gas 
temperatures (i.e. Model~F) produce 
more \ion{C}{4} and less \ion{N}{5}. In order to 
confirm this trend and to find a case closer to the observed value, 
we run an additional NEI simulation with an initial speed of 
$150~\mbox{km}~\mbox{s}^{-1}$ and a hot gas temperature of
$2.0\times10^6~\mbox{K}$. The result of this additional simulation is 
plotted as a filled square in Figure \ref{othermodels_fig} with the
same convention as the other models of our simulations 
($\log[N(\mbox{\ion{N}{5}})/N(\mbox{\ion{O}{6}})]\approx -0.74$ and 
$\log[N(\mbox{\ion{C}{4}})/N(\mbox{\ion{O}{6}})]\approx 0.17$). 
As expected, the filled square is shifted 
toward the star (observed value) from 
the data points of Model~D and Model~F.

It was pointed out in \citet{IndebetouwShull2004bApJ} that 
the collection of halo observations exhibits a wide range of 
column density ratios (their Fig. 4) that cannot 
be explained by a single type of model. However, as discussed 
in \S \ref{uncertainty_S}, the variations of column 
densities in dynamical models may explain the wide range of 
observed column density ratios. The error bars shown in 
Figure \ref{othermodels_fig} are estimated only for the temporal 
variations after the column densities are averaged over all the 
sightlines at a given time. If the variations along sightlines 
are considered, the column density ratios estimated from our 
simulations are further scattered in the plot.

Finally, it is interesting to compare the results of our simulations 
with the ratios of high ion column densities observed with high
velocities (i.e., highly ionized HVCs). For example, 
\citet{Collinsetal2007ApJ} and \citet{Foxetal2004ApJ} measured 
the high ion ratios along three sightlines toward Complex C. 
The combined values from their measurements are 
$\mbox{N(\ion{C}{4})/N(\ion{O}{6})}=0.44^{+0.06}_{-0.06}$ and  
$\mbox{N(\ion{N}{5})/N(\ion{O}{6})}=0.19^{+0.06}_{-0.07}$ for Mrk~279, 
$\mbox{N(\ion{C}{4})/N(\ion{O}{6})}=0.40^{+0.04}_{-0.04}$ and  
$\mbox{N(\ion{N}{5})/N(\ion{O}{6})}<0.11$ for Mrk~876, and 
$\mbox{N(\ion{C}{4})/N(\ion{O}{6})}=0.35^{+0.05}_{-0.06}$ and 
$\mbox{N(\ion{N}{5})/N(\ion{O}{6})}<0.07$ for PG~1259+593. 
(We choose 
the measured values when they are available from each reference 
article. When the measured values are different in different 
articles, we choose smaller values for the measured ion 
ratios and smaller upper limits for 
N(\ion{N}{5})/N(\ion{O}{6}). Note that the difference between 
tow different values are not significant so that choosing 
larger values would not affect the following comparison.) 
These measurements would be close to the average halo observations 
if they were plotted in Figure \ref{othermodels_fig}. However, the 
measured metallicities of carbon, nitrogen, and oxygen in Complex~C 
are much lower than the solar metallicities. \citet{Foxetal2004ApJ} 
measured the abundance of metals in Complex~C along the sightline 
toward PG~1259+593 (see Table \ref{metallicity_T} for 
their measured carbon, nitrogen, and 
oxygen abundances relative to \citet{Allen1973asqu.book}). 
\citet{Collinsetal2007ApJ} measured the average metallicity of 
Complex~C as 0.13 solar metallicity based upon [\ion{O}{1}/\ion{H}{1}]
measurements along 11 sightlines toward Complex~C. 
They also found that there is much less nitrogen in Complex~C by
factor of 0.01 to 0.07 relative to solar abundance. 

When we apply different metallicities to the ion ratios calculated 
from our simulations, we need to shift the data points of our
simulations according to Table \ref{metallicity_T}. 
Applying the measured abundances of metals in Complex~C 
in \citet{Foxetal2004ApJ} 
to our calculations would shift all of our data
points in Figure \ref{othermodels_fig} by $-0.35$ and $-0.97$ 
along the N(\ion{C}{4})/N(\ion{O}{6}) and 
N(\ion{N}{5})/N(\ion{O}{6}) axes, respectively. However, 
other data points from different models also need to be shifted by the 
same amount. This implies that our turbulent mixing model is more 
likely to explain the observations of highly ionized HVCs than 
are other physical phenomena. But, as mentioned 
before, the results presented in this paper pertain to sightlines 
that perpendicularly intersect the turbulent mixing layers and so 
see nearly stationary velocities of calculated column densities. 
In a future study, we will investigate the velocity--resolved 
column densities, particularly focusing on these HVC observations.

\section{Summary}

We investigate the turbulent mixing layer model by running 
2-D numerical simulations. Our simulations include radiative cooling 
and NEI calculations. NEI calculations produce more high ions 
than CIE calculations: 4.6, 2.9, and 2.3 times higher column densities 
for \ion{C}{4}, \ion{N}{5}, and \ion{O}{6}, respectively 
for our standard model 
(Model~A; Tables \ref{models_T} and \ref{nei_cie_ABC_T}). 
We find that in NEI calculations, 
both ionization and recombination for these ions are delayed, 
resulting in more Li--like ions (\ion{C}{4}, \ion{N}{5}, and
\ion{O}{6}) than are present in the CIE calculations 
(Figure \ref{modelA_ioniz_fig} and \S \ref{neicie_S}). 
Many \ion{C}{4} ions, in particular, are produced even in 
radiatively cooled mixed gas ($T\le 10^4~\mbox{K}$) 
at the base of the mixing zone 
because the recombination of \ion{C}{4} is slow 
in this relatively cool gas. 
These results are also valid in our other NEI simulations, 
namely those with higher resolution (Model~B) and 
1/10 smaller computational domain (Model~C).

By changing the model parameters such as the shear speed, 
the initial amplitude of the interface, and the temperature of hot
gas, we study the various configurations of the turbulent mixing 
layer. We find that more \ion{C}{4} is produced when the shear 
speed is smaller (Model~D) and more \ion{C}{4} and \ion{O}{6} are 
produced when the hot reservoir has a higher temperature (Model~F). 
The initial amplitude does not affect the column densities 
significantly (Model~E).

The study of various configurations shows that the
column densities and corresponding ratios between the 
column density of different ions do not vary greatly when we vary the model
parameters within the reasonable conditions for the mixing layers 
(a few times $10^6$ K for hot gas and a few hundreds
$\mbox{km}~\mbox{s}^{-1}$ for the shear speed). When the results of 
our simulations are plotted 
together with previous model calculations in 
log[N(\ion{N}{5})/N(\ion{O}{6})]--log[N(\ion{C}{4})/N(\ion{O}{6})]
space (Figure \ref{othermodels_fig}), they 
are consistent with the analytic estimations of 
\citet{SlavinShullBegelman1993ApJ} and the 3-D MHD calculations of 
\citet{Esquiveletal2006ApJ}. All turbulent mixing layer calculations 
either analytic or numerical predict more \ion{C}{4} and \ion{N}{5} 
than the other models (radiative cooling, cooling SNR, and conductive 
heating).

We compare NEI and CIE predictions for a subset of our models, 
finding that the CIE calculations predict smaller values of 
N(\ion{C}{4})/N(\ion{O}{6}) and N(\ion{N}{5})/N(\ion{O}{6}) 
(see the open circles for Models~A, B, and C in the 
log[N(\ion{N}{5})/N(\ion{O}{6})]--log[N(\ion{C}{4})/N(\ion{O}{6})] 
plot, Figure \ref{othermodels_fig}). However, the deviation of CIE 
from NEI is not as large as that between different phenomenological 
models, especially between the turbulent mixing layer models 
and the other models, or as large as the variation between different 
sightlines or times in a single model.

Because our NEI simulations are capable of calculating the column
densities of various ions along with velocity information, the 
velocity--resolved column densities, especially in highly ionized 
HVCs, can be studied by using our simulations and comparing 
with observations of previous studies such as 
\citet{Collinsetal2007ApJ} and \citet{Foxetal2004ApJ} and 
future observations from STIS and COS newly installed on HST. We will 
investigate the velocity--resolved column densities in a future 
study.

\acknowledgments

The FLASH code used in this work was in part developed 
by the DOE-supported ASC/Alliance Center for Astrophysical 
Thermonuclear Flashes at the University of Chicago. 
The simulations were performed at the Research Computing Center (RCC) 
of the University of Georgia. 
We appreciate the anonymous referee for his or her valuable comments 
on the radiative cooling and time scales. 
This work was supported through 
grant NNX09AD13G through the NASA ATPF program.

\bibliography{apj-jour,ref_mixing}

\end{document}